\documentclass[final]{cvpr}

\usepackage{times}
\usepackage{epsfig}
\usepackage{graphicx}
\usepackage{amsmath}
\usepackage{amssymb}
\usepackage{multirow}

\usepackage[pagebackref=true,breaklinks=true,colorlinks,bookmarks=false]{hyperref}

\def\cvprPaperID{3715} 
\def\confYear{CVPR 2021}

\begin{document}

\title{Unsupervised Alternating Optimization for Blind Hyperspectral Imagery Super-resolution}

\author{Jiangtao Nie$^*$, Lei Zhang$^*$, Wei Wei$^*$, Zhiqiang Lang, Yanning Zhang\\
School of Computer Science, Northwestern Polytechnical University
\thanks{The first two authors contributed equally. The corresponding author is Wei Wei (email: weiweinwpu@nwpu.edu.cn)}
}

\maketitle

\begin{abstract}

Despite the great success of deep model on Hyperspectral imagery (HSI) super-resolution(SR) for simulated data, most of them function unsatisfactory when applied to the real data, especially for unsupervised HSI SR methods. One of the main reason comes from the fact that the predefined degeneration models (\textit{e.g.}~blur in spatial domain) utilized by most HSI SR methods often exist great discrepancy with the real one, which results in these deep models overfit and ultimately degrade their performance on real data. To well mitigate such a problem, we explore the unsupervised blind HSI SR method. Specifically, we investigate how to effectively obtain the degeneration models in spatial and spectral domain, respectively, and makes them can well compatible with the fusion based SR reconstruction model. To this end, we first propose an alternating optimization based deep framework to estimate the degeneration models and reconstruct the latent image, with which the degeneration models estimation and HSI reconstruction can mutually promotes each other. Then, a meta-learning based mechanism is further proposed to pre-train the network, which can effectively improve the speed and generalization ability adapting to different complex degeneration. Experiments on three benchmark HSI SR datasets report an excellent superiority of the proposed method on handling blind HSI fusion problem over other competing methods.

\end{abstract}

\section{Introduction}
Hyperspectral images (HSIs) contain both spatial information and abundant spectral information, which is beneficial for lots of computer vision tasks such as target detection~\cite{8907742,8784403}, tracking~\cite{8960632, 8435971} and classification~\cite{8753677,8439081}. However, due to the hardware limitation, the observed HSIs always have a high spectral resolution but a low spatial resolution~\cite{8099894}, which impedes the widespread use of HSIs in computer vision related tasks.

\begin{figure}[!tbp]
	\centering
	\hspace{0.8cm}
	\begin{minipage}{0.05\textwidth}
		\centerline{\includegraphics[width=1.05in,height=1.05in]{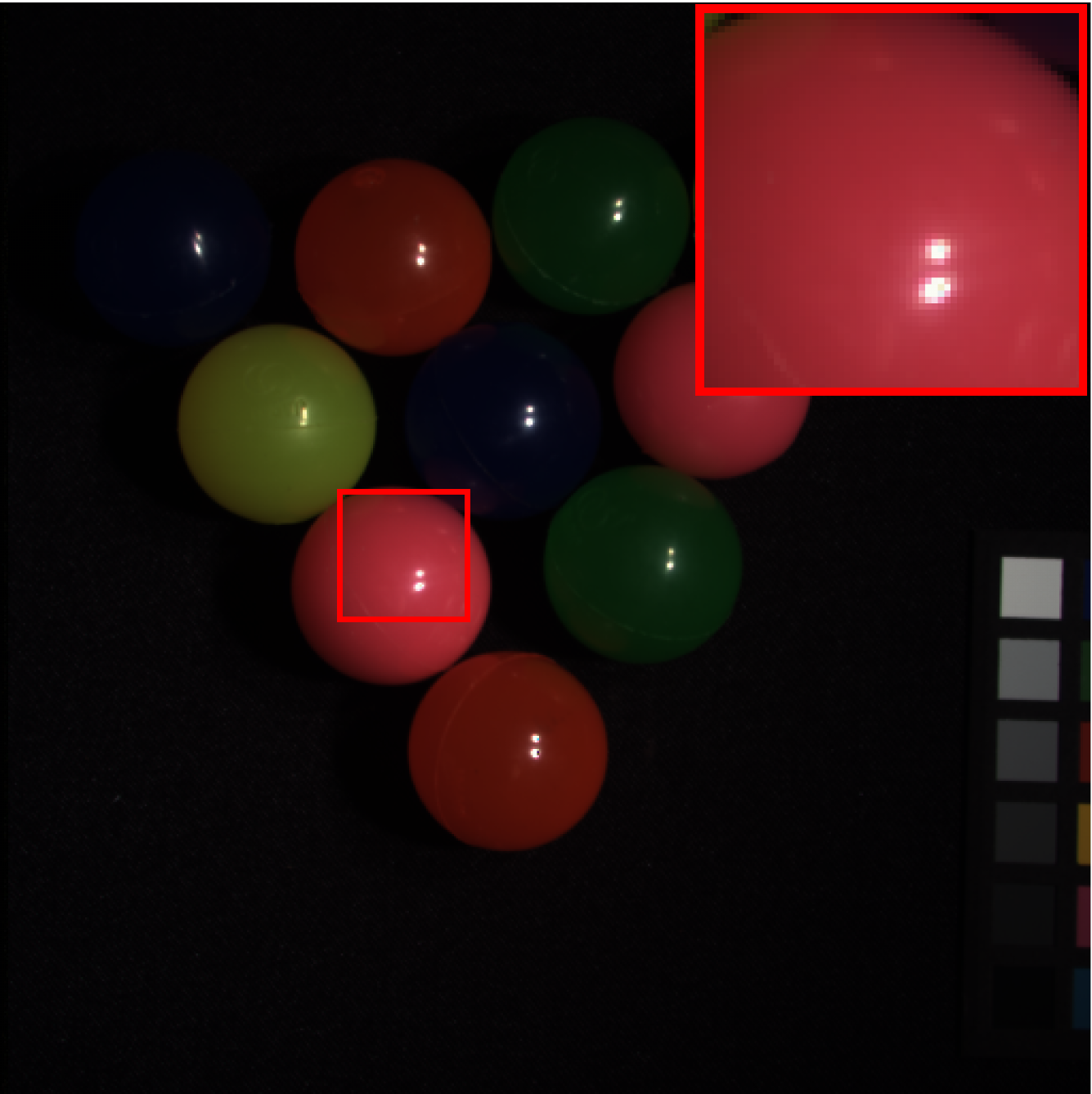}}
	\end{minipage}
	\hspace{1.65cm}
	\begin{minipage}{0.05\textwidth}
		\centerline{\includegraphics[width=1.05in,height=1.05in]{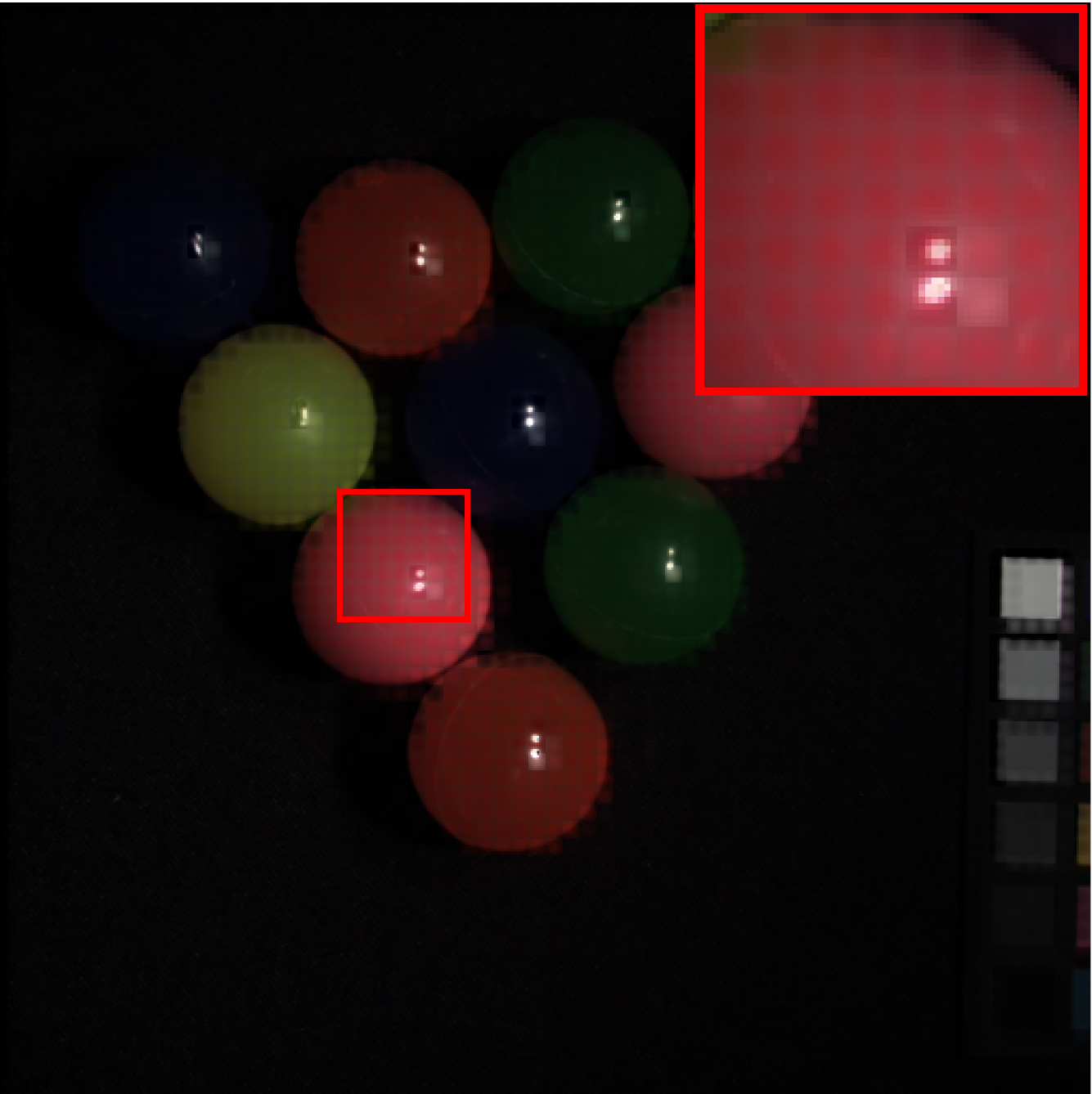}}
	\end{minipage}
	\hspace{1.65cm}
	\begin{minipage}{0.05\textwidth}
		\centerline{\includegraphics[width=1.05in,height=1.05in]{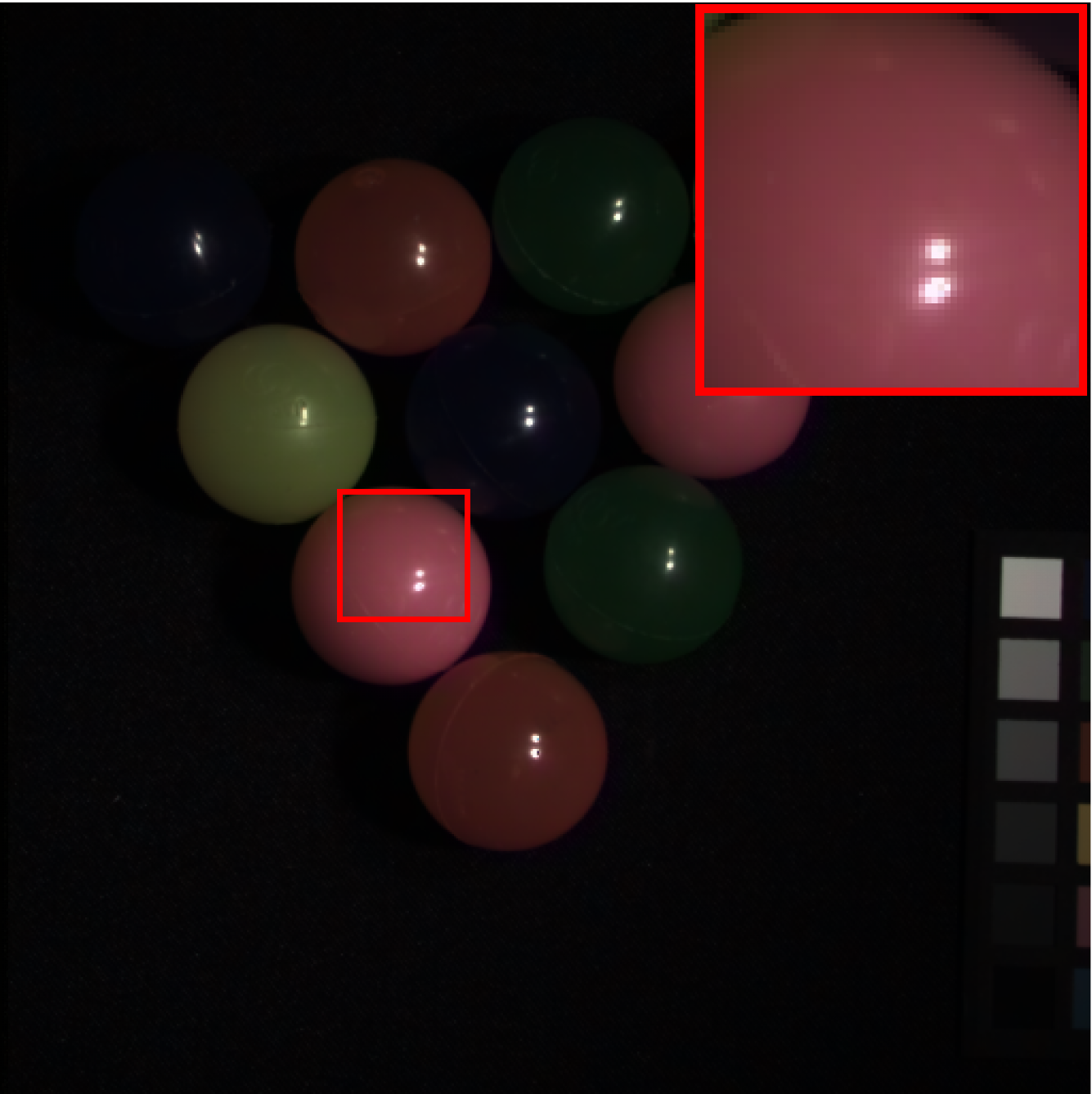}}
	\end{minipage}
	\hspace{0.9cm}
	\begin{minipage}{0.005\textwidth}
		\centerline{\includegraphics[width=0.15in,height=1.05in]{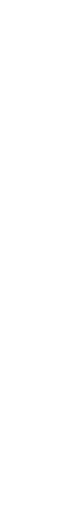}}
	\end{minipage}
	\vfill
	\hspace{0.8cm}
	\begin{minipage}{0.05\textwidth}
		\centerline{\includegraphics[width=1.05in,height=1.05in]{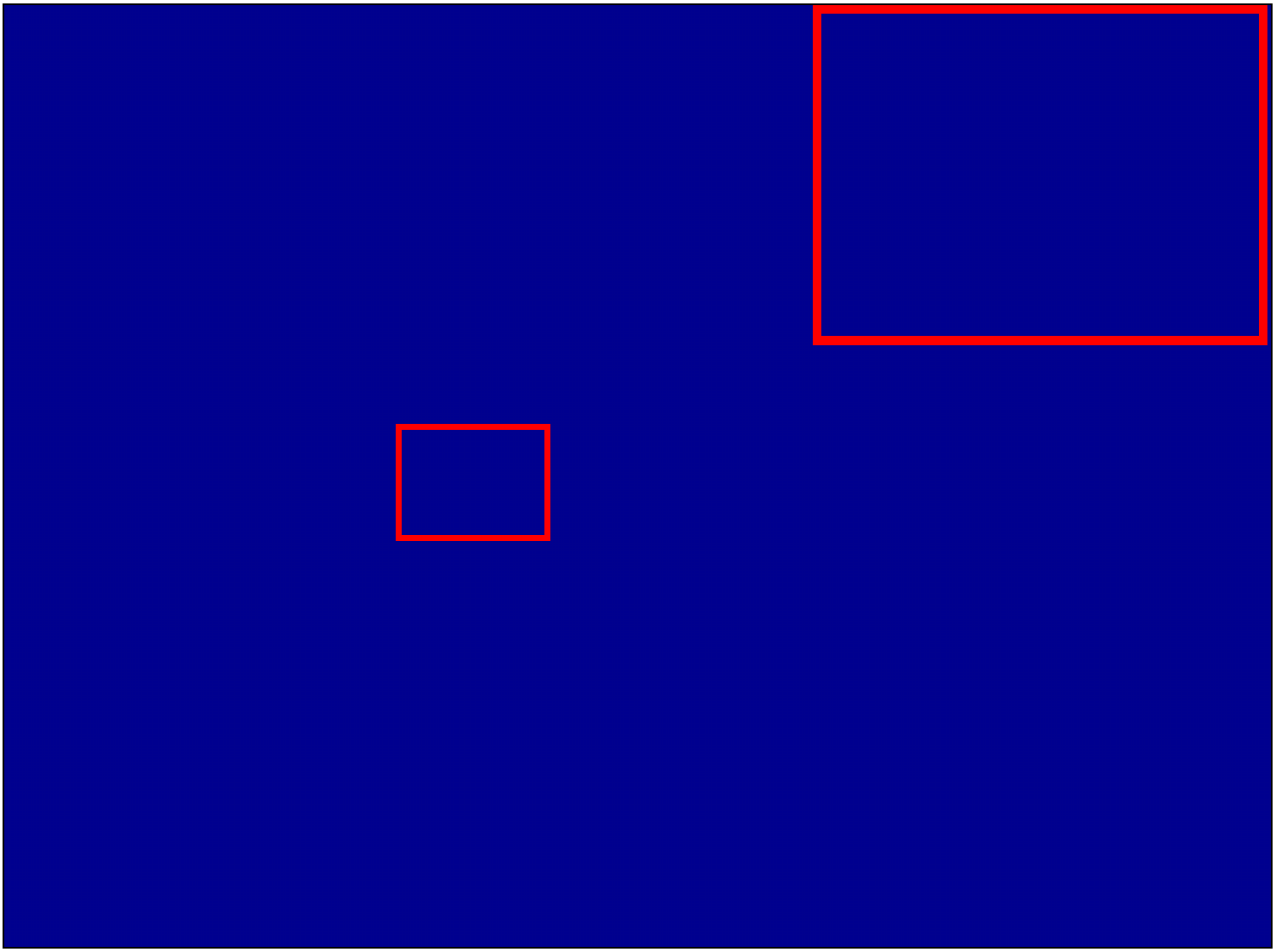}}
		\centerline{{\footnotesize (a) GT}}
	\end{minipage}
	\hspace{1.65cm}
	\begin{minipage}{0.05\textwidth}
		\centerline{\includegraphics[width=1.05in,height=1.05in]{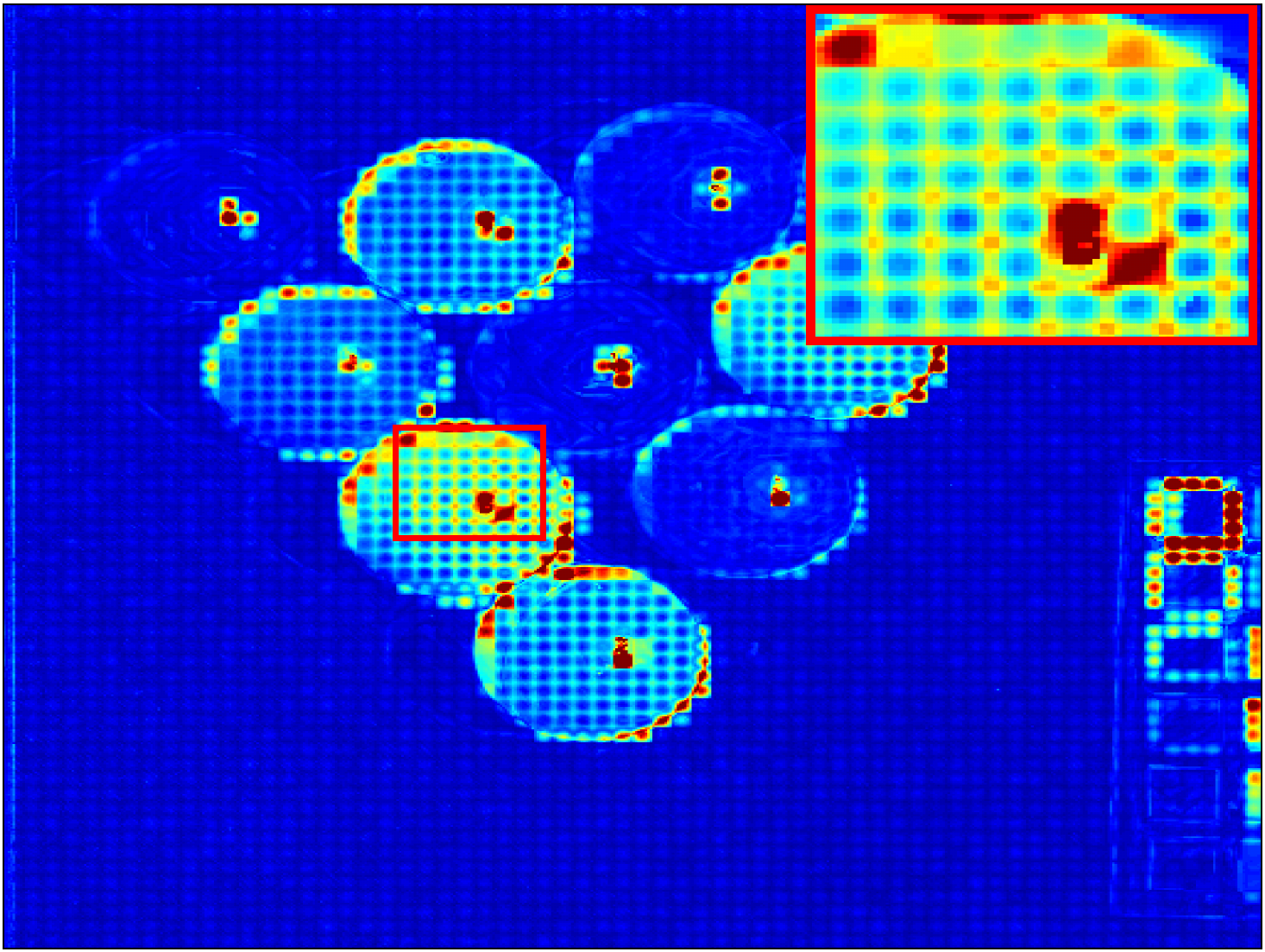}}
		\centerline{{\footnotesize (b) NSSR~\cite{7438864}}}
	\end{minipage}
	\hspace{1.65cm}
	\begin{minipage}{0.05\textwidth}
		\centerline{\includegraphics[width=1.05in,height=1.05in]{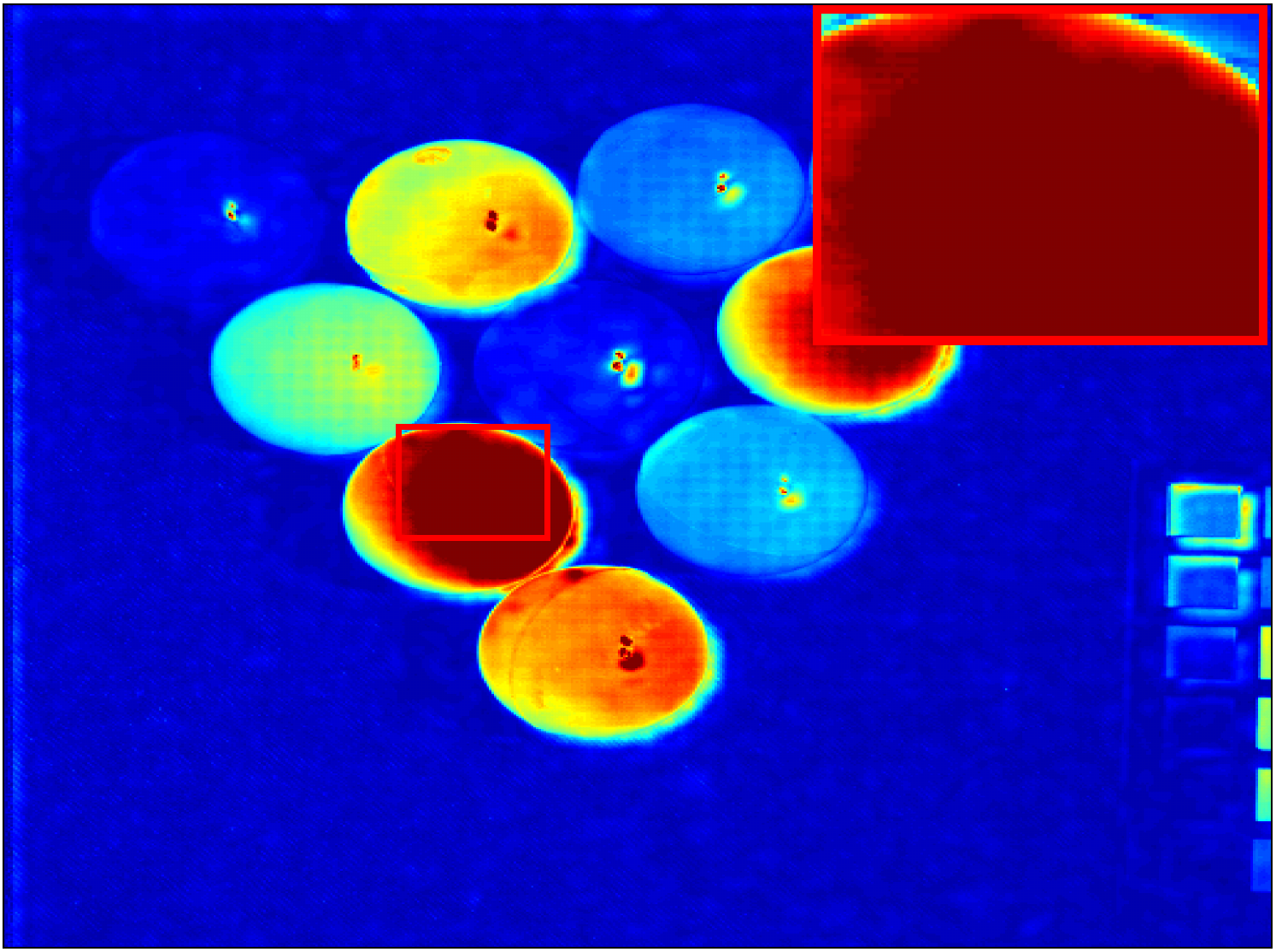}}
		\centerline{{\footnotesize (b) MHFnet~\cite{8953470}}}
	\end{minipage}
	\hspace{0.9cm}
	\begin{minipage}{0.005\textwidth}
		\centerline{\includegraphics[width=0.15in,height=1.05in]{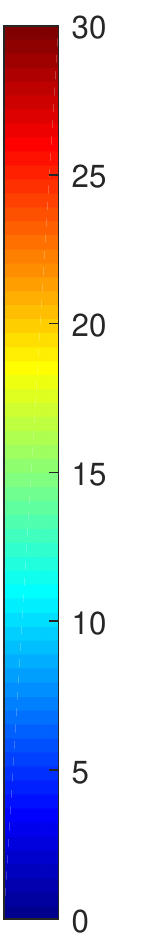}}
		\centerline{}
	\end{minipage}
	\vfill
	\caption{The reconstructed results of NSSR~\cite{7438864} and MHFnet~\cite{8953470} on the HSI '\textit{superballs}' of the CAVE~\cite{5439932} dataset when the  discrepancy between the utilized degeneration models and the real ones exists. 
	The SR scale is 8 and the Signal-to-Noise ratio (SNR) of both observed images are 40dB. The first row provides the pseudo-color maps of the reconstructed HSI and the second row shows the reconstruction error maps. It can be seen that the reconstruction results are not satisfied when the utilized degeneration models bias from the real one.}
	\label{Fig_CAVE_Demo}
	\vspace{-0.3cm}
\end{figure}

Taking advantages of complementarity of HSI and multispectral image (MSI), fusing a spatially low-resolution (LR) HSI $\mathbf{X}$ 
with a spatially high-resolution (HR) MSI $\mathbf{Y}$ 
unsupervisedly has been shown a promising way to obtain an HR HSI $\mathbf{Z}$. 
Different methods are proposed accordingly~\cite{7438864,8019510,8424415,qu2018unsupervised,9044632,Ours_CVPR2020}. Dong \textit{et al.}~develop a clustering-based non-negative structured sparse representation (NSSR~\cite{7438864}) framework to exploit the spatial and spectral characteristics of the observed HSI and MSI simultaneously. Zhang \textit{et al.} propose to fuse the observed images by the clustering manifold structure (CMS~\cite{8424415}), which can well preserve the spatial structure of the input HR MSI. However, the predefined degeneration models utilized by most HSI SR methods often exist great discrepancy with the real one, which results in these methods overfit and ultimately degrade their performance on real data. Specifically, denote the spatial and the spectral degeneration model as 
\begin{equation}
\label{Deg_equ}
\mathbf{X} = (\mathbf{k}*\mathbf{Z})\downarrow_s+ \mathbf{N_X}, \quad  \mathbf{Y} = \mathbf{P}\mathbf{Z} + \mathbf{N_Y}, 
\end{equation}
the blur kernel $\mathbf{k}$ and the spectral response function $\mathbf{P}$ are always assumed given for lots of HSI SR methods, which however, is infeasible for the real applications (within above degeneration models, $\mathbf{N_X}$ and $\mathbf{N_Y}$ are the noise in the observed images, while $\downarrow_s$ indicates the down-sampling operation with a scaling factor $s$). Figure~\ref{Fig_CAVE_Demo} gives an example, in which severe reconstruction errors and color distortion exists for NSSR~\cite{7438864} and MHFnet~\cite{8953470} (two state-of-the-art fusion-based HSI SR method) when inaccurate degeneration models are providing. To fill in this gap, some methods turn to estimate the degeneration model from the data first~\cite{8578442, 9010978,8953757,9156619}, and then utilize it for better HSI SR performance. Despite the advantage estimating the degeneration model from the data, two limitations impede the further improvement of ths kind of HSI SR method. 1) The existing methods estimate the degeneration models with the observed images only, which is hard to obtain an accurate estimation due to the complex and changeable imaging environment. 2) The existing methods estimate the spatial degeneration model only without considering the spectral degeneration model. In these case, these models still will overfit and ultimately degrade their performance on real data.

To address these problems, we investigate how to effectively obtain the degeneration models in both spatial and spectral domain, and makes them can well compatible with the fusion based SR reconstruction model. To this end, we first propose an alternating optimization based deep framework to estimate the degeneration models and reconstruct the latent image, with which the degeneration models estimation and HSI reconstruction can mutually promotes each other. Considering the information can be provided by the observed images is limited, especially for complex blind HSI fusion problem (\textit{e.g.}~HSI SR with a large scaling factor, the given observation images are with heavy noise), we introduce the reconstructed HR HSI with the observed images simultaneously for the degeneration models estimation. Then, a meta-learning based mechanism is further proposed to pre-train the network, which can effectively improve the generalization ability and the speed adapting to different complex degeneration. Extensive experiments report that the proposed method is effective to deal with complex blind HSI fusion problem.



In summary, this study mainly contributes in the following four aspects: 
1) We present an alternating optimization based deep framework to effectively estimate degeneration models as well as reconstruct the latent HR HSI. 
2) We propose a compact reconstruction network, which can further utilize the information in degeneration models to guide the reconstructing of the latent HSI. 
3) We further adopt the meta-learning technology to enhance the generalization ability of the proposed reconstruction network, and improve the speed adapting to different complex degeneration.
4) Extensive experiment results demonstrate the proposed method can achieve state-of-the-art  performance in terms of blind HSI fusion SR problem.

\section{Related works}

\subsection{Blind image SR methods}
Blind SR methods~\cite{8953544,9136736, 9157144, 9156619} aim to alleviate the performance drops that caused by the inexact degeneration model. In traditional blind single image (\textit{e.g.}~RGB image) SR (SISR) methods~\cite{8953544,8953757,9010978,9009805}, they only need to take care of the spatial degeneration model (\textit{i.e.}~blur kernel $\mathbf{k}$ is unknown). In~\cite{9156619}, Hussein \textit{et al.}~propose to utilize a correction filter to match the test images close to the data that generated by known blur kernels (\textit{e.g.}~bicubic). Gu \textit{et al.}~\cite{8953544} develop a supervised end-to-end network to iteratively estimating the blur kernel and reconstructing the latent image. Though these methods incorporating the estimation of blur kernel into SR, they just utilize the observed images to estimate the blur kernel, which is prone to degrade the performance of complex blind HSI fusion problem (\textit{e.g.}~HSI SR with a large scaling factor, the given observation images are with heavy noise). 

\begin{figure*}[!htbp]
	\hspace{4.5cm}
	\begin{minipage}{0.05\textwidth}
		\centerline{\includegraphics[width=4in,height=1.8in]{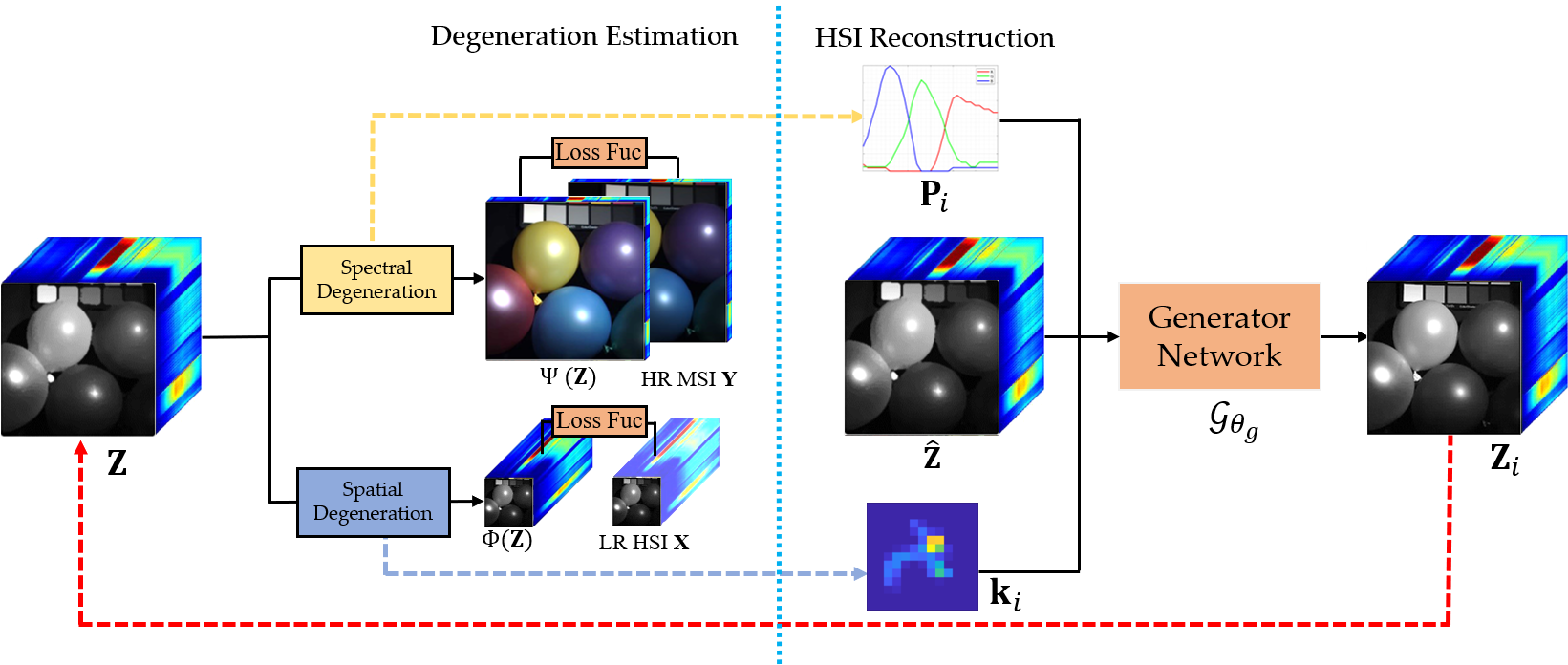}}
	\end{minipage}
	\hspace{8cm}
	\begin{minipage}{0.05\textwidth}
		\centerline{\includegraphics[width=2.4in,height=1.4in]{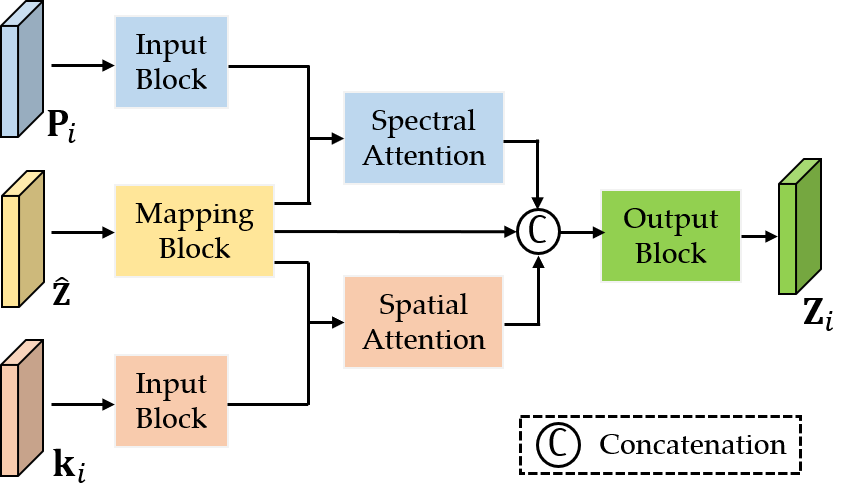}}
	\end{minipage}
	\vfill
	\hspace{4.5cm}
	\begin{minipage}{0.05\textwidth}
		\centerline{{\small (a) The flowchart of the proposed alternating optimization framework.}}
	\end{minipage}
	\hspace{8cm}
	\begin{minipage}{0.05\textwidth}
		\centerline{{\small (b) The structure of the reconstruction network.}}
	\end{minipage}
	\caption{The flowchart of the proposed method and the proposed reconstruction network. In figure (a), the solid lines indicate the data stream of the network interior and the dotted lines denote the information transformation when alternating optimization.}
	\label{Flowchart}
\end{figure*}

As for fusion based method, there are few methods~\cite{9009078,9136736} taking the blind HSI SR problem into consideration. Wang \textit{et al.}~apply a idea similar as ~\cite{8953544,8953757} to train a supervised end-to-end iterative network for solving the blind HSI fusion problem. So far, most of HSI SR methods are supervised which need plenty paired data to train a network. However, the HSI datasets often contain few images which is hard to support large scale supervised training. Therefore, how to unsupervised exploit the image priors~\cite{7438864,8424415,qu2018unsupervised,9044632} that contains in observed images is a promising way to reconstruct the latent HSI. 
In~\cite{9136736}, Zhang \textit{et al.}~propose to joint optimizing the degeneration model and reconstruction model in an unsupervised manner. However, this method demands large amount of computations. In addition, the network builds directly with the observed images without pre-training with the existing available HSI, which thus has a slow training speed and lower generalization ability to complex degradation. In this study, we first parameterize the key components of degeneration model by one single network layer (\textit{e.g.}~2-D convolution layer), and adopt an unsupervised regularization to train the degeneration models. Then, we introduce the alternating optimization mechanism to optimize the degeneration model and reconstruction model, the experiment results demonstrate that the proposed method can well estimate the degeneration model.


\section{Method}

\subsection{Problem formulation} 
Unsupervised fusion-based HSI SR method is to inversely reconstruct the latent HR HSI $\mathbf{Z} \in \mathbb{R}^{B\times N}$ by jointly exploiting the LR HSI $\mathbf{X}\in \mathbb{R}^{B\times n}$ and HR MSI $\mathbf{Y}\in \mathbb{R}^{b\times N}$. $B$, $b$ denote the number of spectral bands, $N$, $n$ represent the number of pixels in spatial domain,  and $b \ll B$, $n \ll N$. With the degeneration model (\textit{i.e.}~the observation model) described in Eq.\eqref{Deg_equ}, a general maximum a posteriori (MAP) estimation is always utilized for fusion-based HSI SR as 
\begin{equation}
\label{MAP_based_equ}
\min_{\mathbf{Z}}\,\,\, \|\mathbf{X} - (\mathbf{k*Z})\downarrow_s\|^{2} + \|\mathbf{Y} - \mathbf{PZ}\|^{2} + \lambda\mathcal{R}(\mathbf{Z}),
\end{equation}
where the first two terms are the data fidelity terms, and $\mathcal{R}(\mathbf{Z})$ is the image prior (regularizer) on $\mathbf{Z}$ for optimization. However,      
the premise of Eq.\eqref{MAP_based_equ} is that the utilized degeneration models are well consistent with the real ones. Thus, the HSI SR is regarded as a non-blind optimization problem, \textit{i.e.,}~the optimization of $\mathbf{Z}$ is accomplished in Eq.\eqref{MAP_based_equ} without considering the degeneration models.

\subsection{Alternating optimization based learning}
Recent studies on traditional blind SISR~\cite{8953544,Luo2020UnfoldingTA,9010978} reveals that adding a step of estimating the degeneration model is benefit to promote the performance of blind SR. Following this idea, we can model the blind fusion-based HSI SR method as
\begin{equation}
\label{Two_Step}
\left\{
\begin{array}{l}
\mathbf{k} = E_k(\mathbf{X}), \quad \mathbf{P} = E_p(\mathbf{Y}), \\
\begin{split}
\mathbf{Z} = \mathop{\arg\min}\limits_{\mathbf{Z}}\, &\|\mathbf{X} - (\mathbf{k*Z})\downarrow_s\|^{2} + \\
&\|\mathbf{Y} - \mathbf{PZ}\|^{2} +\lambda\mathcal{R}(\mathbf{Z}).
\end{split}
\end{array}
\right.
\end{equation}
Within this model, we can first estimate the degeneration models by $E_k(\cdot)$ and $E_p(\cdot)$ from the observed data $\mathbf{X, Y}$, respectively. Then, we can utilize the estimated degeneration models to reconstruct the HR HSI $\mathbf{Z}$. Despite the advantage Eq.\eqref{MAP_based_equ} estimates both the spatial and spectral degeneration models (i.e., $E_k(\cdot)$ and $E_p(\cdot)$), the following two limitations impede the further improvement of such a blind fusion-based HSI SR method.  
First, $E_k(\cdot)$ and $E_p(\cdot)$ are estimated with the observed images $\mathbf{X, Y}$ only, which is insufficient to obtain an accurate estimation of the degeneration models, especially for complex blind HSI fusion problem (\textit{e.g.}~HSI SR with a large scaling factor, the given observation images are with heavy noise). Second, the estimation of $E_k(\cdot)$ and $E_p(\cdot)$ is independent with the estimation of $\mathbf{Z}$, which makes the estimated degeneration models still cannot well fit the estimation of $\mathbf{Z}$.

To boost the performance of HSI fusion method, we further propose a blind fusion-based HSI SR method within alternating optimization based framework~\cite{Luo2020UnfoldingTA}. First, we reformulate the degeneration models into learnable networks as Eq.\eqref{Deg_equ_Net} for better estimation. Specifically, the spatial degeneration model is reformulate by a network $\Phi(\cdot)$ parameterized by $\mathbf{k}$, which can be implemented by a 2-D convolutional layer with kernel as $\mathbf{k}$. While the spectral degeneration model is reformulated by a network $\Phi(\cdot)$ parameterized by $\mathbf{P}$, which is implemented by a fully connection layer with $\mathbf{P}$ as its parameter. It is noticable that both $\Phi(\cdot)$ and $\Phi(\cdot)$ 
relate to the HR HSI $\mathbf{Z}$, which provides richer information for estimating the degeneration model as well as integrating the estimation of the degeneration model and estimating $\mathbf{Z}$ into a close loop.   
\begin{equation}
\label{Deg_equ_Net} 
\mathbf{X} = \Phi(\mathbf{Z; k}) + \mathbf{N_X}, \quad
\mathbf{Y} = \Psi(\mathbf{Z; P}) + \mathbf{N_Y}.
\end{equation}

Then, we can optimize $\mathbf{k}$, $\mathbf{P}$ as well as $\mathbf{Z}$ jointly in one unified framework taking advantages of modeling $\Phi(\cdot)$ and $\Phi(\cdot)$. Considering simultaneously optimizing $\mathbf{k}$, $\mathbf{P}$ as well as $\mathbf{Z}$ is difficult, an alternating optimization based blind HSI SR framework is formulated as    

\begin{equation}
\label{AlterOpri_Eq}
\left\{
\begin{array}{l}
\mathbf{k}_{i+1} = \mathop{\arg\min}\limits_{\mathbf{k}} \|\mathbf{X} - \Phi(\mathbf{Z}_i)\|^{2} + \eta\|\mathbf{k}\|^2,\\

\mathbf{P}_{i+1} = \mathop{\arg\min}\limits_{\mathbf{P}} \|\mathbf{Y} - \Psi(\mathbf{Z}_i)\|^{2} + \xi \|\mathbf{P}\|^2, \\

\begin{split}
\mathbf{Z}_{i+1} = \mathop{\arg\min}\limits_{\mathbf{Z}}\, &\|\mathbf{X} - \Phi_i(\mathbf{Z})\|^{2} + \\
&\|\mathbf{Y} - \Psi_i(\mathbf{Z})\|^{2} + \lambda\mathcal{R}(\mathbf{Z}).
\end{split}
\end{array}
\right.
\end{equation}

By alternatively optimizing the degeneration models and the reconstruction model (\textit{i.e.}~optimization of $\mathbf{Z}$), the errors between the estimated degeneration model and the real one can be reduced gradually. Therefore, the reconstruction model can benefit from the exactly estimated degeneration models and thus improve the performance of blind HSI SR.

\subsection{Efficient deep HSI reconstruction network with deep image prior}
Eq.\eqref{AlterOpri_Eq} provides an effective strategy to incorporate the estimation of degeneration models into the reconstruction model. However, the reconstruction model (\textit{i.e.}~optimization of $\mathbf{Z}$) is still challenging due to the ill-posed nature. To well address such an ill-posed inverse problem, proper image priors such as sparsity prior~\cite{7438864,8019510} are modeled to regularize the solution of $\mathbf{Z}$ during its optimization. Recently, deep image prior (DIP)~\cite{8579082}  is proposed to utilize a generation network to represent the image prior, which shows the state-of-the-art performance for image restoration. The advantage of DIP is that it can capture plenty of low-level statistics which can better reflect the image characteristics and thus well regularize the inverse problem. However, it consumes too much time during optimizing. We analyze the reason as follows. For DIP based SR method~\cite{8900117,9136736}, it utilizes the random noise for initialization, and then gradually approximates the optimum solution (\textit{i.e.}~the final reconstructed image). Due to the huge differences between the random noise and the optimum solution, the method with DIP will need a complex and deep network as the generator network, which consumes too much time on optimizing. Thus, a natural idea is the difficulty of training as well as the time for optimization can be tremendously decreased if we can provide a good estimated image instead of random noise for initialization.

Therefore, we propose to pre-train a network which can provide a rough reconstructed image as the initial input of the DIP based HSI reconstruction network. Instead of using the observed images for pre-training, we pre-train the network with the existing available amounts of HSI data in the external dataset (\textit{e.g.}~CAVE dataset) since they can provide rich low-level features which always repeatedly appear in different images. For this purpose, we first pre-train a backbone network $\mathcal{F}$ to extract the general image priors within external dataset for roughly reconstructing the latent HSI, which is trained via solving the following problem.
\begin{equation}
\label{Fusion_Train}
\begin{split}
\min_{\theta_f} \sum_{a,b,c} &\|\mathbf{Z}_a -\hat{\mathbf{Z}}_{abc}\|^2 ,\,\,
{\rm{s.t.}}, \hat{\mathbf{Z}}_{abc}=\mathcal{F}(\mathbf{X}_{ab}, \mathbf{Y}_{ac};\theta_f),\\ 
&\mathbf{X}_{ab}=(\mathbf{k}_b*\mathbf{Z}_a)\downarrow_s, \,\, \mathbf{Y}_{ac} =\mathbf{P}_c\mathbf{Z}_a, 
\end{split}
\end{equation} 
${\mathbf{Z}_a}$ denotes the HSI within the external datasets, while ${\mathbf{X}_{ab}}$ and ${\mathbf{Y}_{ac}}$ are the degenerated images obtained by applying $\mathbf{k}_b$ and $\mathbf{P}_c$ on ${\mathbf{Z}_a}$. $\hat{\mathbf{Z}}_{abc}$ is the roughly reconstructed HR HSI and $\theta_f$ is the parameter of the backbone network. Since $\mathcal{F}$ is used for generating a rough reconstructed HSI, most existing HSI fusion networks structure can be adopted as $\mathcal{F}$. 

Once obtaining the roughly reconstructed HSI via $\mathcal{F}$, we then utilize it as the initialized input for the generation network of DIP based method. The generation network in this study is termed as the reconstruction network since it is utilized to generate $\mathbf{Z}$. Considering HSI contains both spectral and spatial information, we develop a compact reconstruction network shown as Figure~\ref{Flowchart}(b) to further exploit the additional information contains in $\mathbf{k}$ and $\mathbf{P}$, in which $\mathbf{k}$ and $\mathbf{P}$ provide extra guidance information for the reconstruction network. Specifically, $\mathbf{k}$ is utilized with $\hat{\mathbf{Z}}$ to generate features focusing on spatial information, while $\mathbf{P}$ together with $\hat{\mathbf{Z}}$ aims to generate features relate with spectral information. 

Based on above defined backbone as well as the reconstruction network, the alternating optimization based blind HSI fusion model (depicted in Eq.\eqref{AlterOpri_Eq}) can be illustrated in Figure~\ref{Flowchart}(a) and further expressed as 
\begin{equation}
\label{Final_Eq}
\left\{
\begin{array}{l}
\mathbf{k}_{i+1} = \mathop{\arg\min}\limits_{\mathbf{k}} \|\mathbf{X} - \Phi(\mathbf{Z}_i)\|^{2} + \eta\|\mathbf{k}\|^2,\\

\mathbf{P}_{i+1} = \mathop{\arg\min}\limits_{\mathbf{P}} \|\mathbf{Y} - \Psi(\mathbf{Z}_i)\|^{2} + \xi\|\mathbf{P}\|^2, \\

\begin{split}
\mathbf{Z}_{i+1} = &\mathop{\arg\min}\limits_{\mathbf{Z}, \theta_g}\, \|\mathbf{X} - \Phi_i(\hat{\mathbf{Z}} )\|^{2} + \|\mathbf{Y} - \Psi_i(\hat{\mathbf{Z}})\|^{2}, \\
&s.t., \mathbf{Z}=\mathcal{G}(\hat{\mathbf{Z}}, \mathbf{k},\mathbf{P};\theta_g), \hat{\mathbf{Z}}=\mathcal{F}(\mathbf{X}, \mathbf{Y};\theta_f),
\end{split}
\end{array}
\right.
\end{equation}
where $\mathcal{G}$ denotes the proposed reconstruction network parameterized by $\theta_g$. 

Considering the potential $\mathbf{k}$ and $\mathbf{P}$ within Eq.\eqref{Final_Eq} is diverse, we finally introduce the meta-learning mechanism~\cite{9157144} to enhance the learning ability of the reconstruction network to different $\mathbf{k}$ and $\mathbf{P}$. To this end, we synthesize plenty of degenerated images from HR HSIs by combining different $\mathbf{k}$ and $\mathbf{P}$, and forming these data into a meta-learning training dataset $\mathcal{M}$. Then, we pre-train the reconstruction network via the Model-Agnostic Meta-Learning (MAML) method proposed in~\cite{MAML_2020_CVPR}, in which the meta-objective function is defined as
\begin{equation}
\mathop{\arg\min}\limits_{\theta_g} \sum_{\mathcal{M}_j\sim p(\mathcal{M})} \mathcal{L}_{\mathcal{M}_j}^{te}(\theta_g - \alpha\nabla_{\theta_g}\mathcal{L}_{\mathcal{M}_j}^{tr}(\theta_g)),
\end{equation}
$\mathcal{L}_{\mathcal{M}_j}^{te}$ and $\mathcal{L}_{\mathcal{M}_j}^{tr}$ denote the loss of test and training on sub-task $\mathcal{M}_j$, and $\alpha$ is the task-level learning rate. With the meta-learning mechanism, we can obtain a better initialization value of the reconstruction network $\mathcal{G}$, which results in 
better generalization ability and faster learning speed. The details analysis can be seen from the ablation study. 

\section{Experimental results and analysis}

\subsection{Experimental settings}
\noindent \textbf{Datasets}\quad
In this study, we utilize two benchmark HSI datasets and one real HSI dataset to testify the effectiveness of the proposed method, including CAVE~\cite{5439932}, Harvard~\cite{5995660} and HypSen~\cite{rs10050800}. Specifically, the CAVE dataset contains 32 indoor HSIs and each of which has 31 spectral bands covering the wavelength range from 400 $nm$ to 700 $nm$ with a 10 $nm$ interval. Meanwhile, each HSI in the CAVE dataset contains $512 \times 512$ pixels in the spatial domain. The Harvard dataset includes 50 HSIs containing both indoor and outdoor scenes. Each HSI in the Harvard dataset also has 31 bands in a wavelength range from 420 $nm$ to 720 $nm$, and $1392 \times 1300$ pixels in spatial domain. Different with the above two datasets, HypSen is a real fusion-based HSI dataset containing a 10m-resolution MSI and a 30m-resolution HSI. After removing the noisy bands as well as the water-absorbed bands as~\cite{rs10050800}, the resulted LR HSI includes 84 spectral bands and HR MSI consists of 13 spectral bands. In the experiment on the HypSen dataset, a sub-image with $250 \times 330$ spatial size is cropped from the LR HSI and the one with $750 \times 990$ spatial size is cropped from the HR MSI.

\noindent \textbf{Comparison methods and Evaluation Metrics}\quad
In this study, due to the lack of blind fusion-based HSI SR method, five state-of-the-art fusion-based HSI SR methods including two non-blind method, two semi-blind method and one blind method are utilized for comparison. The first two methods are non-blind HSI fusion methods, in which NSSR~\cite{7438864} is a non-deep unsupervised method with optimization, and MHFnet~\cite{8953470} is a supervised deep learning based method. Yong~\cite{8019510} and UAL~\cite{Ours_CVPR2020} are semi-blind methods, in which $\mathbf{P}$ is provided and only $\mathbf{k}$ of the spatial degeneration is unknown. 
The last method DBSR~\cite{9136736} is a DIP~\cite{8579082} based unsupervised blind method, which simultaneous optimize the degeneration estimation as well as the HSI reconstruction within a jointly framework.

Four quantitative metrics including root-mean-square error (RMSE), peak signal-to-noise ratio (PSNR), spectral angle mapper (SAM) and structural similarity index (SSIM) is utilized to evaluate the SR performance.

\noindent \textbf{Implementation details} \quad
In the experiments, the HSIs in CAVE and Harvard datasets are regarded as the ground truth of the latent HR HSI $\mathbf{Z}$. The corresponding LR HSI and HR MSI for pre-training in these two datasets are generated from the latent HR HSI $\mathbf{Z}$ based on Eq~\eqref{Deg_equ}, among which the spectral response function of Nikon D700 camera in~\cite{7438864,8019510} and Gaussian kernel is adopted as $\mathbf{k}$ and  $\mathbf{P}$. 

\begin{equation}
\mathbf{P}_c = Softmax(\mathbf{P} + c).
\label{Gen_P}
\end{equation}

For the observed images $\mathbf{X}$ and $\mathbf{Y}$, we generate them with the blur kernels and the spectral response function, different with $\mathbf{k}$ and $\mathbf{P}$ utilized for the pre-training data. Specifically, four different motion kernels includes $\mathbf{k_1}$, $\mathbf{k_2}$, $\mathbf{k_3}$ and $\mathbf{k_4}$ (shown in the Figure~\ref{Fig_K_Estimation}) instead of Gaussian kernel to generate the observed LR HSI $\mathbf{X}$. While, Eq~\eqref{Gen_P} is utilized to generate the spectral response matrix $\mathbf{P_c}$ of the observed HR MSI, in which $c$ is a transform coefficient and $\mathbf{P}$ is the spectral response matrix for pre-train. In addition, to simulate the noise disturbance, we add Additive White Gaussian Noise with different intensities into the observed image $\mathbf{X}$ and $\mathbf{Y}$. Some other detailed settings are given in the supplement due to the limitations of page length.

\begin{table}[!htbp]\small
	\centering
	\caption{The performance of each method on the CAVE dataset with different settings. SNR of both two observed images are 40dB, and the SR scale is 8. In experiments, the input LR HSI and HR MSI are generated by $\mathbf{k}_1$ and $\mathbf{P}_{0.02}$, respectively. The best results are in bold.}
	\vspace{0.1cm}
	\setlength{\tabcolsep}{1.5mm}{\begin{tabular}{c|c|c|c|c}
			\hline\hline
			Methods&RMSE&PSNR&SAM&SSIM\\
			\hline\hline
			Ours\_Sep&5.72&34.13&8.98&0.9629\\
			\hline
			Ours\_Joint&3.87&37.06&6.95&0.9828\\
			\hline
			Ours\_Alter&2.97&39.38&6.75&0.9838\\
			\hline
			Ours\_Basic&3.03&39.19&6.80&0.9841\\
			\hline
			Ours&\textbf{2.86}&\textbf{39.74}&\textbf{6.64}&\textbf{0.9845}\\
			\hline\hline
	\end{tabular}}
	\label{Tab_CAVE_Ablation}
	\vspace{-0.3cm}
\end{table}

\begin{table*}[!htbp]\small
	\centering
	\caption{The performance of each method on the CAVE dataset with different SR scale factors. 
		The best results are in bold.}
	\setlength{\tabcolsep}{2.4mm}{\begin{tabular}{c|c|c|c|c|c|c|c|c|c|c|c|c}
			\hline
			\multirow{2}*{Methods}& \multicolumn{4}{c|}{s = 8}&\multicolumn{4}{c|}{s = 16}& \multicolumn{4}{c}{s = 32}\\
			\cline{2-13}&RMSE&PSNR&SAM&SSIM&RMSE&PSNR&SAM&SSIM&RMSE&PSNR&SAM&SSIM\\
			\hline	
			NSSR~\cite{7438864}&8.11&30.53&18.85&0.8533&8.31&30.27&14.75&0.8976&7.87&30.68&15.89&0.8287\\
			MHFnet~\cite{8953470}&4.53&35.67&14.01&0.9409&4.74&35.44&11.75&0.9547&5.17&34.80&12.85&0.9524\\
			\hline
			Yong~\cite{8019510}&4.46&35.72&11.29&0.9491&4.98&34.88&11.80&0.9468&6.34&32.72&14.04&0.9318\\
			UAL~\cite{Ours_CVPR2020}&3.93&37.07&7.57&0.9724&4.12&36.59&8.06&0.9704&4.59&35.61&9.21&0.9652\\
			\hline
			DBSR~\cite{9136736}&4.01&36.66&\textbf{5.53}&0.9739&5.10&34.32&9.80&0.9482&8.96&29.49&13.95&0.9054\\
			\hline
			Ours&\textbf{2.40}&\textbf{41.12}&6.53&\textbf{0.9854}&\textbf{2.67}&\textbf{40.27}&\textbf{6.93}&\textbf{0.9838}&\textbf{3.47}&\textbf{38.41}&\textbf{8.46}&\textbf{0.9755}\\
			\hline
	\end{tabular}}
	\label{Tab_CAVE_Result_Diff_Scale}
\end{table*}

\begin{figure*}[!ht]
	\centering
	\vspace{-0.4cm}
	\hspace{0.55cm}
	\begin{minipage}{0.05\textwidth}
		\centerline{\includegraphics[width=0.94in,height=0.94in]{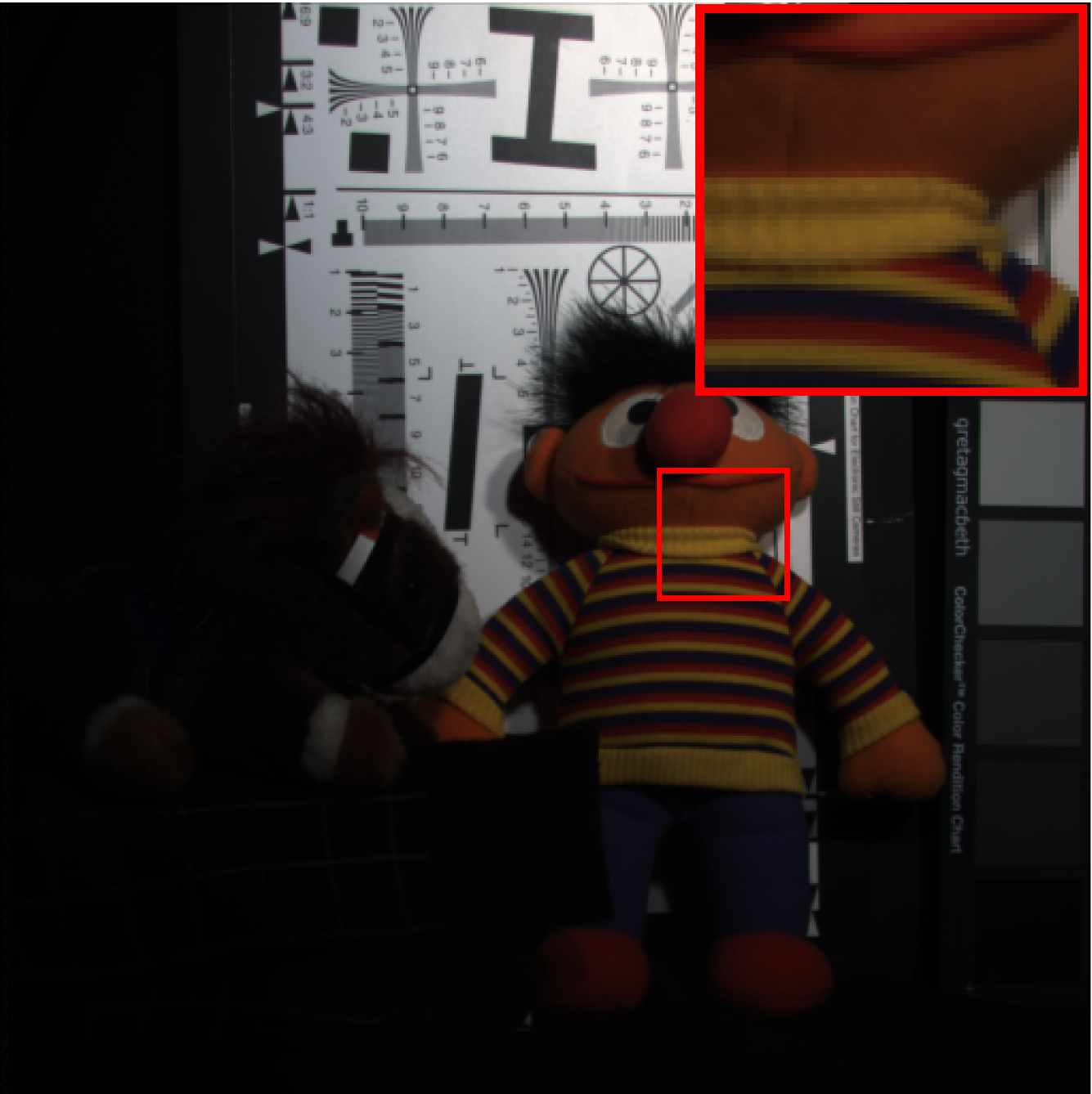}}
	\end{minipage}
	\hspace{1.4cm}
	\begin{minipage}{0.05\textwidth}
		\centerline{\includegraphics[width=0.94in,height=0.94in]{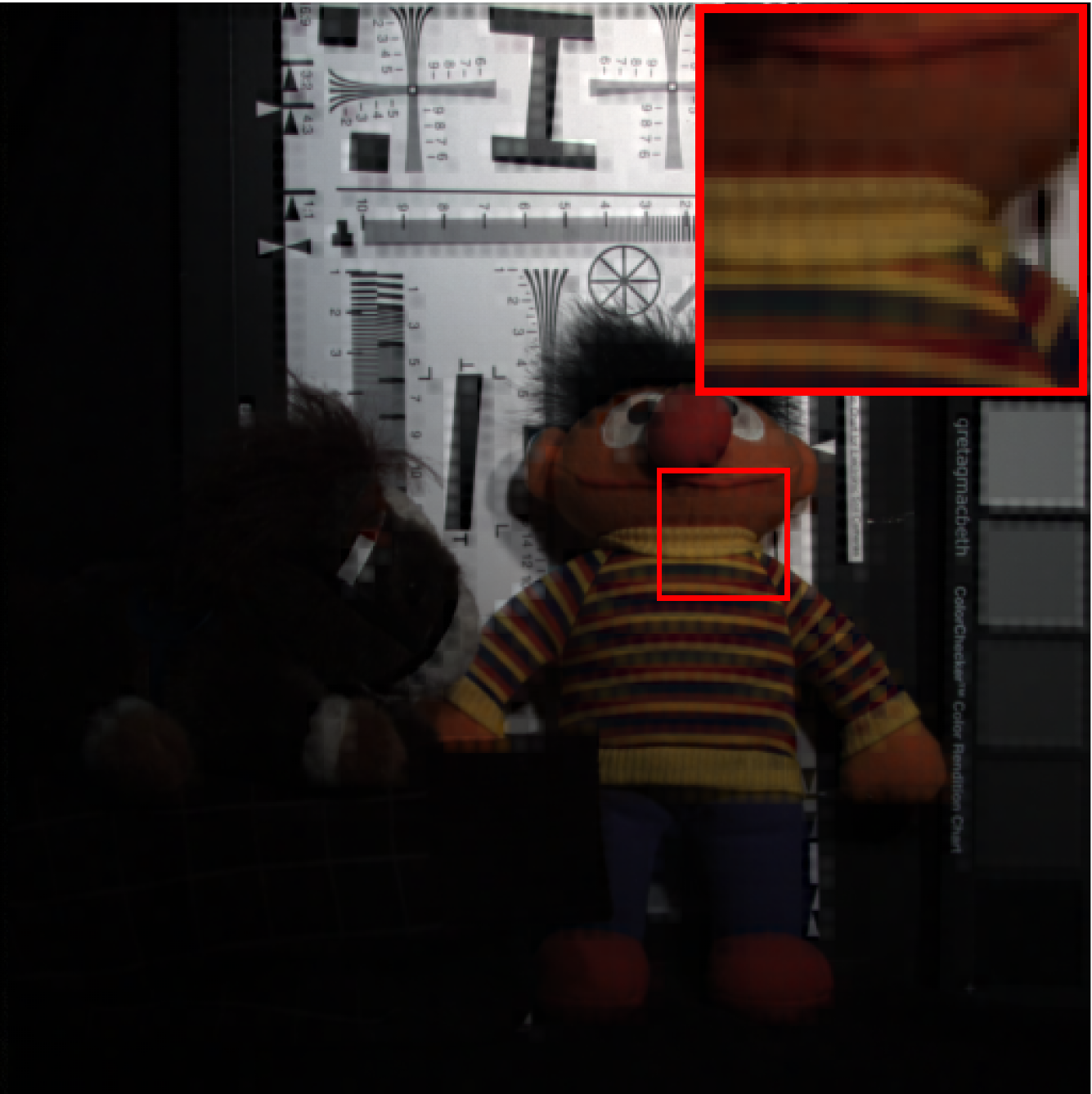}}
	\end{minipage}
	\hspace{1.4cm}
	\begin{minipage}{0.05\textwidth}
		\centerline{\includegraphics[width=0.94in,height=0.94in]{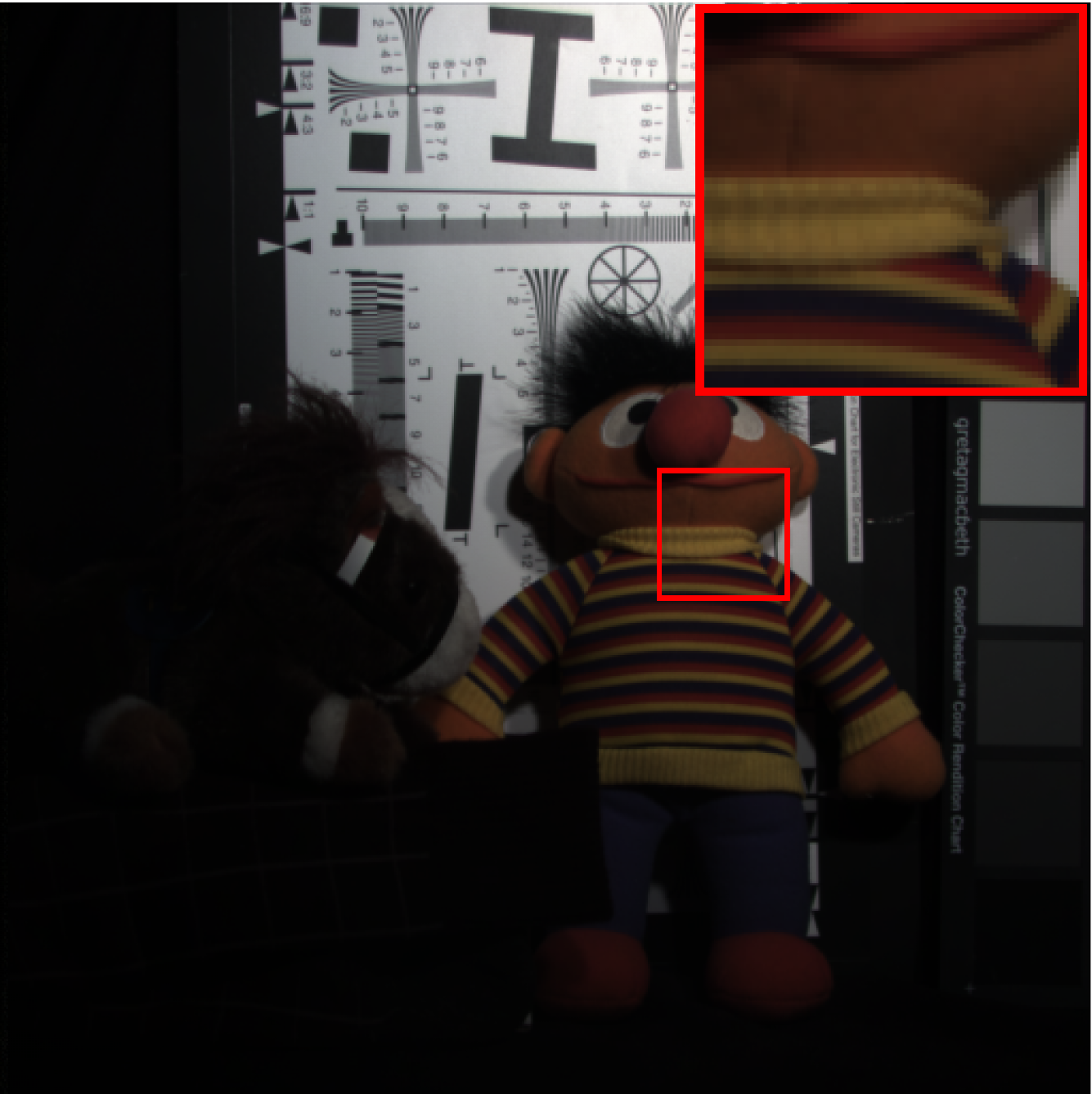}}
	\end{minipage}
	\hspace{1.4cm}
	\begin{minipage}{0.05\textwidth}
		\centerline{\includegraphics[width=0.94in,height=0.94in]{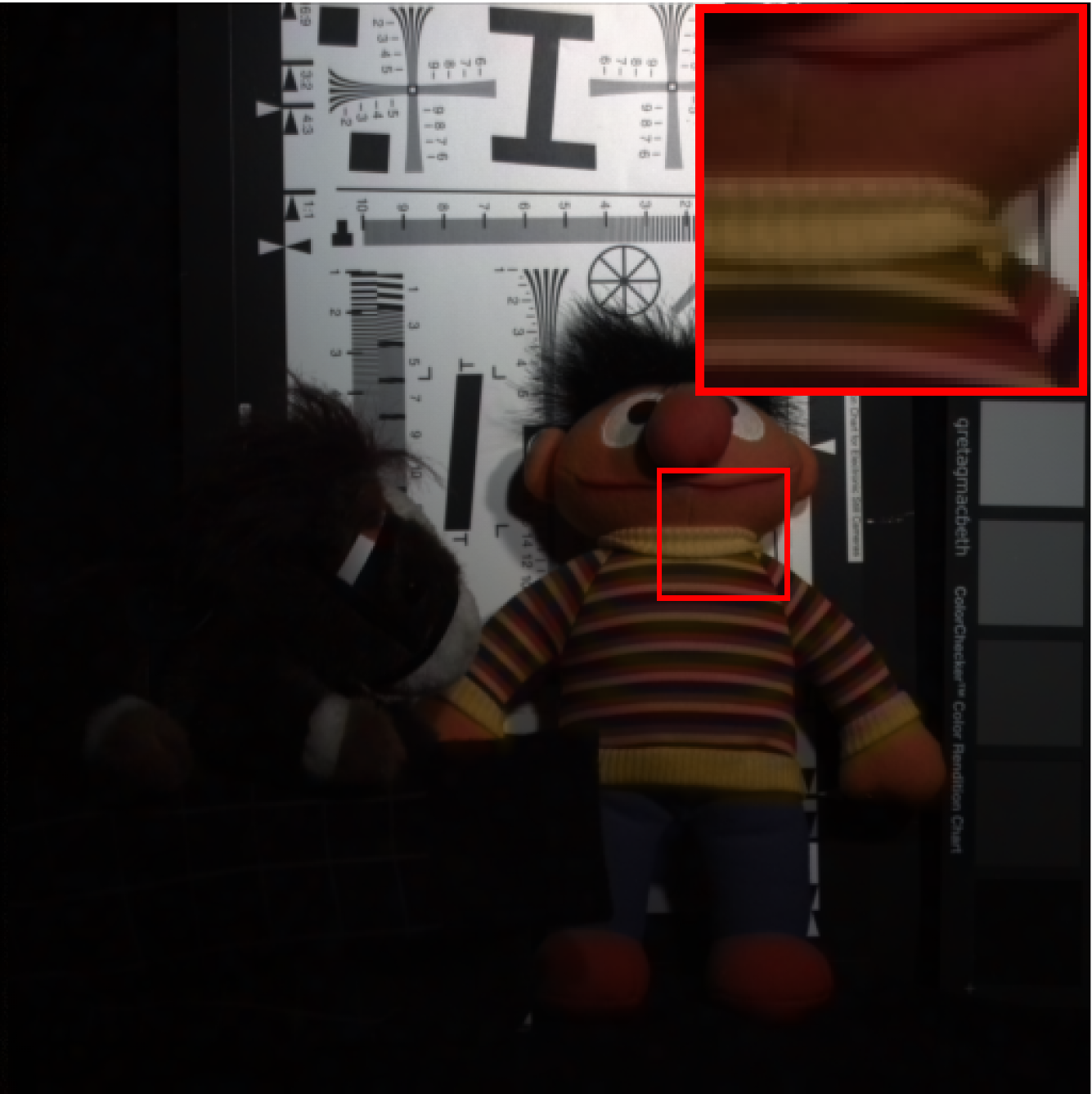}}
	\end{minipage}
	\hspace{1.4cm}
	\begin{minipage}{0.05\textwidth}
		\centerline{\includegraphics[width=0.94in,height=0.94in]{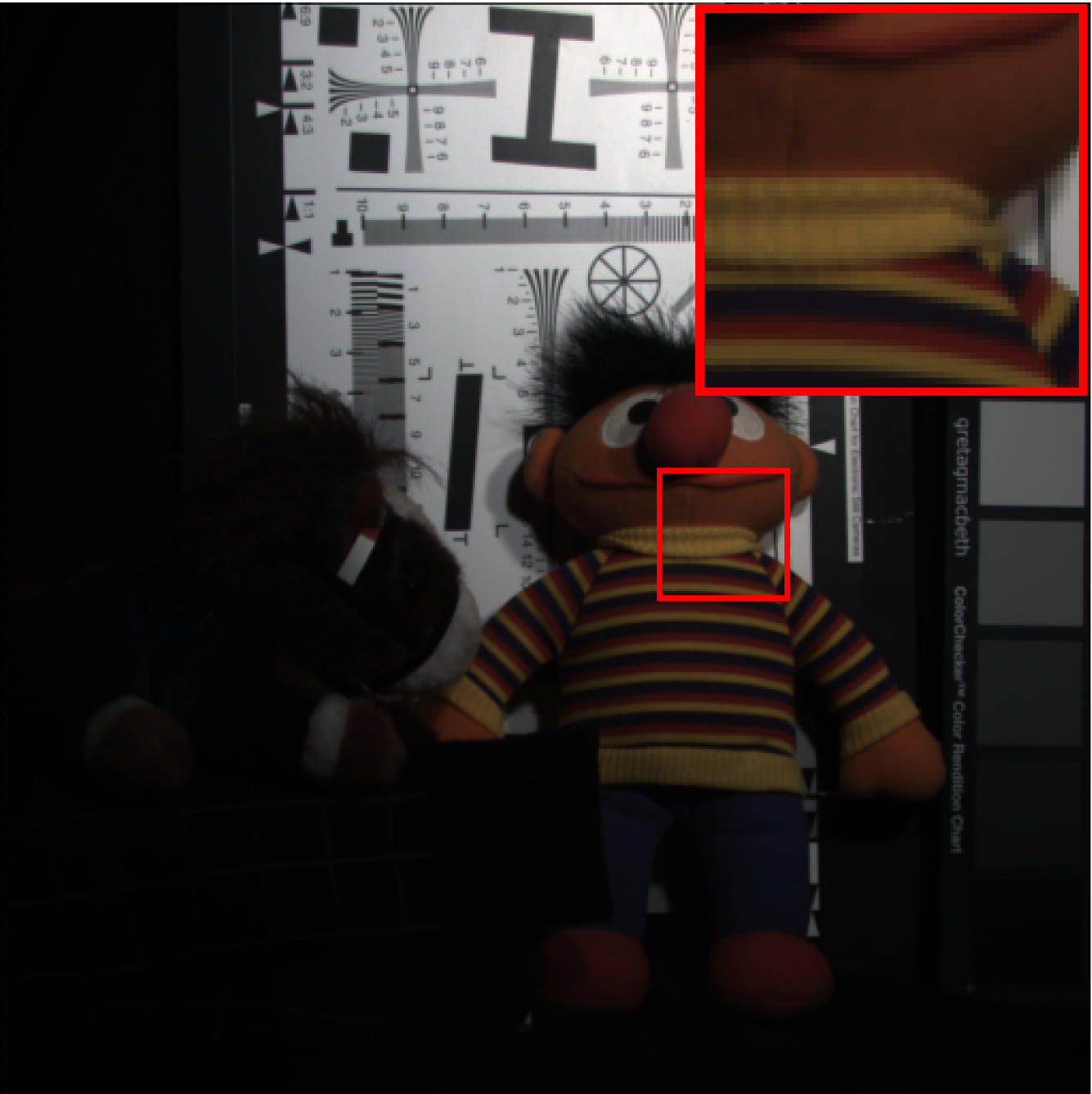}}
	\end{minipage}
	\hspace{1.4cm}
	\begin{minipage}{0.05\textwidth}
		\centerline{\includegraphics[width=0.94in,height=0.94in]{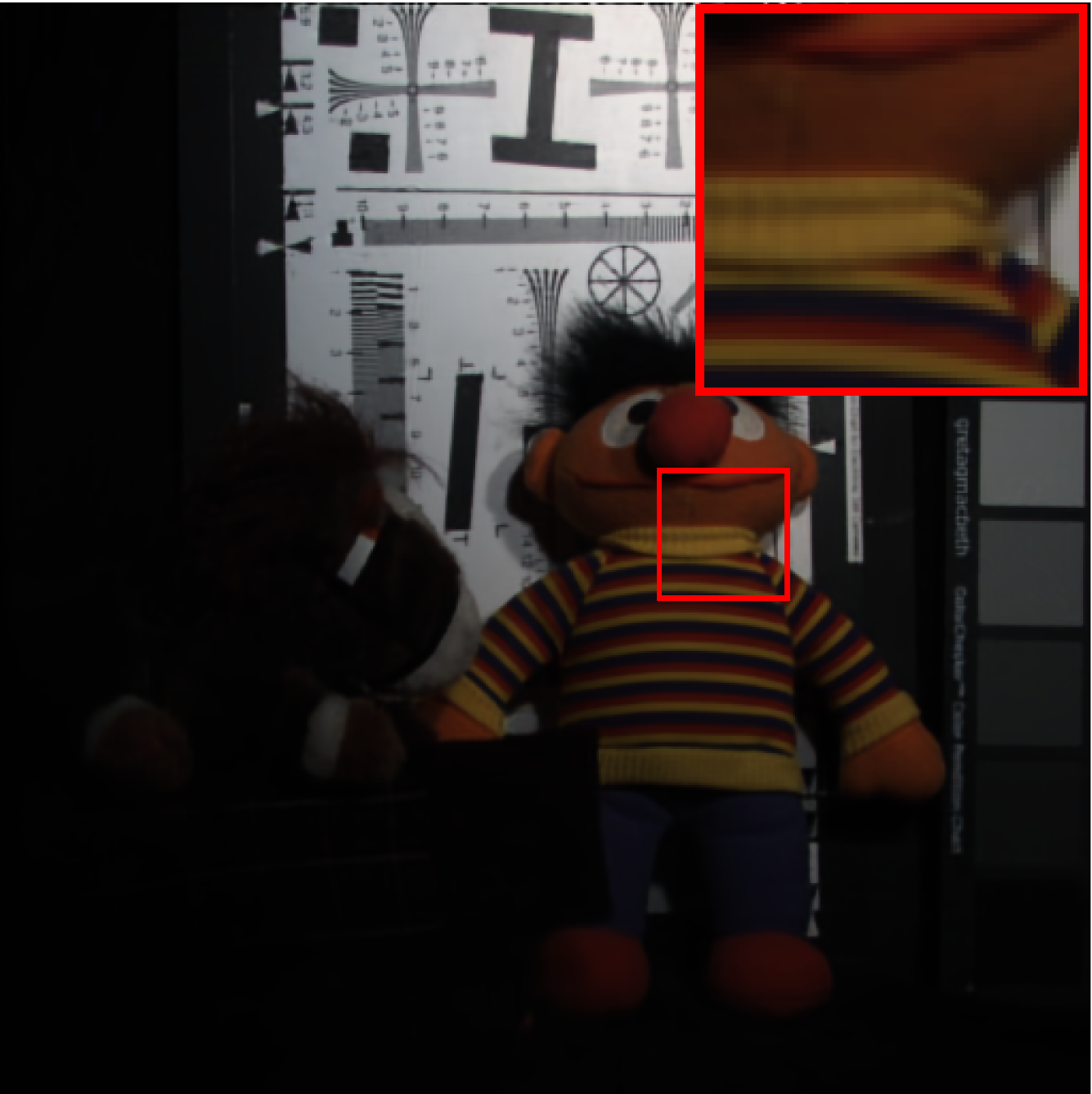}}
	\end{minipage}
	\hspace{1.4cm}
	\begin{minipage}{0.05\textwidth}
		\centerline{\includegraphics[width=0.94in,height=0.94in]{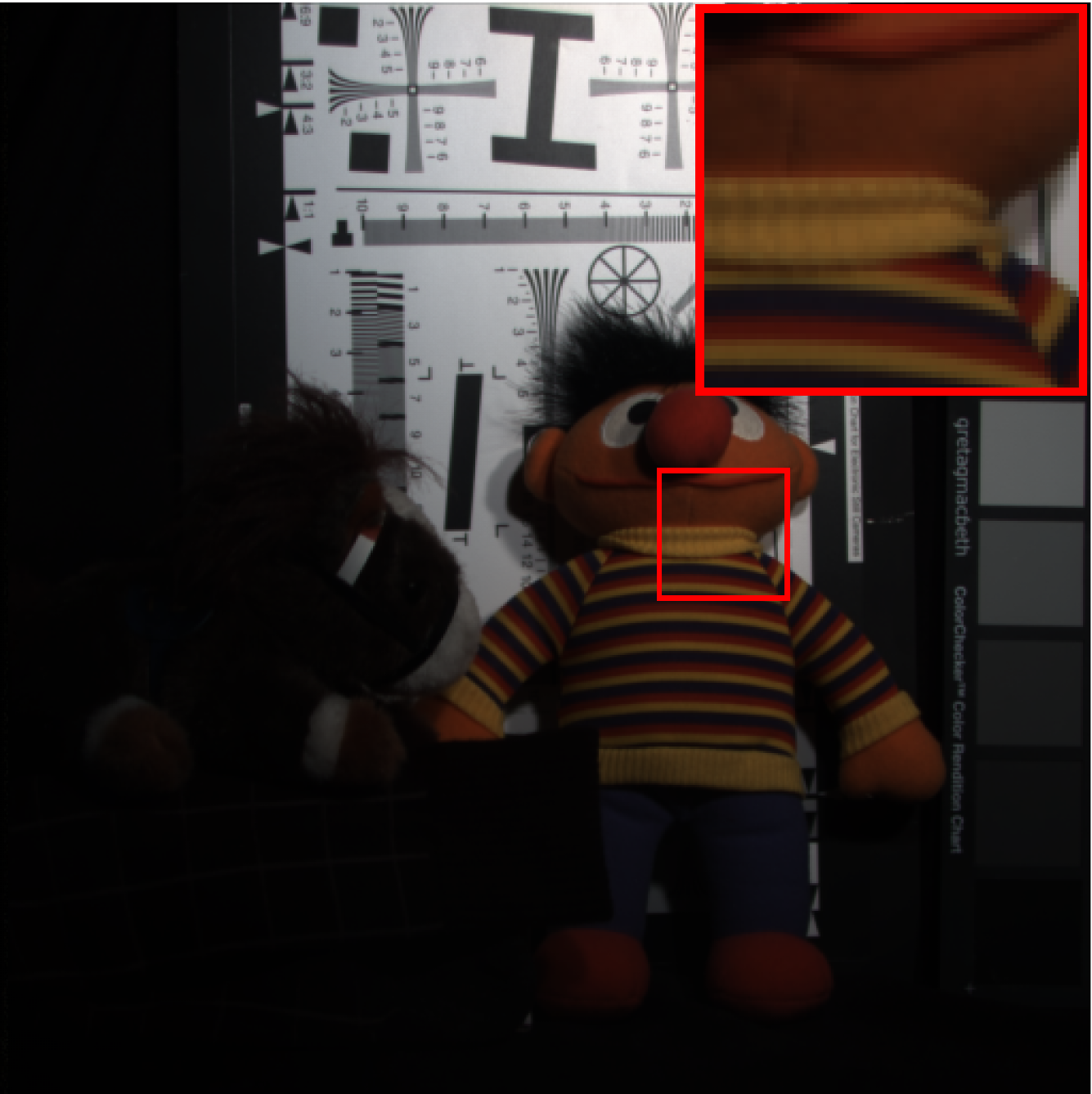}}
	\end{minipage}
	\hspace{0.8cm}
	\begin{minipage}{0.005\textwidth}
		\centerline{\includegraphics[width=0.1in,height=0.94in]{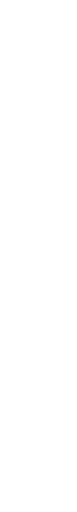}}
	\end{minipage}
	\vfill
	\hspace{0.55cm}
	\begin{minipage}{0.05\textwidth}
		\centerline{\includegraphics[width=0.94in,height=0.94in]{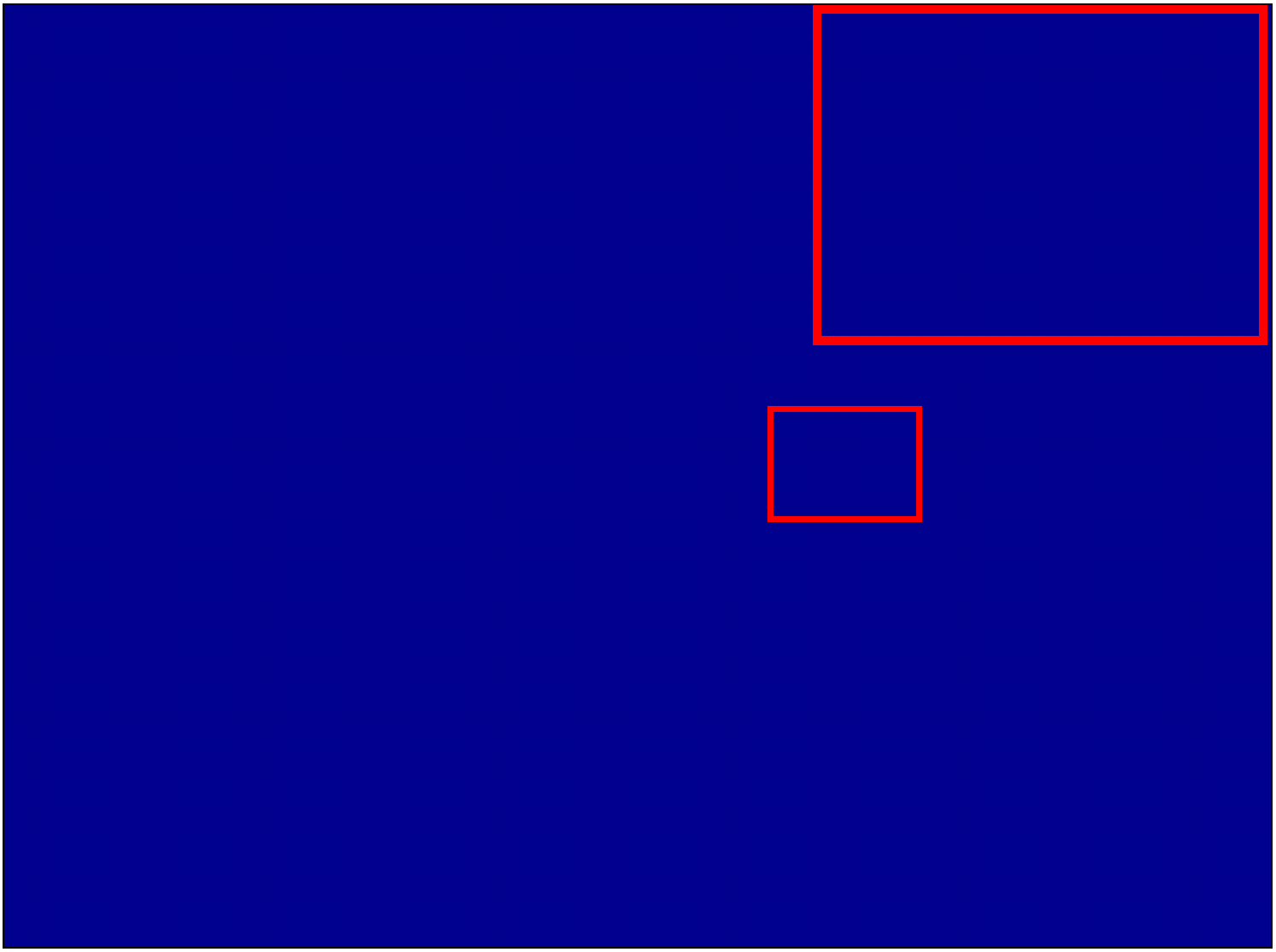}}
		\centerline{{\scriptsize (a) GT}}
	\end{minipage}
	\hspace{1.4cm}
	\begin{minipage}{0.05\textwidth}
		\centerline{\includegraphics[width=0.94in,height=0.94in]{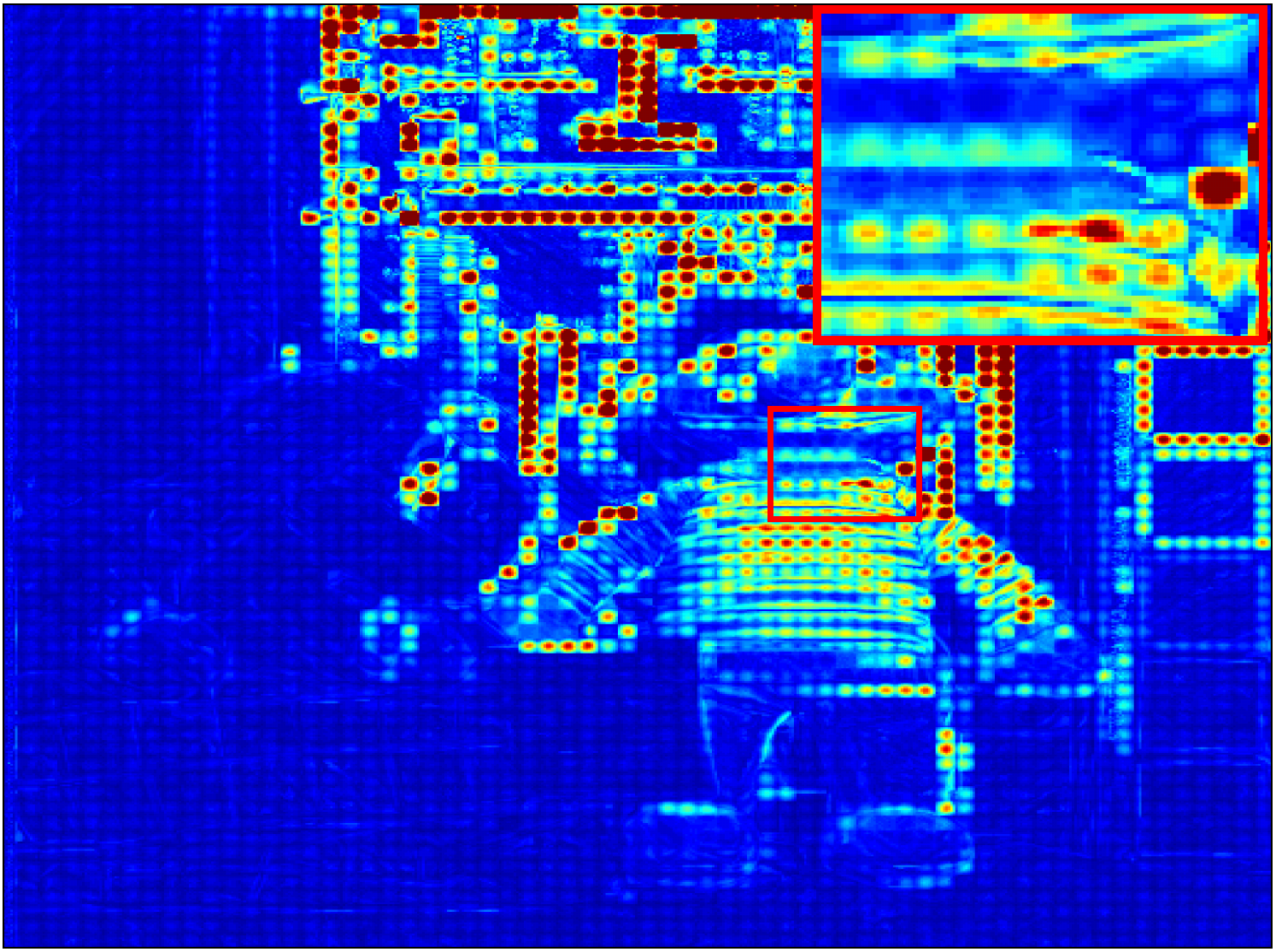}}
		\centerline{{\scriptsize (b) NSSR~\cite{7438864}}}
	\end{minipage}
	\hspace{1.4cm}
	\begin{minipage}{0.05\textwidth}
		\centerline{\includegraphics[width=0.94in,height=0.94in]{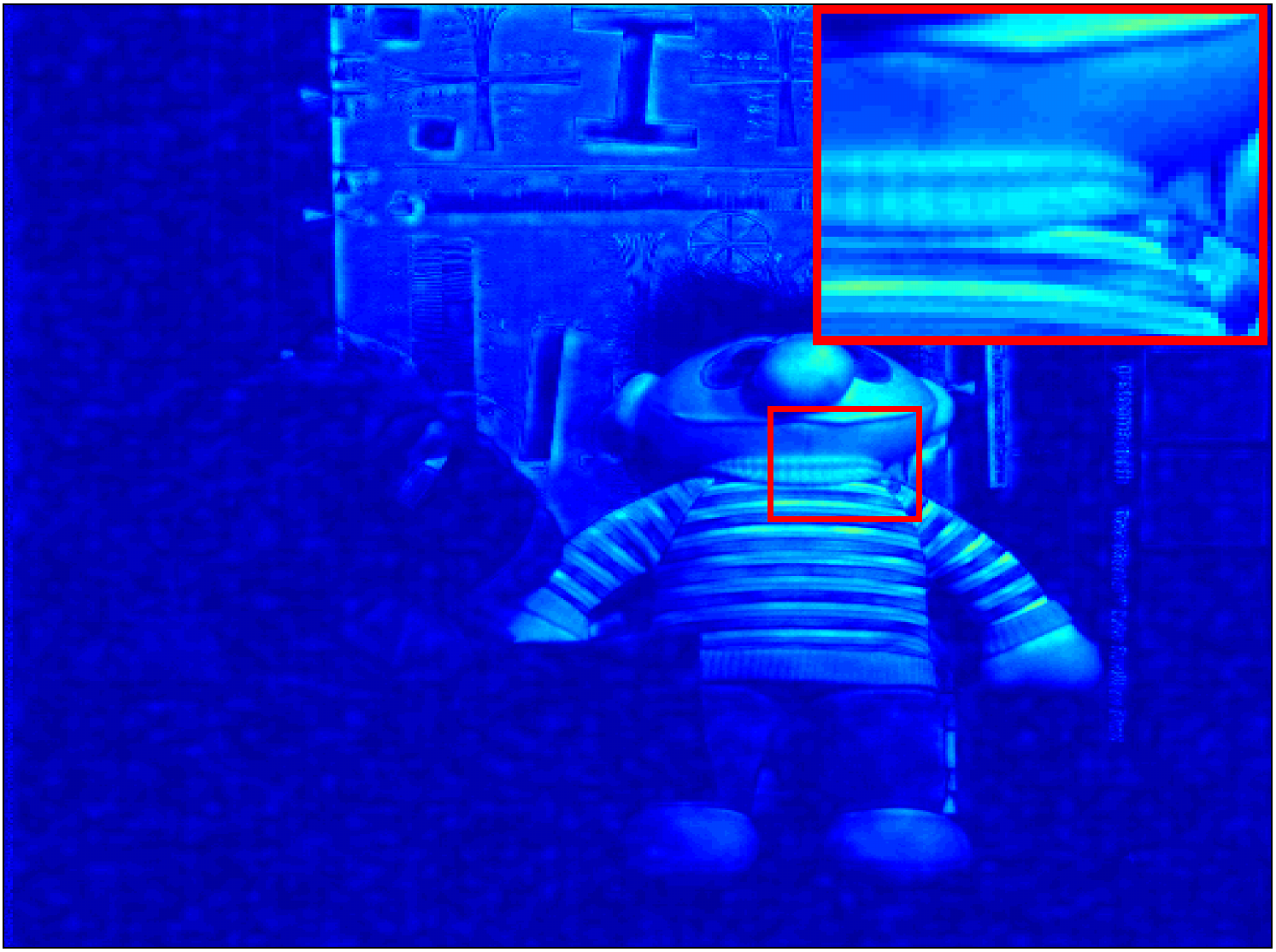}}
		\centerline{{\scriptsize (c) MHFnet~\cite{8953470}}}
	\end{minipage}
	\hspace{1.4cm}
	\begin{minipage}{0.05\textwidth}
		\centerline{\includegraphics[width=0.94in,height=0.94in]{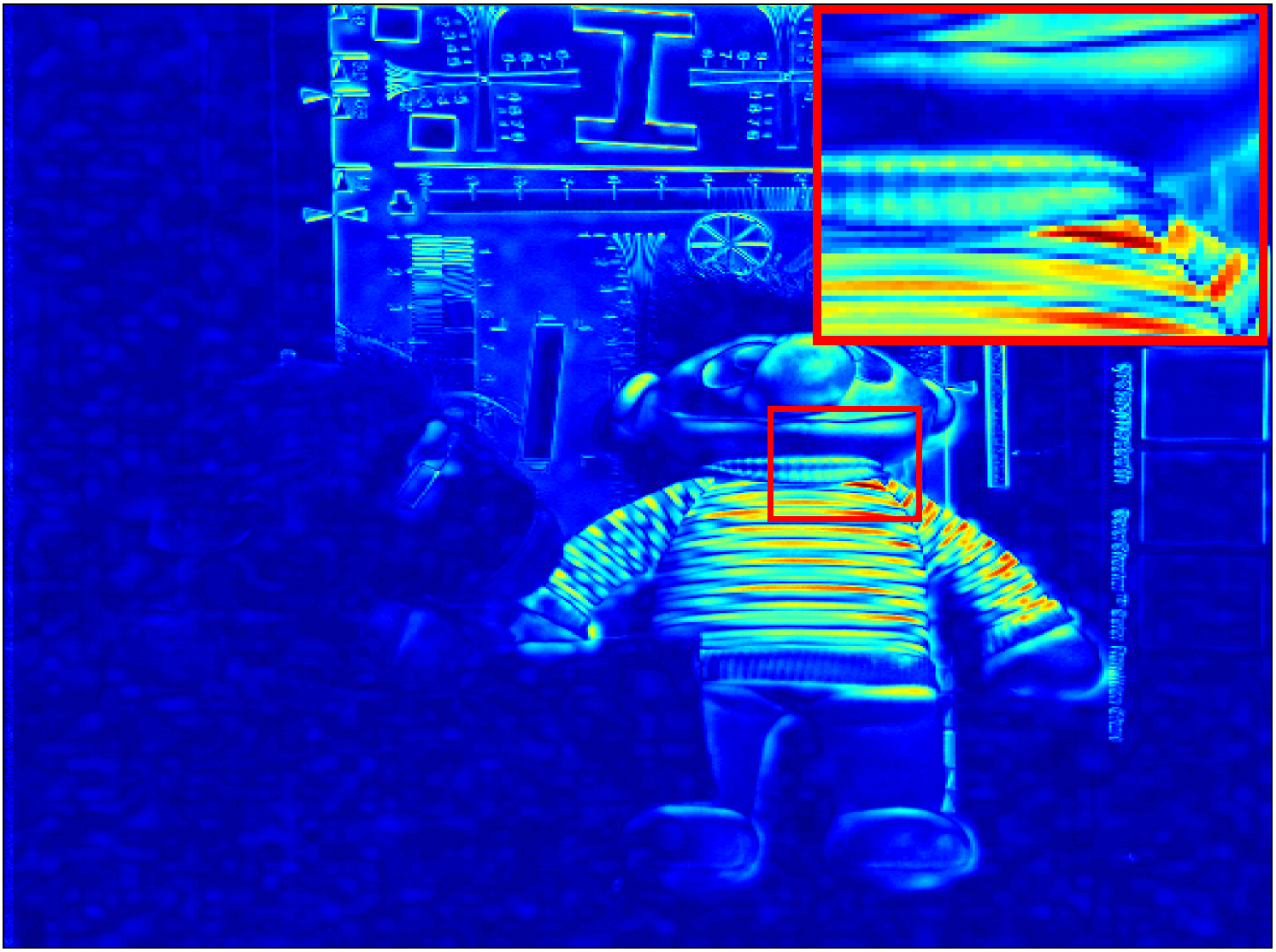}}
		\centerline{{\scriptsize (d) Yong~\cite{8019510}}}
	\end{minipage}
	\hspace{1.4cm}
	\begin{minipage}{0.05\textwidth}
		\centerline{\includegraphics[width=0.94in,height=0.94in]{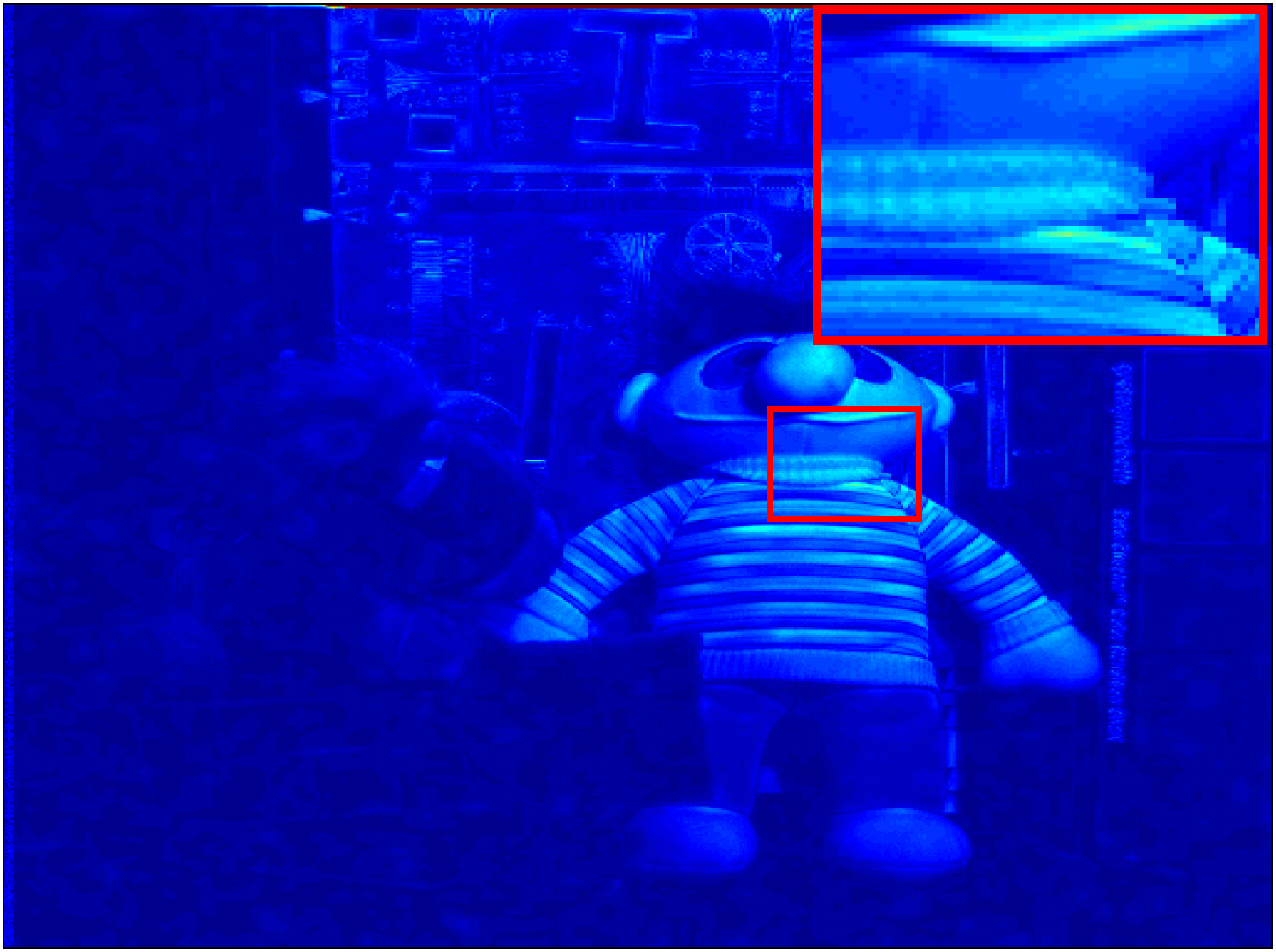}}
		\centerline{{\scriptsize (e) UAL~\cite{Ours_CVPR2020}}}
	\end{minipage}
	\hspace{1.4cm}
	\begin{minipage}{0.05\textwidth}
		\centerline{\includegraphics[width=0.94in,height=0.94in]{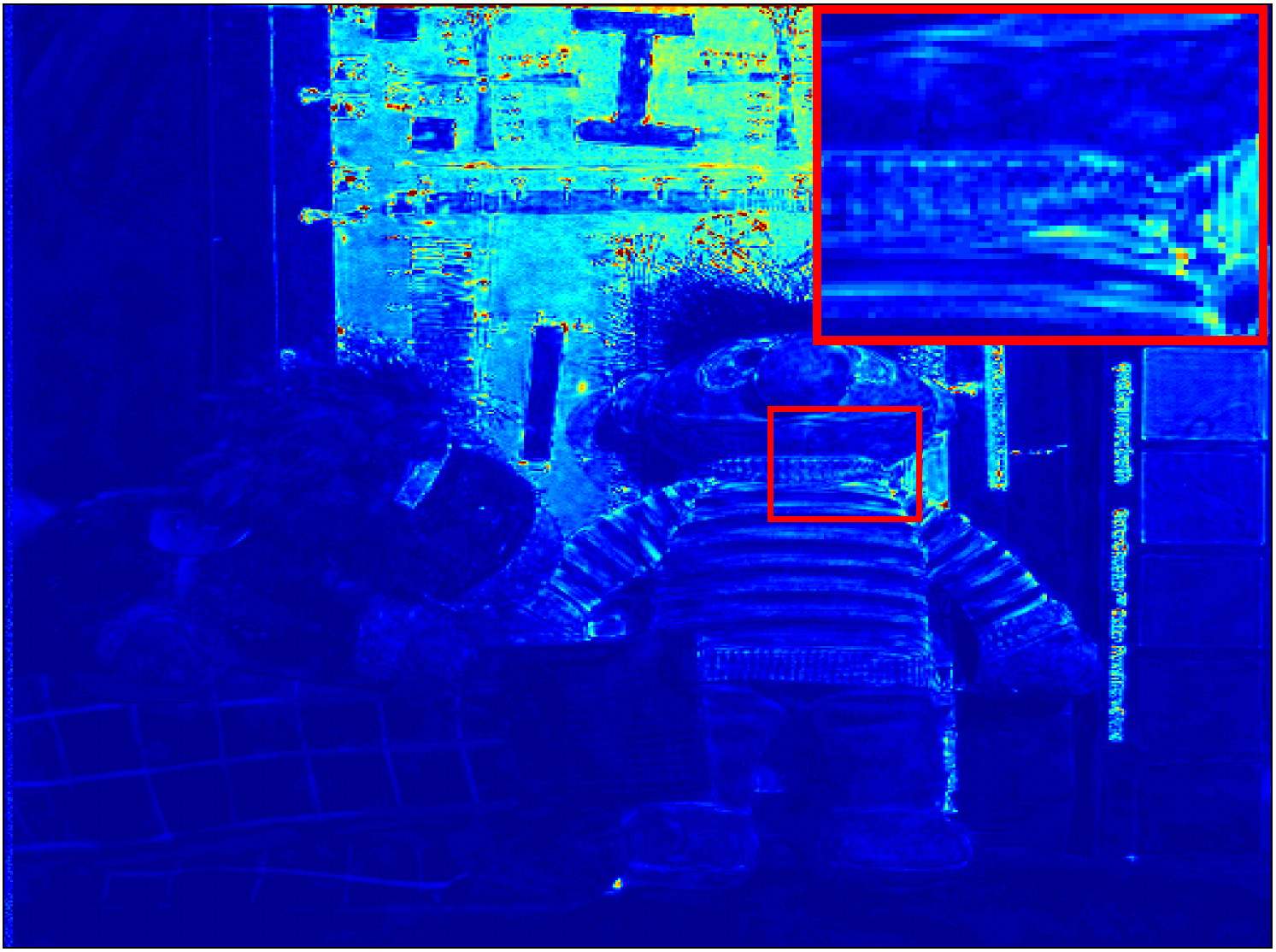}}
		\centerline{{\scriptsize (f) DBSR~\cite{9136736}}}
	\end{minipage}
	\hspace{1.4cm}
	\begin{minipage}{0.05\textwidth}
		\centerline{\includegraphics[width=0.94in,height=0.94in]{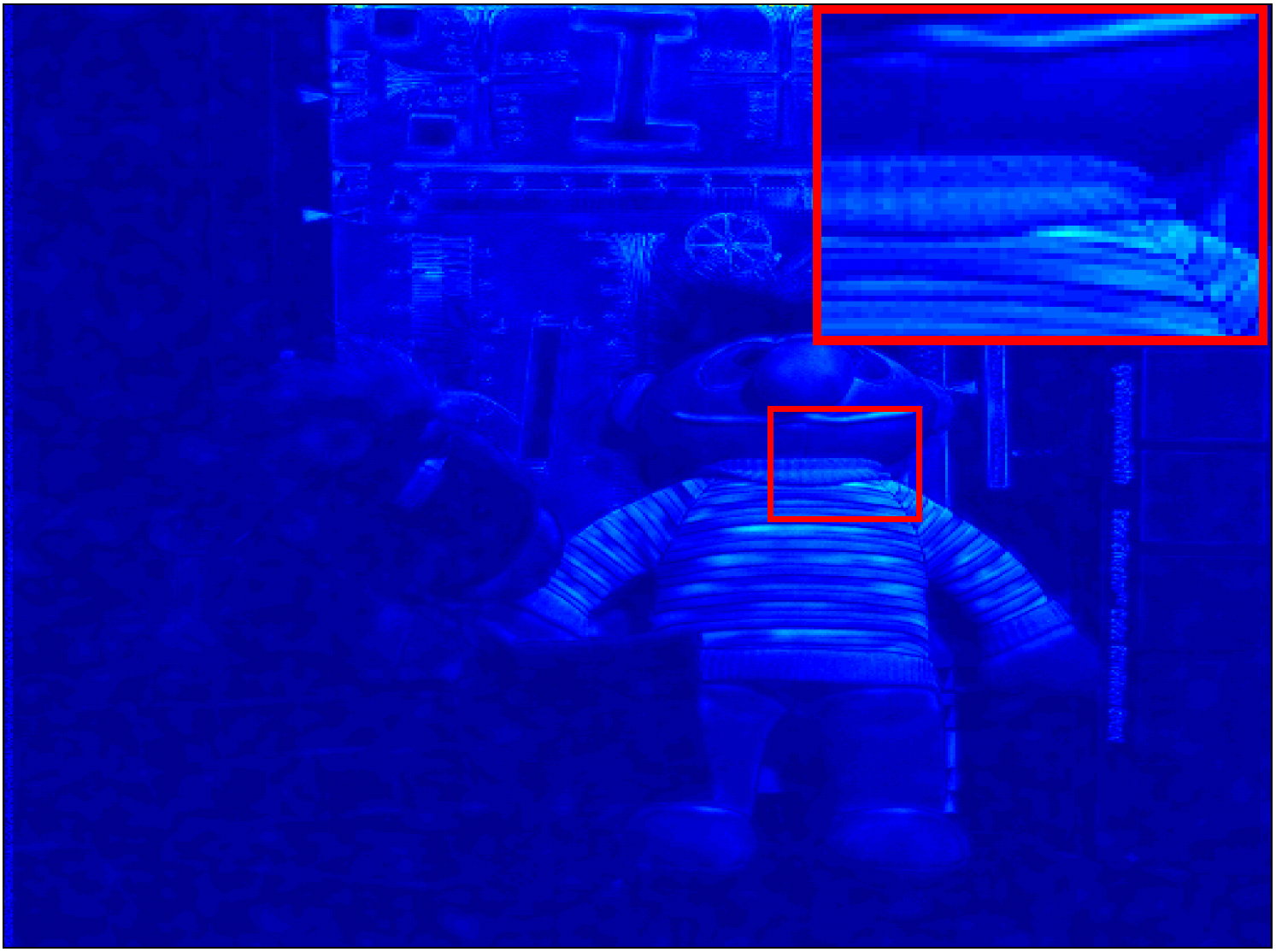}}
		\centerline{{\scriptsize (g) Ours}}
	\end{minipage}
	\hspace{0.8cm}
	\begin{minipage}{0.005\textwidth}
		\centerline{\includegraphics[width=0.15in,height=0.94in]{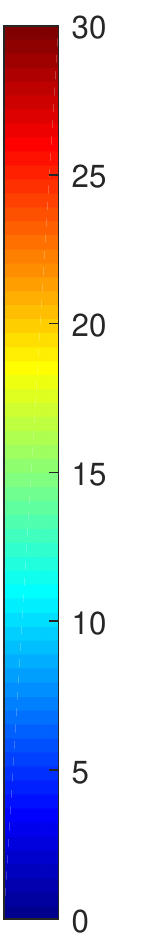}}
		\centerline{}
	\end{minipage}
	\caption{The visual SR results of all methods on the CAVE dataset. The input LR HSI and HR MSI are generated by $\mathbf{k}_1$ and $\mathbf{P}_{0.01}$, respectively. SNRs of both two observed images are 40dB, and the SR scale is 8.}
	\label{Fig_CAVE_Visible}
\end{figure*}

\subsection{Ablation studies}
The proposed method contains three main components: 1) alternating optimization based framework; 2) the modeled compact reconstruction network (shown as Figure~\ref{Flowchart}(b)) incorporating the information contains $\mathbf{k}$ and $\mathbf{P}$; 3) meta-learning for pre-training the reconstruction network.

To demonstrate the advantage of the utilized alternating optimization framework, we compare it with other two commonly used optimization manners, including a two-step based optimization method and jointly optimization method. 
Specifically, without using the meta-learning strategy for pre-training the reconstruction network, we separately utilize these two kind of optimization manners and the proposed alternating framework to optimize the degeneration model as well as the reconstruction network, and summarize the results on CAVE dataset in Table~\ref{Tab_CAVE_Ablation}. These three kind of methods are termed as 'Ours\_Sep', 'Ours\_Joint' and 'Ours\_Alter' in this study respectively. As can be seen, the results from 'Ours\_Sep' is the worst among three kind of methods. This is because separately estimating the degeneration model and the reconstruction model provides an inexactly estimated degeneration model for the reconstruction model, which thus inferiors to those from 'Ours\_Joint' which jointly optimizing the degeneration model and the reconstruction model. 'Ours\_Alter' obtains better results compared with those from 'Ours\_Joint', which demonstrates that the alternating optimization method proposed in this study is more beneficial for blind HSI SR. 
 
To testify the modeled compact reconstruction network, we compare the proposed method with its variant (termed as Ours\_Basic), in which $\mathbf{k}$ and $\mathbf{P}$ are removed from the input of the reconstruction network. It can be seen that the proposed reconstruction network suffers obvious degradation without the guidance information provided by $\mathbf{k}$ and $\mathbf{P}$. 'Ours' is the proposed method. Since the only difference between 'Ours' and 'Ours\_Alter' is the meta-learning strategy for pre-training, the advantage of meta-learning can be seen by comparing 'Ours' with 'Ours\_Alter'.



\subsection{Performance comparison on the CAVE dataset}
In this Subsection, we compare the proposed method with five competing state-of-the-art methods on the CAVE dataset under different settings including experiments with different SR scale, experiments with different spectral response matrix $\mathbf{P}_c$, experiments with different blur kernels. 

\noindent \textbf{Different SR scales}\quad
In this experiment, we generate the observed LR HSI $\mathbf{X}$ and HR MSI $\mathbf{Y}$ with the kernel $\mathbf{k}_1$ and the spectral response matrix $\mathbf{P}_{0.01}$, respectively, and add noise into $\mathbf{X}$ and $\mathbf{Y}$, in which SNRs of both observed images are 40dB. We then testify the applicability of all methods with different SR scale factors (\textit{i.e.}~up-sampling the LR HSI for 8, 16, 32 times) and summarize the numerical results in Table~\ref{Tab_CAVE_Result_Diff_Scale}. It can be seen that the proposed method has obvious advantage over other competing methods on all three SR scales. The PSNR of the proposed method even surpasses the second best method almost for 3dB. Non-blind methods including NSSR~\cite{7438864} and MHFnet~\cite{8953470} present severe performance degradation on the blind settings, especially with a larger SR scale. In contrast, the semi-blind methods including Yong~\cite{8019510} and UAL~\cite{Ours_CVPR2020} have better performance. In addition, the performance of the blind competing method DBSR~\cite{9136736} inferiors to the proposed method. We attribute the reason to the optimization method DBSR~\cite{9136736} utilized, which cannot provide a good degeneration model and consequently resulting in the performance degradation. For better comparison, we also provide the visual results of all methods in Figure~\ref{Fig_CAVE_Visible}. The first row of Figure~\ref{Fig_CAVE_Visible} provides the pseudo color image of reconstructed HSIs and the second row shows the reconstruction error maps of all methods. It can be seen that the reconstructed HSI of the proposed method has lower reconstruction errors, which is consistent with those in Table~\ref{Tab_CAVE_Result_Diff_Scale}.

\begin{table}[!htbp]\small
	\centering
	\caption{The performance of each method on the CAVE dataset with HR MSI generated by different $\mathbf{P}_c$ ($c= 0.01, 0.015, 0.02$). 
		The best results are in bold.}
	\setlength{\tabcolsep}{1.5mm}{\begin{tabular}{c|c|c|c|c|c}
			\hline\hline
			&Method&RMSE&PSNR&SAM&SSIM\\
			\hline
			\multirow{6}*{$\mathbf{P}_{0.01}$}&NSSR~\cite{7438864}&8.11&30.53&18.85&0.8533\\	
			\cline{2-6}&MHFnet~\cite{8953470}&4.53&35.67&14.01&0.9409\\
			\cline{2-6}&Yong~\cite{8019510}&4.46&35.72&11.29&0.9491\\
			\cline{2-6}&UAL~\cite{Ours_CVPR2020}&3.93&37.07&7.57&0.9724\\
			\cline{2-6}&DBSR~\cite{9136736}&4.01&36.66&\textbf{5.53}&0.9739\\
			\cline{2-6}&Ours&\textbf{2.40}&\textbf{41.12}&6.53&\textbf{0.9854}\\
			\hline
			\hline
			\multirow{6}*{$\mathbf{P}_{0.015}$}&NSSR~\cite{7438864}&8.34&30.28&19.08&0.8481\\	
			\cline{2-6}&MHFnet~\cite{8953470}&7.00&32.07&16.50&0.9121\\
			\cline{2-6}&Yong~\cite{8019510}&5.07&34.70&12.05&0.9421\\
			\cline{2-6}&UAL~\cite{Ours_CVPR2020}&4.96&35.23&8.40&0.9671\\
			\cline{2-6}&DBSR~\cite{9136736}&4.71&35.65&\textbf{5.65}&0.9733\\
			\cline{2-6}&Ours&\textbf{2.59}&\textbf{40.51}&6.66&\textbf{0.9848}\\
			\hline
			\hline
			\multirow{6}*{$\mathbf{P}_{0.02}$}&NSSR~\cite{7438864}&8.56&30.05&19.28&0.8438\\	
			\cline{2-6}&MHFnet~\cite{8953470}&7.43&31.54&16.94&0.9072\\
			\cline{2-6}&Yong~\cite{8019510}&5.50&34.04&12.79&0.9349\\
			\cline{2-6}&UAL~\cite{Ours_CVPR2020}&5.76&34.03&9.06&0.9629\\
			\cline{2-6}&DBSR~\cite{9136736}&5.22&34.79&\textbf{5.83}&0.9722\\
			\cline{2-6}&Ours&\textbf{2.86}&\textbf{39.74}&6.64&\textbf{0.9845}\\
			\hline\hline
	\end{tabular}}
	\label{Tab_CAVE_Result_Diff_PScales}
\end{table}

\begin{figure}[!htbp]
	\vspace{-0.5cm}
	\centering
	\begin{minipage}{0.05\textwidth}
		\centerline{\includegraphics[width=1.65in,height=1.15in]{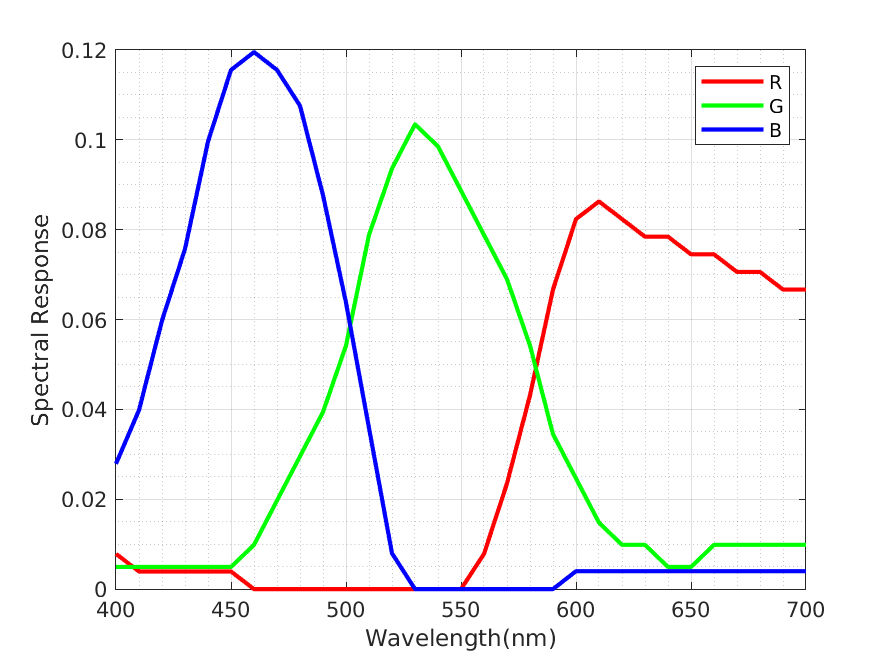}}
		\centerline{{\scriptsize (a) The default spectral response matrix}}
	\end{minipage}
	\hspace{3cm}
	\begin{minipage}{0.05\textwidth}
		\centerline{\includegraphics[width=1.65in,height=1.15in]{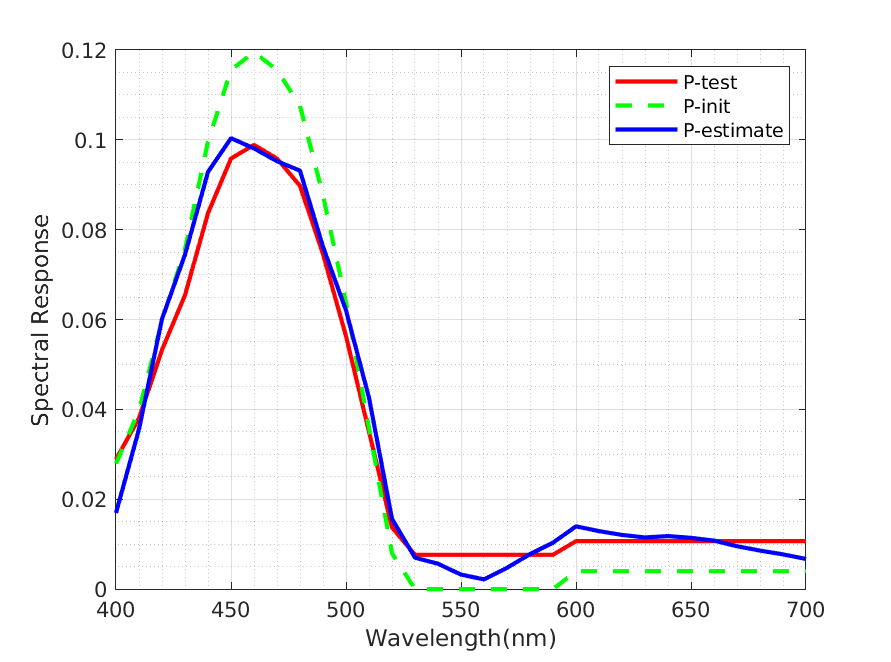}}
		\centerline{{\scriptsize (b) The spectral response of blue band}}
	\end{minipage}
	\vfill
	\begin{minipage}{0.05\textwidth}
		\centerline{\includegraphics[width=1.65in,height=1.15in]{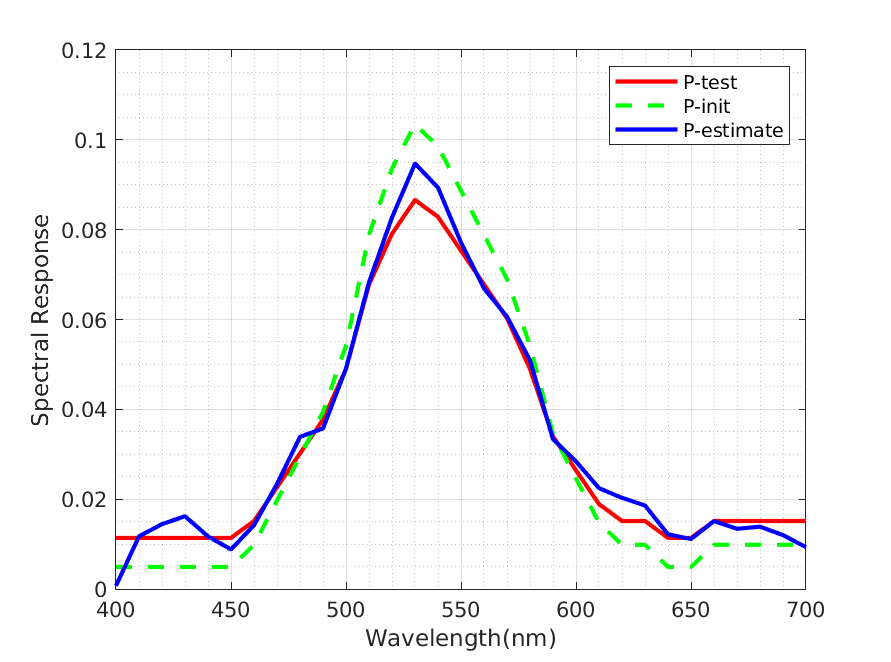}}
		\centerline{{\scriptsize (c) The spectral response of green band}}
	\end{minipage}
	\hspace{3cm}
	\begin{minipage}{0.05\textwidth}
		\centerline{\includegraphics[width=1.65in,height=1.15in]{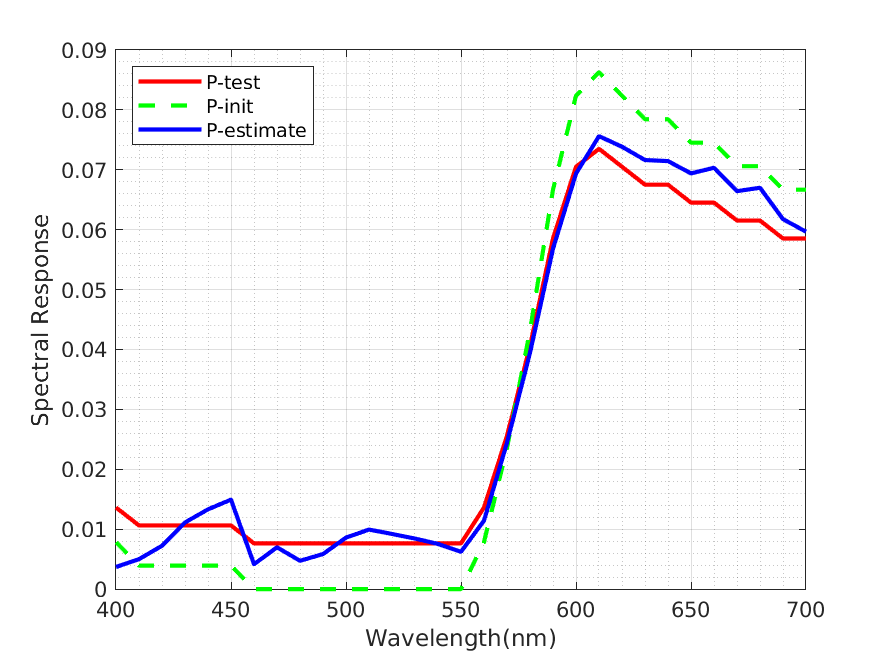}}
		\centerline{{\scriptsize (d) The spectral response of red band}}
	\end{minipage}
	\vfill
	\vspace{0.05cm}
	\caption{The default spectral response matrix $\mathbf{P}$ as well as the estimated spectral response matrix by the proposed method. (a) the default spectral response matrix $\mathbf{P}$. (b)(c)(d) represent the estimate spectral response metrics for the blue, green and red band. }
	\label{Fig_P_Estimation}
\end{figure}

\noindent \textbf{Different spectral response matrix}\quad
In this experiment, we further explore the performance of the proposed method influenced by the spectral response matrix $\mathbf{P}_c$ with different transform coefficient. The blur kernel is $\mathbf{k}_1$ and the SR scale is 8. SNRs of both the observed LR HSI and HR MSI are 40dB. The detailed numerical results are summarized in Table~\ref{Tab_CAVE_Result_Diff_PScales}. It can be be seen that the difficulty of HSI SR is increased with larger transformation coefficient $c$. With the increasing of $c$, the performance of both non-blind methods (\textit{i.e.}~NSSR~\cite{7438864} and MHFnet~\cite{8953470}) and semi-blind methods (\textit{i.e.}~Yong~\cite{8019510} and UAL~\cite{Ours_CVPR2020} which consider the degeneration only in spatial domain) degrade severely. The blind method DBSR~\cite{9136736} as well as the proposed method present much robust performance when $c$ increases. Nevertheless, the proposed method still has advantage over DBSR~\cite{9136736} on three out of all four evaluation metrics. The true spectral response matrix utilized as well as the one estimated by the proposed method are visualized in Figure~\ref{Fig_P_Estimation}. In Figure~\ref{Fig_P_Estimation},  P-test denotes the $\mathbf{P}_c$ that used to generate the observed HR MSI, $c$ is 0.01. P-init is the default spectral response matrix $\mathbf{P}$ and is utilized as the initial parameter of $\Psi$. P-estimate is the estimated spectral response matrix. It can be seen that the proposed method can well estimate the spectral response matrix. 
\textbf{\begin{table}[!tbp]\small
		\centering
		\caption{The performance of each method on the CAVE dataset with different blur kernel $\mathbf{k}$. 
			The best results are in bold.}
		\setlength{\tabcolsep}{2.3mm}{\begin{tabular}{c|c|c|c|c|c}
				\hline\hline
				&Method&RMSE&PSNR&SAM&SSIM\\
				\hline
				\multirow{6}*{K1}&NSSR~\cite{7438864}&8.11&30.53&18.85&0.8533\\	
				\cline{2-6}&MHFnet~\cite{8953470}&4.53&35.67&14.01&0.9409\\
				\cline{2-6}&Yong~\cite{8019510}&4.46&35.72&11.29&0.9491\\
				\cline{2-6}&UAL~\cite{Ours_CVPR2020}&3.93&37.07&7.57&0.9724\\
				\cline{2-6}&DBSR~\cite{9136736}&4.01&36.66&\textbf{5.53}&0.9739\\
				\cline{2-6}&Ours&\textbf{2.40}&\textbf{41.12}&6.53&\textbf{0.9854}\\
				\hline
				\hline
				\multirow{6}*{K2}&NSSR~\cite{7438864}&9.26&29.44&19.24&0.8384\\	
				\cline{2-6}&MHFnet~\cite{8953470}&4.59&35.59&14.05&0.9403\\
				\cline{2-6}&Yong~\cite{8019510}&4.46&35.77&10.78&0.9510\\
				\cline{2-6}&UAL~\cite{Ours_CVPR2020}&3.94&37.05&7.59&0.9424\\
				\cline{2-6}&DBSR~\cite{9136736}&4.39&36.26&\textbf{5.54}&0.9740\\
				\cline{2-6}&Ours&\textbf{2.44}&\textbf{41.17}&6.48&\textbf{0.9858}\\
				\hline
				\hline
				\multirow{6}*{K3}&NSSR~\cite{7438864}&7.50&31.23&18.50&0.8592\\	
				\cline{2-6}&MHFnet~\cite{8953470}&4.52&35.69&14.06&0.9405\\
				\cline{2-6}&Yong~\cite{8019510}&4.42&35.84&11.17&0.9486\\
				\cline{2-6}&UAL~\cite{Ours_CVPR2020}&3.94&37.03&7.58&0.9722\\
				\cline{2-6}&DBSR~\cite{9136736}&4.40&36.41&\textbf{5.62}&0.9745\\
				\cline{2-6}&Ours&\textbf{2.45}&\textbf{41.03}&6.46&\textbf{0.9862}\\
				\hline\hline
				\multirow{6}*{K4}&NSSR~\cite{7438864}&8.30&30.37&18.76&0.8503\\	
				\cline{2-6}&MHFnet~\cite{8953470}&4.55&35.65&14.02&0.9406\\
				\cline{2-6}&Yong~\cite{8019510}&4.50&35.72&11.16&0.9472\\
				\cline{2-6}&UAL~\cite{Ours_CVPR2020}&3.95&37.02&7.59&0.9722\\
				\cline{2-6}&DBSR~\cite{9136736}&4.74&35.71&\textbf{5.53}&0.9738\\
				\cline{2-6}&Ours&\textbf{2.57}&\textbf{40.68}&6.51&\textbf{0.9857}\\
				\hline\hline
		\end{tabular}}
		\label{Tab_CAVE_Result_Diff_Kernels}
\end{table}}

\begin{figure}[!htbp]
	\vspace{-0.3cm}
	\centering
	\begin{minipage}{0.05\textwidth}
		\centerline{\includegraphics[width=0.65in,height=0.65in]{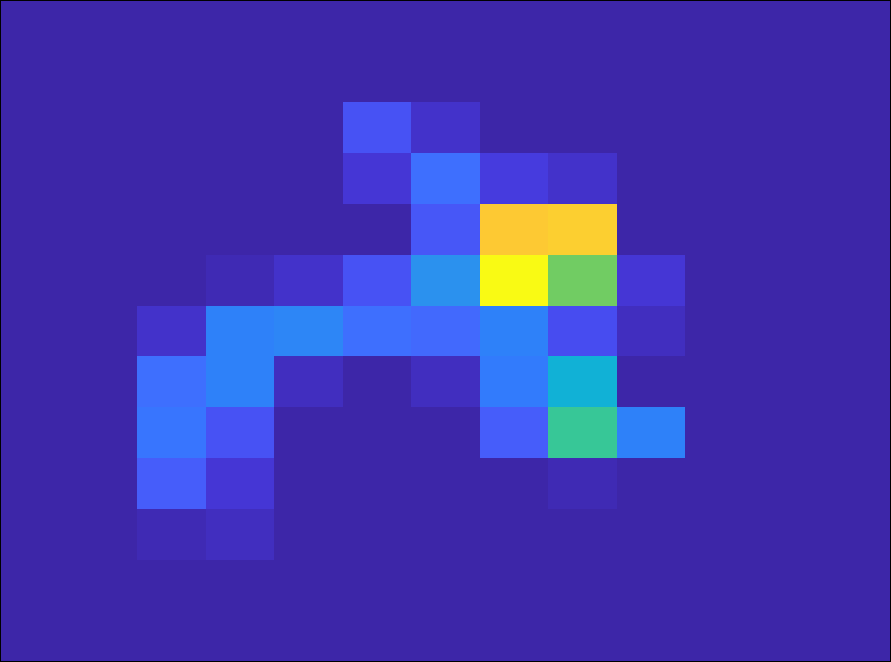}}
	\end{minipage}
	\hspace{0.8cm}
	\begin{minipage}{0.05\textwidth}
		\centerline{\includegraphics[width=0.65in,height=0.65in]{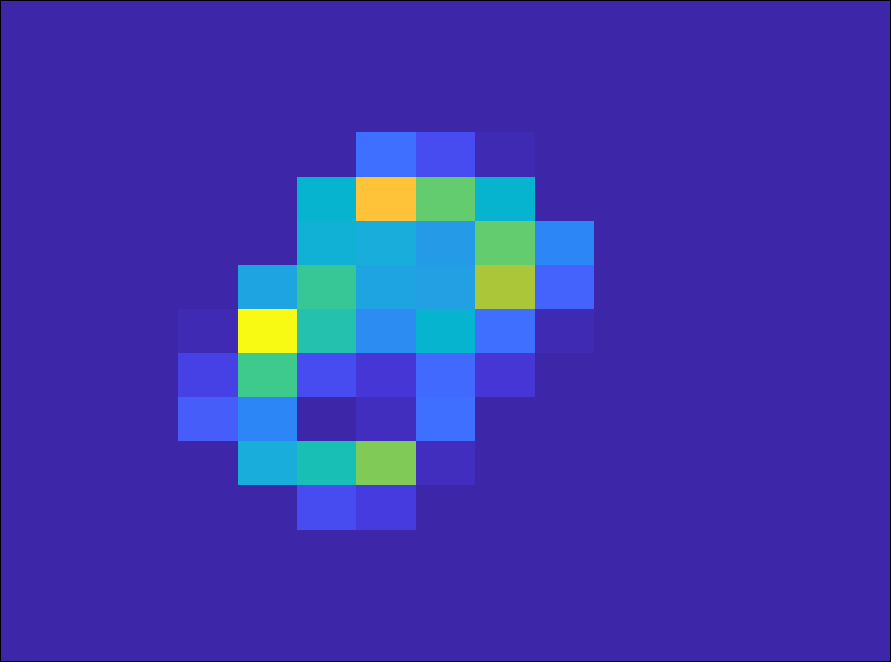}}
	\end{minipage}
	\hspace{0.8cm}
	\begin{minipage}{0.05\textwidth}
		\centerline{\includegraphics[width=0.65in,height=0.65in]{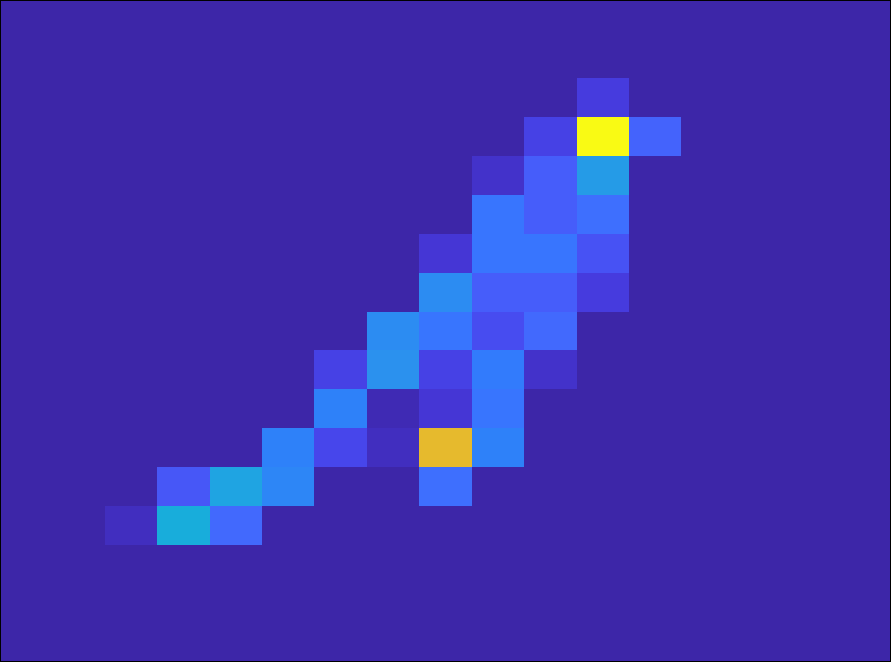}}
	\end{minipage}
	\hspace{0.8cm}
	\begin{minipage}{0.05\textwidth}
		\centerline{\includegraphics[width=0.65in,height=0.65in]{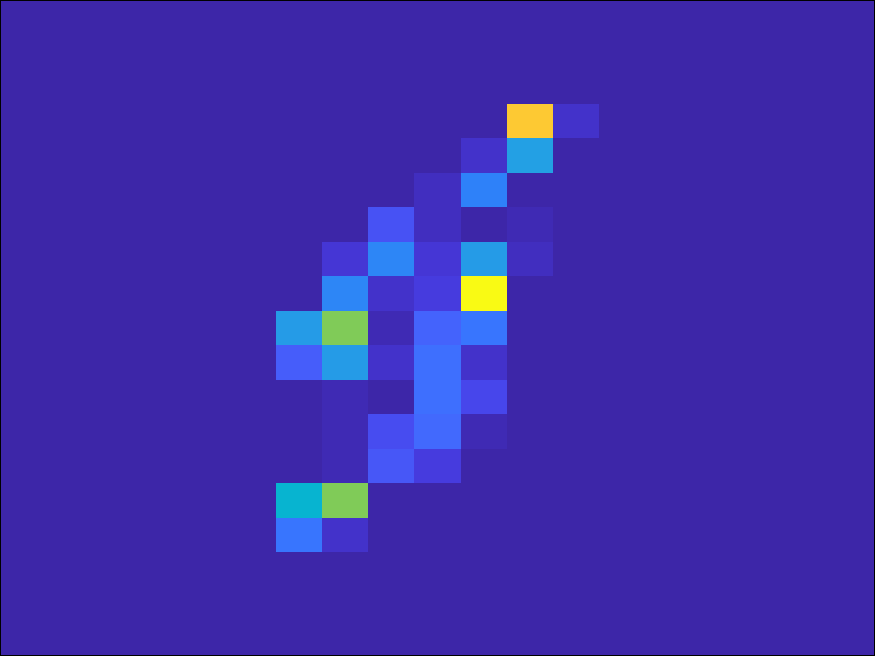}}
	\end{minipage}
	\vfill
	\vspace{0.05cm}
	\begin{minipage}{0.05\textwidth}
		\centerline{\includegraphics[width=0.65in,height=0.65in]{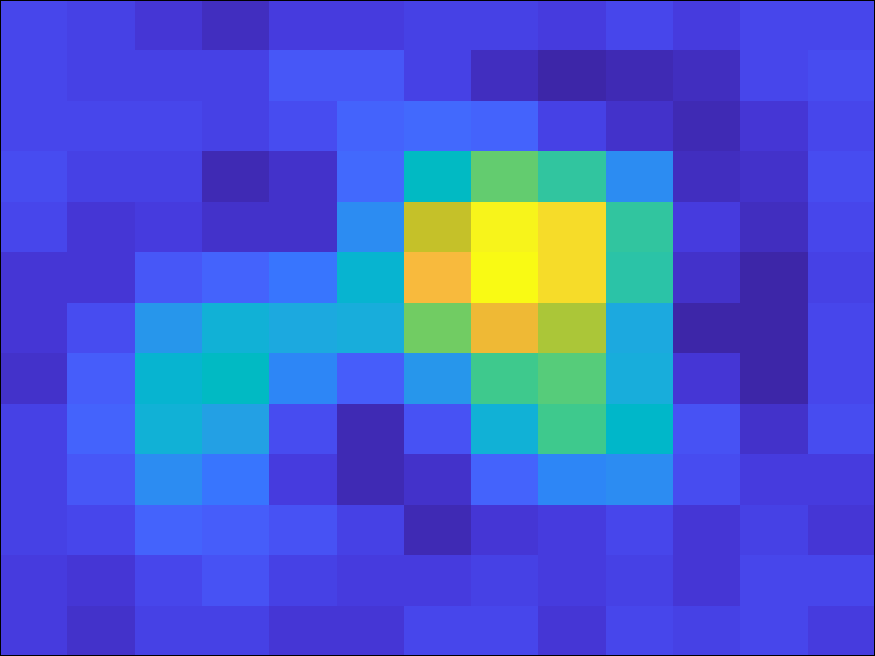}}
		\centerline{{\scriptsize (a) $\mathbf{k}_1$}}
	\end{minipage}
	\hspace{0.8cm}
	\begin{minipage}{0.05\textwidth}
		\centerline{\includegraphics[width=0.65in,height=0.65in]{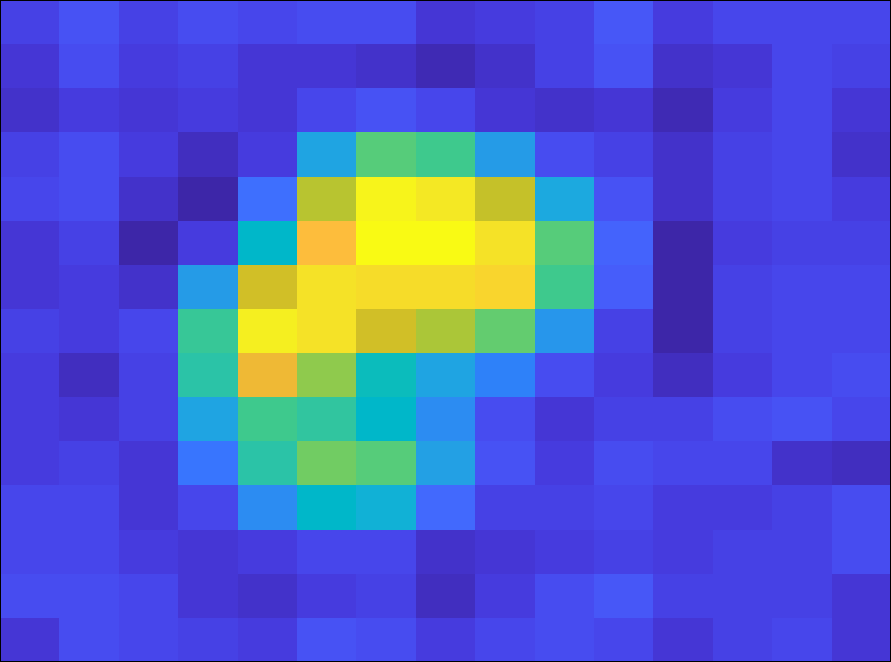}}
		\centerline{{\scriptsize (b) $\mathbf{k}_2$}}
	\end{minipage}
	\hspace{0.8cm}
	\begin{minipage}{0.05\textwidth}
		\centerline{\includegraphics[width=0.65in,height=0.65in]{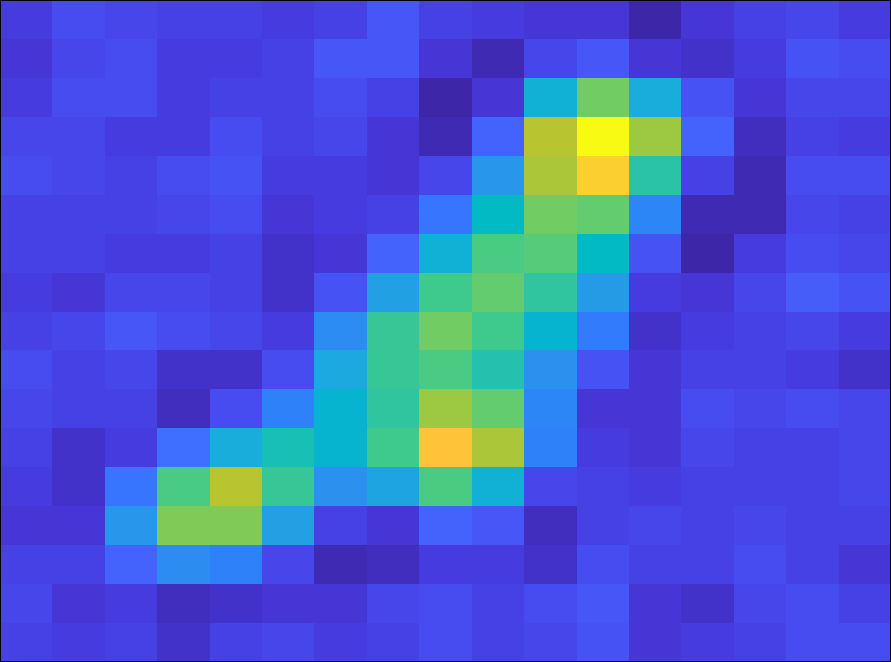}}
		\centerline{{\scriptsize (c) $\mathbf{k}_3$}}
	\end{minipage}
	\hspace{0.8cm}
	\begin{minipage}{0.05\textwidth}
		\centerline{\includegraphics[width=0.65in,height=0.65in]{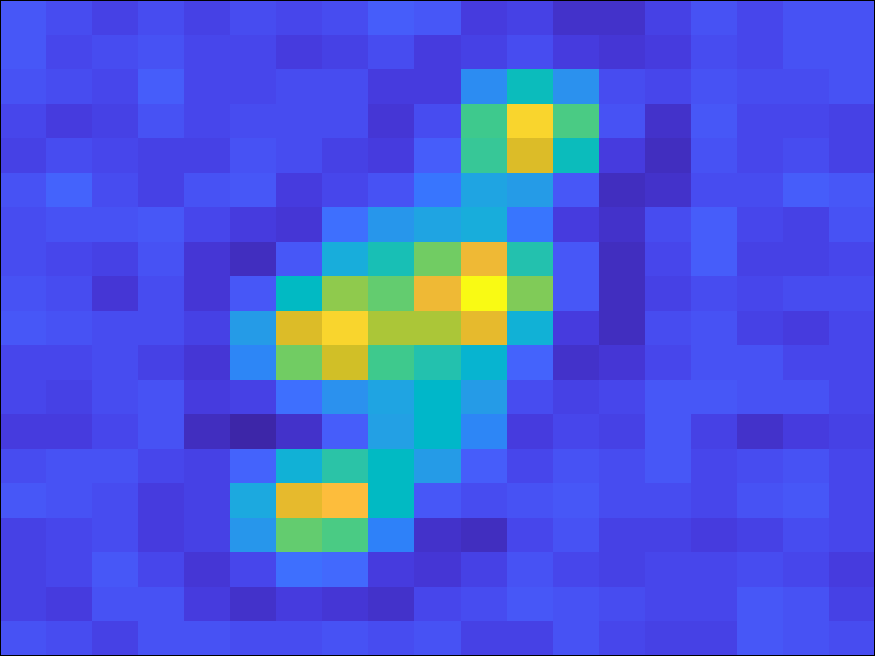}}
		\centerline{{\scriptsize (d) $\mathbf{k}_4$}}
	\end{minipage}
	\vfill
	\caption{The utilized true blur kernels (the first row) as well as the estimated blur kernels (the second row) by the proposed method. 
	}
	\label{Fig_K_Estimation}
\end{figure}
\begin{table*}[!htbp]\small
	\centering
	\caption{The performance of each method on the Harvard dataset with different SR scale factors. The best results are in bold.}
	\setlength{\tabcolsep}{2.4mm}{\begin{tabular}{c|c|c|c|c|c|c|c|c|c|c|c|c}
			\hline
			\multirow{2}*{Methods}& \multicolumn{4}{c|}{s = 8}&\multicolumn{4}{c|}{s = 16}& \multicolumn{4}{c}{s = 32}\\
			\cline{2-13}&RMSE&PSNR&SAM&SSIM&RMSE&PSNR&SAM&SSIM&RMSE&PSNR&SAM&SSIM\\
			\hline	
			NSSR~\cite{7438864}&6.80&32.01&3.85&0.9320&7.05&31.73&5.47&0.9298&6.69&32.25&6.03&0.9463\\
			MHFnet~\cite{8953470}&4.80&35.23&5.47&0.9712&4.84&35.18&5.66&0.9698&5.03&34.85&5.78&0.9688\\
			\hline
			Yong~\cite{8019510}&3.15&38.73&3.47&0.9670&3.37&38.20&3.72&0.9675&3.68&37.49&4.20&0.9685\\
			UAL~\cite{Ours_CVPR2020}&7.06&32.19&6.00&0.9547&6.92&32.25&6.06&0.9530&6.57&32.63&6.11&0.9492\\
			\hline
			DBSR~\cite{9136736}&4.40&36.14&\textbf{2.48}&0.9660&5.58&34.31&\textbf{2.87}&0.9592&10.06&28.93&3.59&0.9359\\
			\hline
			Ours&\textbf{2.37}&\textbf{41.22}&2.53&\textbf{0.9803}&\textbf{2.59}&\textbf{40.40}&3.06&\textbf{0.9787}&\textbf{3.04}&\textbf{39.10}&\textbf{3.40}&\textbf{0.9757}\\
			\hline
	\end{tabular}}
	\label{Tab_Harvard_Result_Diff_Scale}
\end{table*}
\begin{figure*}[!htbp]
	\centering
	\vspace{-0.4cm}
	\begin{minipage}{0.05\textwidth}
		\centerline{\includegraphics[width=0.94in,height=0.94in]{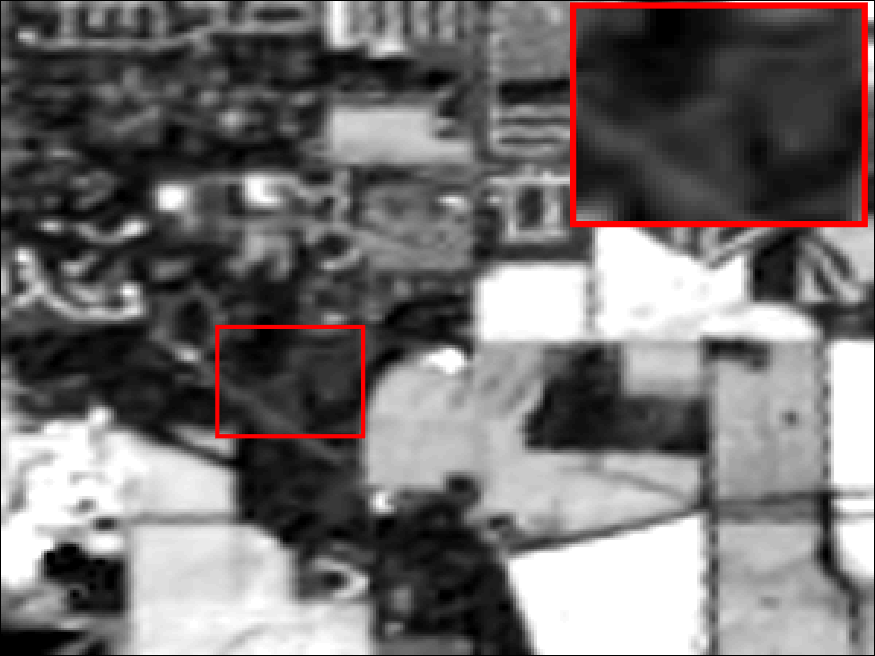}}
		\centerline{{\scriptsize (a) Bicubic}}
	\end{minipage}
	\hspace{1.4cm}
	\begin{minipage}{0.05\textwidth}
		\centerline{\includegraphics[width=0.94in,height=0.94in]{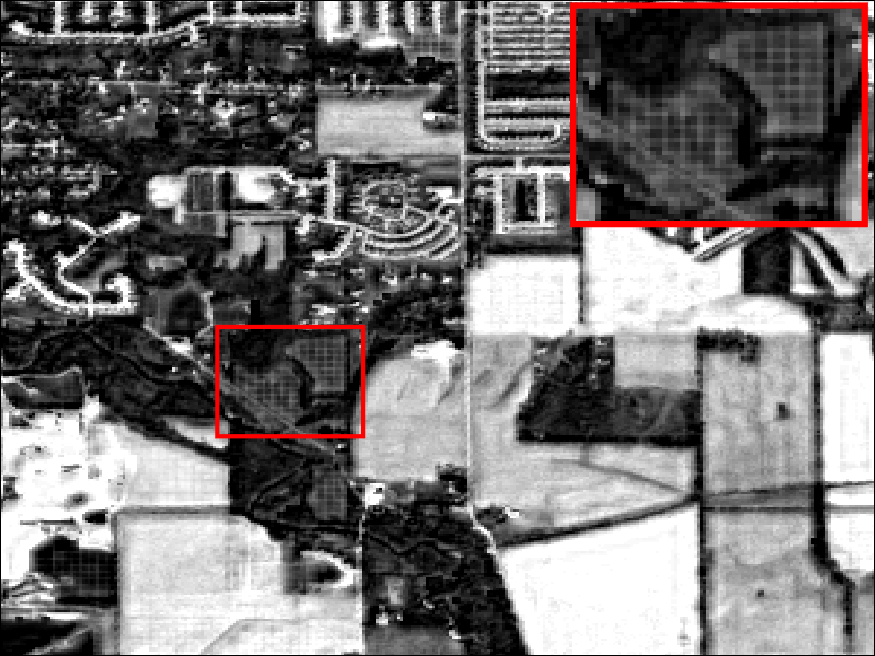}}
		\centerline{{\scriptsize (b) NSSR~\cite{7438864}/\textbf{12.84}}}
	\end{minipage}
	\hspace{1.4cm}
	\begin{minipage}{0.05\textwidth}
		\centerline{\includegraphics[width=0.94in,height=0.94in]{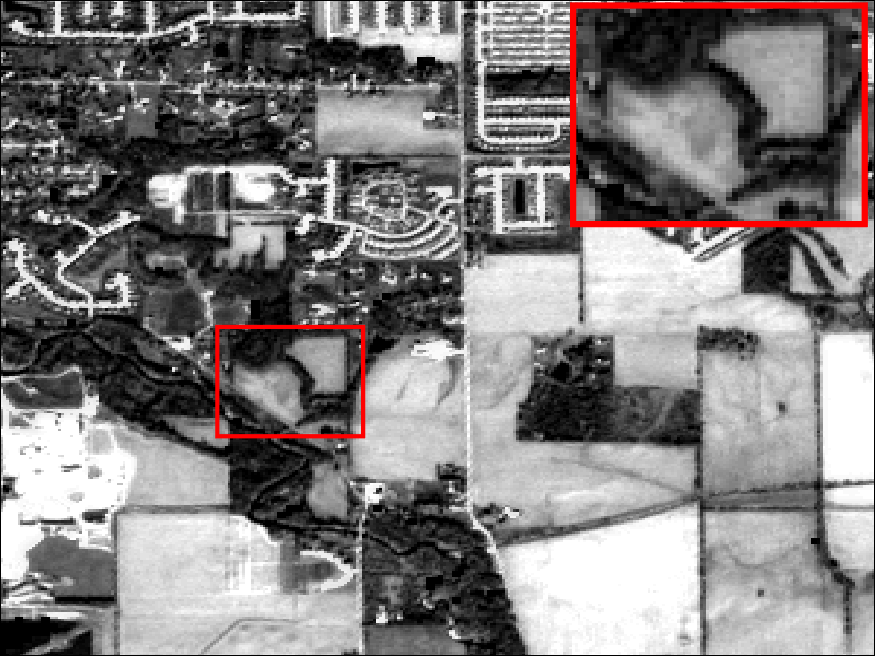}}
		\centerline{{\scriptsize (c) MHFnet~\cite{8953470}/\textbf{11.19}}}
	\end{minipage}
	\hspace{1.4cm}
	\begin{minipage}{0.05\textwidth}
		\centerline{\includegraphics[width=0.94in,height=0.94in]{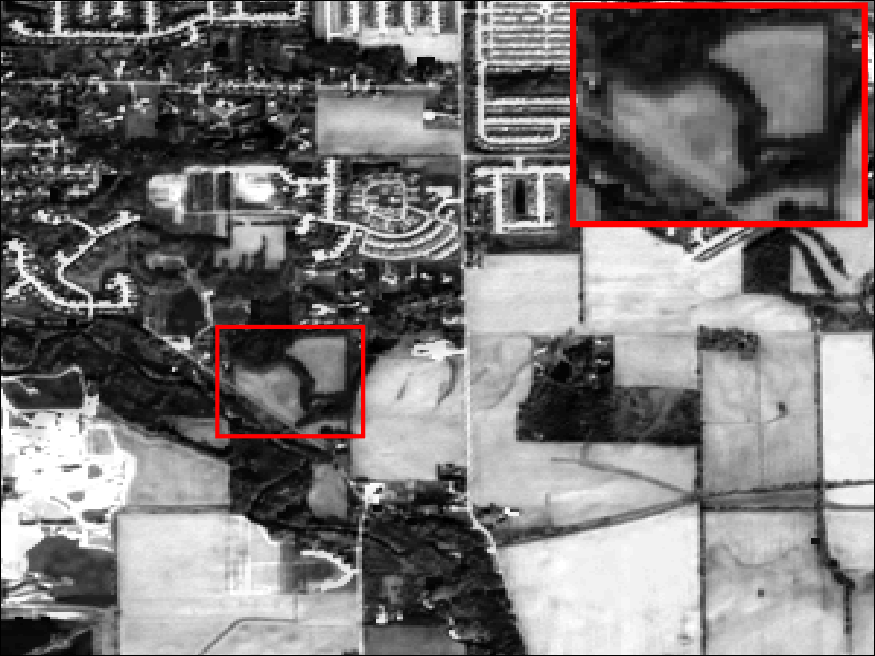}}
		\centerline{{\scriptsize (d) Yong~\cite{8019510}/\textbf{12.47}}}
	\end{minipage}
	\hspace{1.4cm}
	\begin{minipage}{0.05\textwidth}
		\centerline{\includegraphics[width=0.94in,height=0.94in]{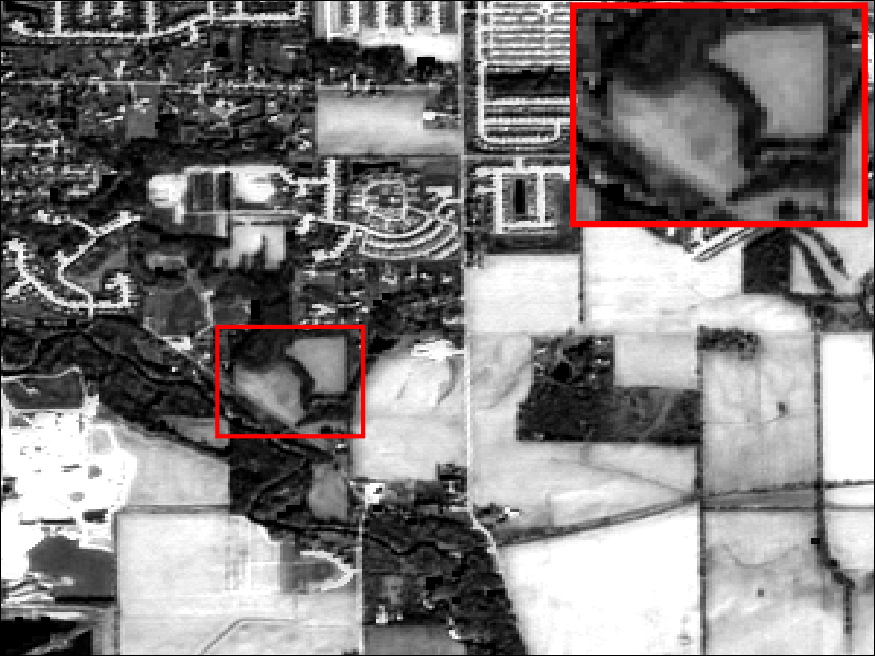}}
		\centerline{{\scriptsize (e) UAL~\cite{Ours_CVPR2020}/\textbf{10.98}}}
	\end{minipage}
	\hspace{1.4cm}
	\begin{minipage}{0.05\textwidth}
		\centerline{\includegraphics[width=0.94in,height=0.94in]{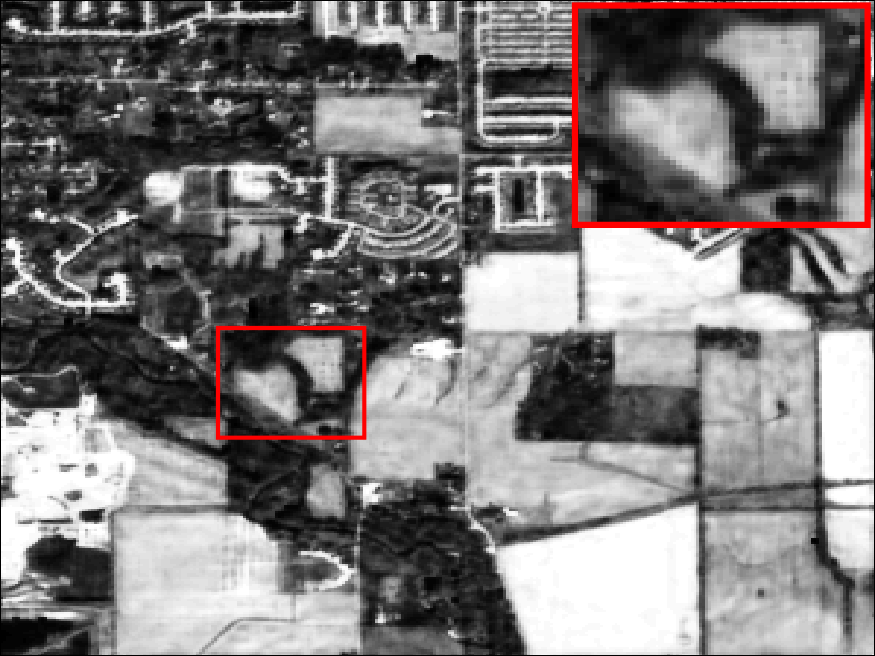}}
		\centerline{{\scriptsize (f) DBSR~\cite{9136736}/\textbf{12.40}}}
	\end{minipage}
	\hspace{1.4cm}
	\begin{minipage}{0.05\textwidth}
		\centerline{\includegraphics[width=0.94in,height=0.94in]{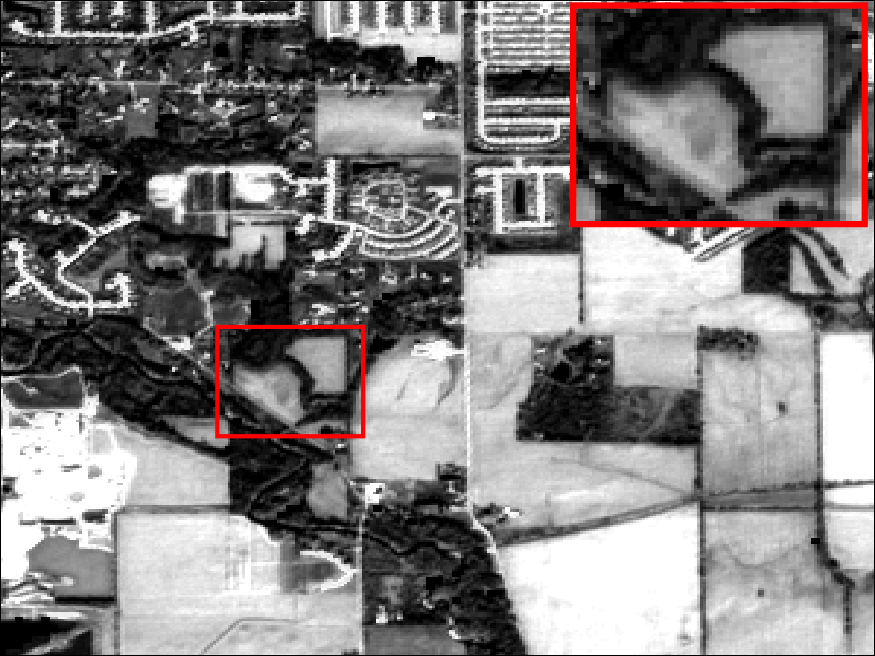}}
		\centerline{{\scriptsize (g) Ours/\textbf{10.87}}}
	\end{minipage}
	\caption{The visual SR results of all methods on the real dataset HypSen.}
	\label{Fig_Real_Visible}
\end{figure*}

\noindent \textbf{Different blur kernels}\quad
In this experiment, we further testify the performance of the proposed method influenced by different blur kernels and summarize the experimental results in Table~\ref{Tab_CAVE_Result_Diff_Kernels}. To visualize the accuracy of the estimated kernel, we further display the utilized true blur kernels as well as the estimated blur kernels in Figure~\ref{Fig_K_Estimation}. In experiments, $\mathbf{P}_{0.01}$ is adopted as the spectral response matrix. SNRs of both the observed LR HSI and HR MSI are 40dB. From the experimental results, we can see that the proposed method can better reconstruct the HR HSI as well as accurately estimate the blur kernels, compared with those from other competing methods. 

Besides these three experiments, another experiment is further conducted to verify the performance of the proposed method influenced by the noisy observation images with different intensity of noise. The proposed method still has obvious superiority over other methods, which can be seen from the supplement for detail. From all these experiments, we can conclude the proposed method obtains best performance in both accuracy and robustness.


\subsection{Performance comparison on the other datasets}

In this Subsection, we further conduct the experiments on the Harvard dataset and a real dataset HypSen dataset to verify the applicability of the proposed method to other datasets.

\noindent \textbf{Performance on the Harvard dataset}\quad
In this experiment, we compare the proposed method with other competing methods with different SR scales. The observed LR HSI and HR MSI are generated by $\mathbf{k}_1$ and $\mathbf{P}_{0.01}$, respectively. SNRs of both these two observed images are 40dB. The detailed numerical results are reported in Table~\ref{Tab_Harvard_Result_Diff_Scale}. Similar as the results on CAVE dataset, the results of the proposed method on Harvard dataset also has obvious advantages over the comparison methods. The visualization results of all methods on the Harvard dataset is consistent with the result given in Table~\ref{Tab_Harvard_Result_Diff_Scale}, which can be seen from the supplement for the details. 

\noindent \textbf{Performance on the Real dataset}\quad
In this experiment, we verify the proposed method on the real HypSen dataset without knowing the HR HSI. The visualization of reconstructed HSIs of all methods are shown in Figure~\ref{Fig_Real_Visible}, in which Figure~\ref{Fig_Real_Visible}(a) is the result directly up-sampling from the LR HSI via bicubic interpolation. 
It can be seen that the proposed method can well reconstruct the HSI and preserve the image details. In addition, we also measure the reconstruction quality for each method via the no-reference HSI evaluation metric~\cite{rs9040305} and list the obtained score in the legend of Figure~\ref{Fig_Real_Visible}. Smaller score represents better results. Thus, it can be seen that the proposed method obtains best performance from both the reconstructed image and the obtained scores. 

In summary, we can conclude that the proposed method can applicable to different datasets, which further demonstrates the effectiveness of the proposed method for HSI SR.

\subsection{Runtime analysis}
In the Table~\ref{Tab_CAVE_RunningTime} we summarize the average runtime of all the methods. The non-deep methods (\textit{i.e.}~NSSR~\cite{7438864} and Yong~\cite{8019510}) are run on a workstation with Intel Xeon E5-2640 CPU and 128G memory. The other methods are deep learning based methods, we run them on the same workstation with GeForce GTX 1080ti GPU. The Table~\ref{Tab_CAVE_RunningTime} demonstrates that the proposed method is faster than most unsupervised methods, especially DBSR~\cite{9136736} which requires almost 2900 second to reconstruct one single HSI.

\begin{table}[!htbp]\scriptsize 
	\centering
	\caption{Average runtime of all methods on CAVE dataset when SR scale is 8.}
	\hspace{0.005cm}
	\setlength{\tabcolsep}{1.2mm}{\begin{tabular}{c|c|c|c|c|c|c}
			\hline\hline
			Method&NSSR~\cite{7438864}& MHFnet~\cite{8953470}&Yong~\cite{8019510}&UAL~\cite{Ours_CVPR2020}&DBSR~\cite{9136736}&Ours\\
			\hline
			Time (s)&98.70&2.77&373.98&20.64&2899.80&32.40\\
			\hline\hline
	\end{tabular}}
	\label{Tab_CAVE_RunningTime}
\end{table}

\section{Conclusion}

In this study, we explore an unsupervised blind HSI fusion SR method, which can effectively estimate the degeneration models in spatial and spectral domain, respectively. To this end, we first propose an alternating optimization based deep framework to estimate the degeneration models and reconstruct the latent image, which makes them can mutually promotes each other. A compact reconstruction network is further designed to utilize the information containing in degeneration models. Additionally, we use a meta-learning based training mechanism to further pre-train the reconstruction network to make the proposed reconstruction network can fast adapt to different complex degeneration. Experiments on two benchmark HSI SR datasets and one real dataset demonstrate the proposed method is effective for the blind HSI SR.

\newpage
	
	\section{Implementation details}	
	In this material, we further supplement the details of the training and testing, as well as more experimental results of the proposed method.

	
	\subsection{Training details of backbone network}
	In the submitted manuscript, we propose to pre-train a backbone network $\mathcal{F}$ first, aiming to exploit the general image priors contained in the existing available amounts of HSI data. Since the pre-trained network $\mathcal{F}$ we utilized is to generate a rough reconstructed HSI as the initial input for the reconstruction network, most of the existing fusion based HSI network can be adopted as the backbone network. In this work, we select a multi-branch network~\cite{Ours_CVPR2020} as the $\mathcal{F}$ to jointly exploit the general spatial-spectral information that contain in the available HSI data. 
	
	We separate each benchmark dataset into training set and testing set. Specifically, we select 12 HSIs, 15 HSIs, 20 HSIs from the CAVE dataset, the Harvard dataset and NTIRE~\cite{8575282} dataset as the training data, respectively, and the remaining data within these datasets are left for test. NTIRE dataset is a newly introduced dataset. It has 255 HSIs each has $1392 \times 1300$ pixels in spatial domain and 31 bands covering the wavelength range from 400nm to 700nm.   
	
	To pre-train the backbone network, we crop the HSIs from the training set into overlapping patches with spatial size as $128 \times 128$ and stride as 64, and then utilize cropped HSIs to generate the corresponding LR HSI and HR MSI. It is noticeable that the spectral and spatial degeneration models we utilized to generate the LR HSI and HR MSI are different with those utilized in test. Specifically, the bluer kernel $\mathbf{k}$ is initialed with the Gaussian distribution, which kernel size and standard deviation are randomly chosen within the ranges [5, 15] and [0.5, 2], respectively. In addition, $c$ used for the spectral response matrix (i.e., spectral degeneration model) is chosen within the range [5e-3, 8e-3]. Once obtaining the LR HSI and HR MSI, we train the backbone network $\mathcal{F}$ by setting a $\ell_1$ norm based loss and Adam optimizer~\cite{Kingma2014Adam}.  The learning rate is initialized as 1e-4 and decreased every 10 epochs by 0.7. The maximum training epoch is 150 and the batchsize is set as 6. All deep models for the proposed method including the backbone network are implemented by the Pytorch~\cite{NEURIPS2019_9015} deep learning framework.
	
	\subsection{Training details of meta-learning}
	We first utlize the LR HSI and HR MSI for pre-training $\mathcal{F}$ in order to generate the rough reconstructed HSI $\hat{\mathbf{Z}}$. Then the $\hat{\mathbf{Z}}$ will be input to the alternating optimization based framework for meta-learning. The manner of generating the meta-learning dataset $\mathcal{M}$ is same as that for the backbone network. The meta-learning is conducted for 100 epochs, with initial task-level learning rate as 1e-3 and halved after every 10 epochs. Due to the spatial and spectral degeneration networks only contain few learnable parameters, thus, the meta-learning mechanism can not brings obvious improvements on it. Therefore, we adopt the default blur kernel and spectral response matrix as the initial parameters of them, respectively.
	
	\subsection{The details for the test}
	In the test phase, we first input the test LR HSI and HR MSI (without cropped) into the pre-trained backbone network $\mathcal{F}$, then input the obtained $\hat{\mathbf{Z}}$ into the degeneration network as the $\mathbf{Z}_0$, from which the parameters $\mathbf{k}_0$ and $\mathbf{P}_0$ related with the spatial and spectral degeneration network can be extracted. The obtained $\mathbf{k}_0$ and $\mathbf{P}_0$ are then combined with $\hat{\mathbf{Z}}$ be utilized to reconstruct the $\mathbf{Z}_1$ via the reconstruction network, from which the degeneration models and the reconstruction network are related. Finally, we optimize the degeneration models and the reconstruction network via the proposed alternating optimization framework.
	In the test phase, We also adopt the Adam algorithm~\cite{Kingma2014Adam} as the optimizer and $\ell_1$ norm based loss function to optimize the degeneration network and the meta-learning pre-trained reconstruction network. The learning rate of these two parts are 1e-4 and 1e-3, respectively. The total iteration number is 400, in which the outer-loop iteration number and the inner-loop iteration number is set as 40 and 10 in the experiment, respectively. 
	\section{More Experiments}
	\begin{table}[!htbp]\small
		\centering
		\caption{The performance of each method on the CAVE dataset with input images, which suffers from noise with different intensities. 
			The best results are in bold.}
		\setlength{\tabcolsep}{1.3mm}{\begin{tabular}{c|c|c|c|c|c}
				\hline\hline
				Noise intensity&Method&RMSE&PSNR&SAM&SSIM\\
				\hline
				\multirow{3}*{k1}&Ours\_Sep&5.72&34.13&8.98&0.9629\\	
				\cline{2-6}&Ours\_Joint&3.87&37.06&6.95&0.9828\\
				\cline{2-6}&Ours\_Alter&\textbf{2.97}&\textbf{39.38}&\textbf{6.75}&\textbf{0.9838}\\
				\hline
				\hline
				\multirow{3}*{k2}&Ours\_Sep&5.84&33.97&9.10&0.9628\\	
				\cline{2-6}&Ours\_Joint&3.96&36.84&7.08&0.9819\\
				\cline{2-6}&Ours\_Alter&\textbf{3.01}&\textbf{39.27}&\textbf{6.80}&\textbf{0.9837}\\
				\hline
				\hline
				\multirow{3}*{k3}&Ours\_Sep&5.69&34.21&8.99&0.9625\\	
				\cline{2-6}&Ours\_Joint&4.30&36.26&7.03&0.9809\\
				\cline{2-6}&Ours\_Alter&\textbf{3.33}&\textbf{38.58}&\textbf{6.75}&\textbf{0.9833}\\
				\hline\hline
				\multirow{3}*{k4}&Ours\_Sep&5.74&34.16&8.99&0.9625\\	
				\cline{2-6}&Ours\_Joint&4.18&36.48&7.00&0.9816\\
				\cline{2-6}&Ours\_Alter&\textbf{3.03}&\textbf{39.41}&\textbf{6.76}&\textbf{0.9832}\\
				\hline\hline
		\end{tabular}}
		\label{Tab_Diff_Opti}
	\end{table}
	\subsection{Ablation study}
\begin{figure}[!htbp]
		\centering
		\begin{minipage}{0.05\textwidth}
			\centerline{\includegraphics[width=0.65in,height=0.65in]{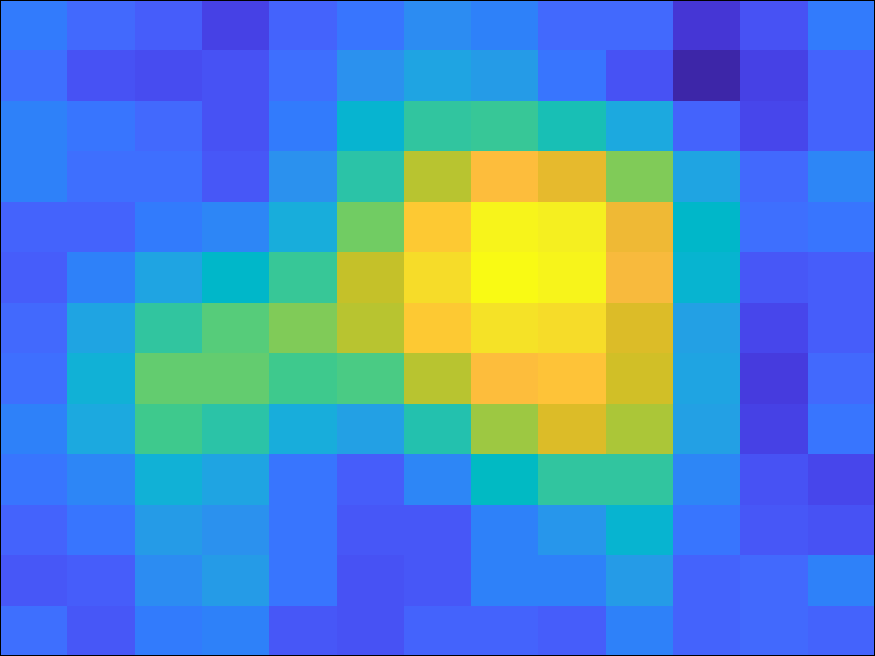}}
		\end{minipage}
		\hspace{0.65cm}
		\begin{minipage}{0.05\textwidth}
			\centerline{\includegraphics[width=0.65in,height=0.65in]{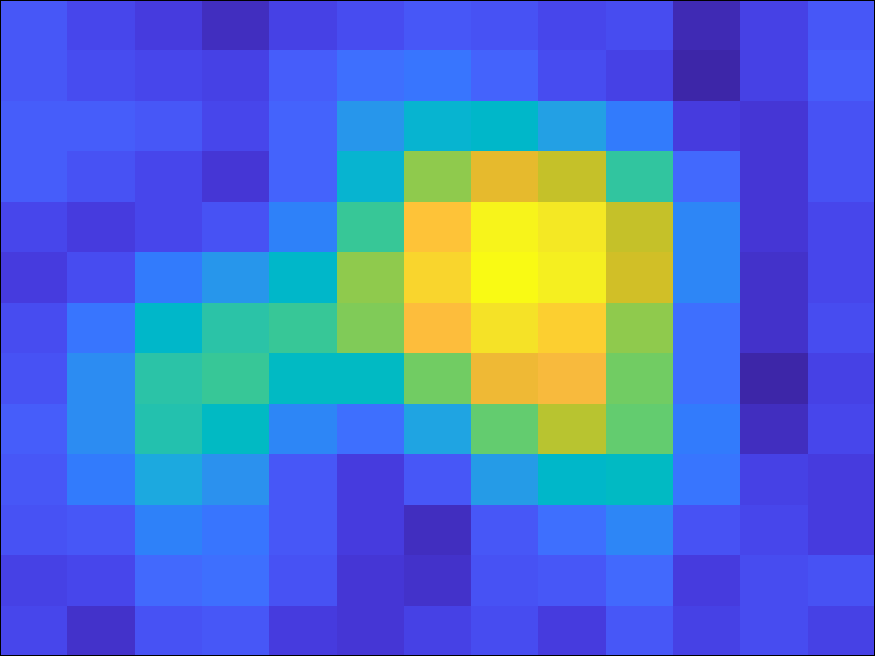}}
		\end{minipage}
		\hspace{0.65cm}
		\begin{minipage}{0.05\textwidth}
			\centerline{\includegraphics[width=0.65in,height=0.65in]{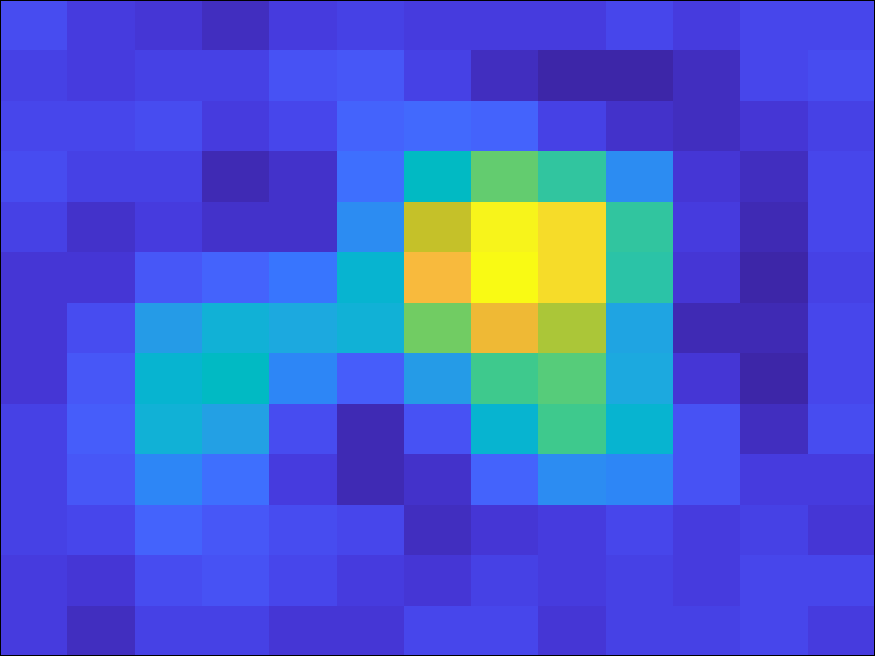}}
		\end{minipage}
		\hspace{0.65cm}
		\begin{minipage}{0.05\textwidth}
			\centerline{\includegraphics[width=0.65in,height=0.65in]{./Picture_Result/KernelEstimation/K1/cloth_GT.png}}
		\end{minipage}
		\vfill
		\begin{minipage}{0.05\textwidth}
			\centerline{\includegraphics[width=0.65in,height=0.65in]{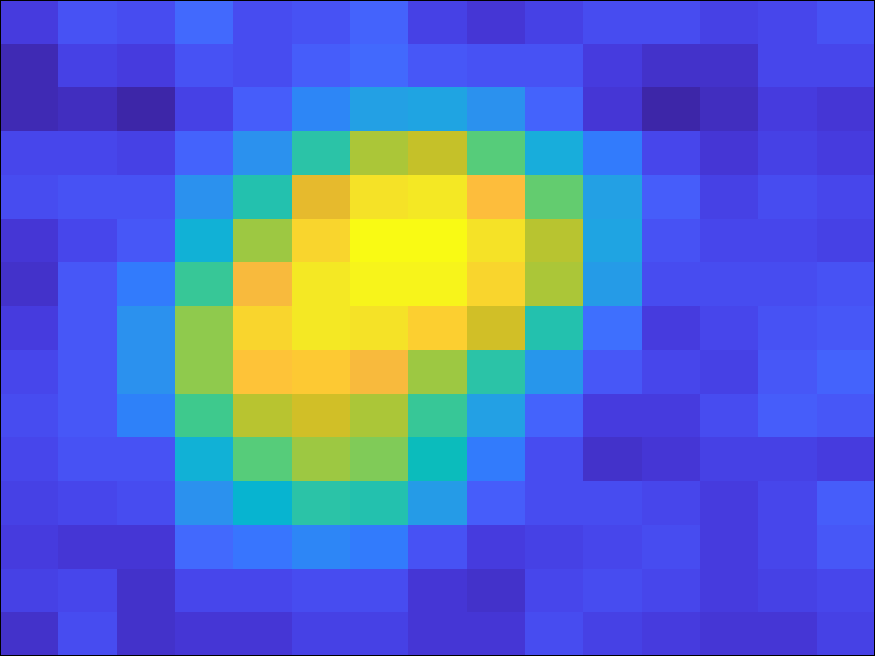}}
		\end{minipage}
		\hspace{0.65cm}
		\begin{minipage}{0.05\textwidth}
			\centerline{\includegraphics[width=0.65in,height=0.65in]{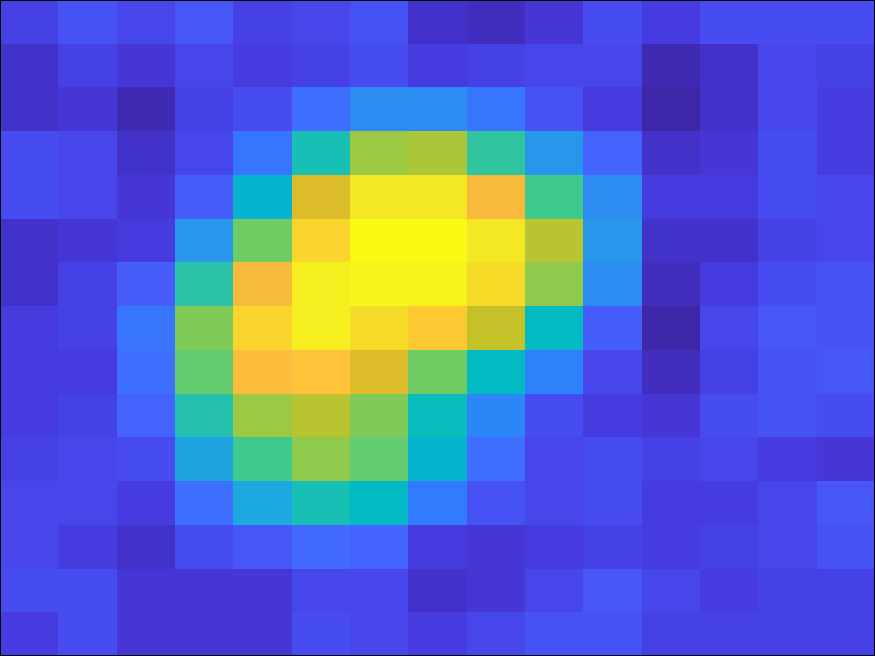}}
		\end{minipage}
		\hspace{0.65cm}
		\begin{minipage}{0.05\textwidth}
			\centerline{\includegraphics[width=0.65in,height=0.65in]{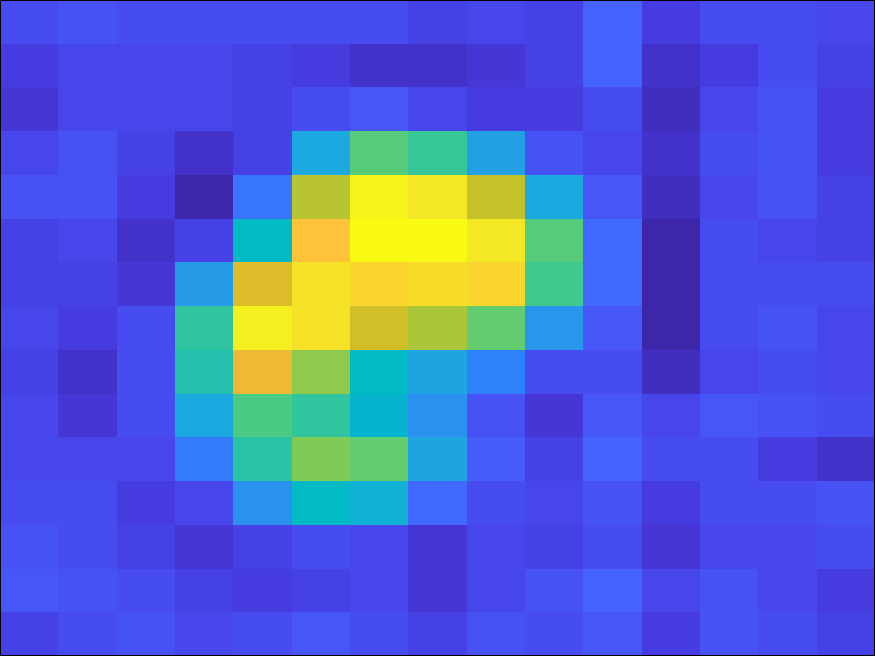}}
		\end{minipage}
		\hspace{0.65cm}
		\begin{minipage}{0.05\textwidth}
			\centerline{\includegraphics[width=0.65in,height=0.65in]{./Picture_Result/KernelEstimation/K2/cloth_GT.png}}
		\end{minipage}
		\vfill
		\begin{minipage}{0.05\textwidth}
			\centerline{\includegraphics[width=0.65in,height=0.65in]{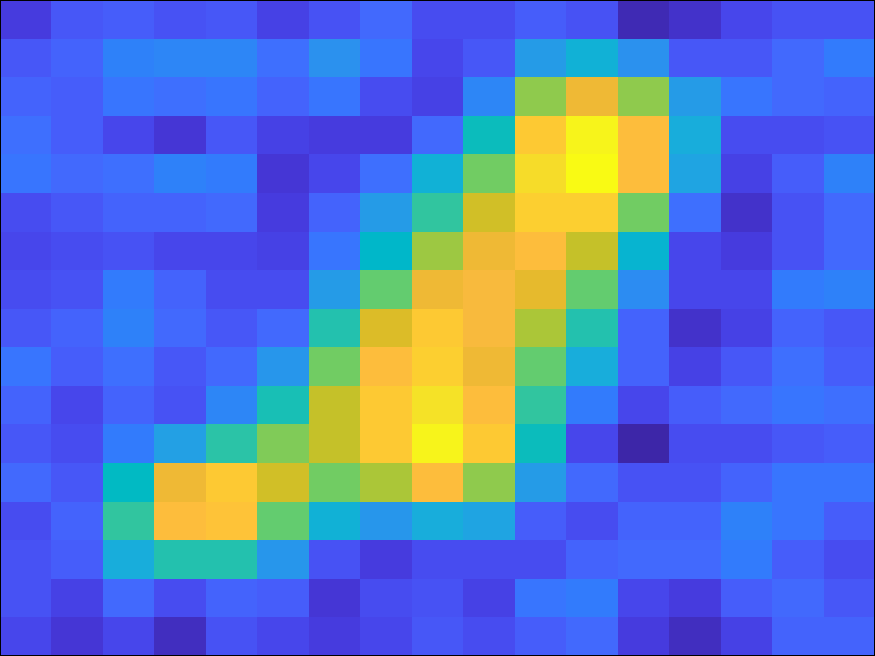}}
		\end{minipage}
		\hspace{0.65cm}
		\begin{minipage}{0.05\textwidth}
			\centerline{\includegraphics[width=0.65in,height=0.65in]{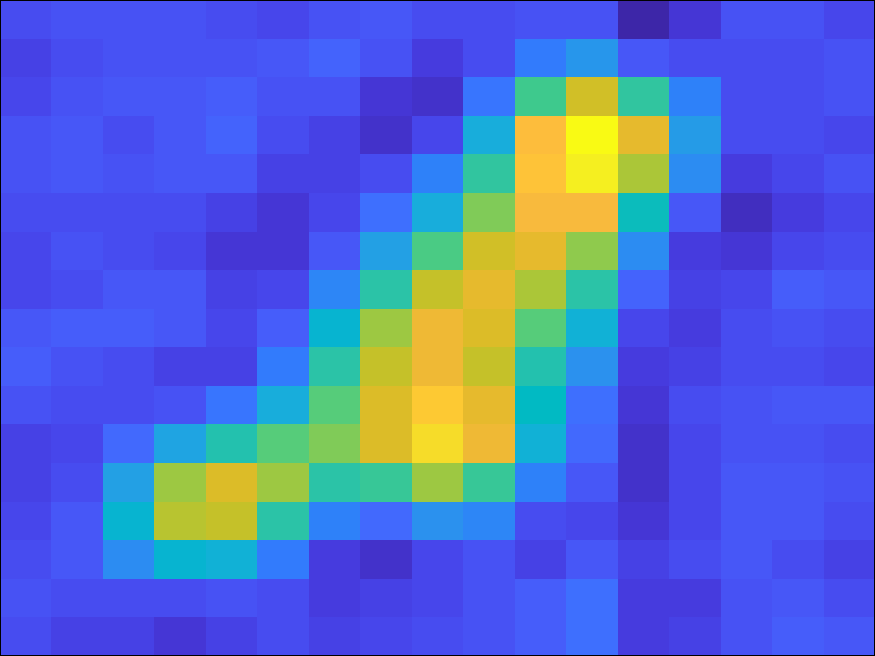}}
		\end{minipage}
		\hspace{0.65cm}
		\begin{minipage}{0.05\textwidth}
			\centerline{\includegraphics[width=0.65in,height=0.65in]{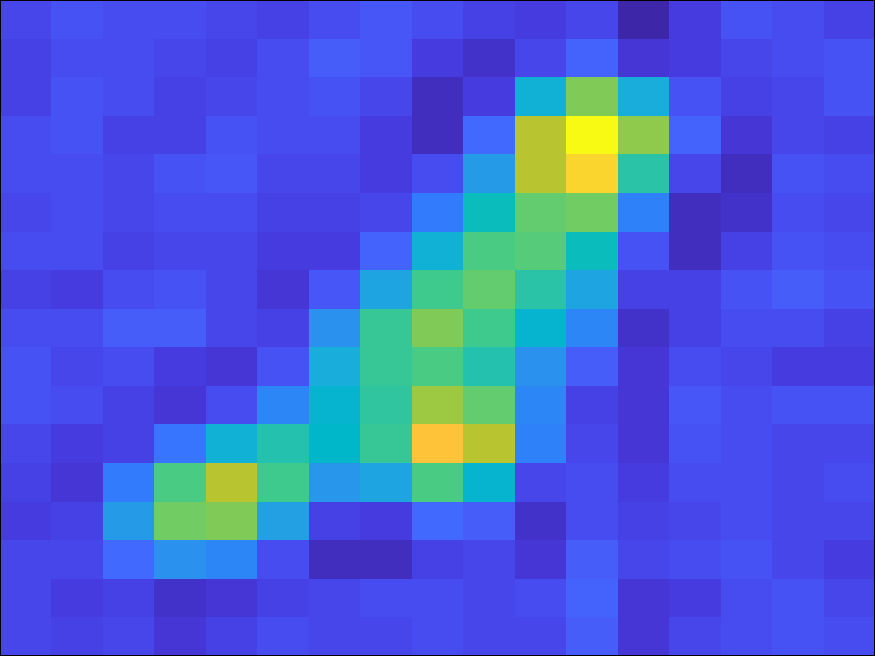}}
		\end{minipage}
		\hspace{0.65cm}
		\begin{minipage}{0.05\textwidth}
			\centerline{\includegraphics[width=0.65in,height=0.65in]{./Picture_Result/KernelEstimation/K3/cloth_GT.png}}
		\end{minipage}
		\vfill
		\begin{minipage}{0.05\textwidth}
			\centerline{\includegraphics[width=0.65in,height=0.65in]{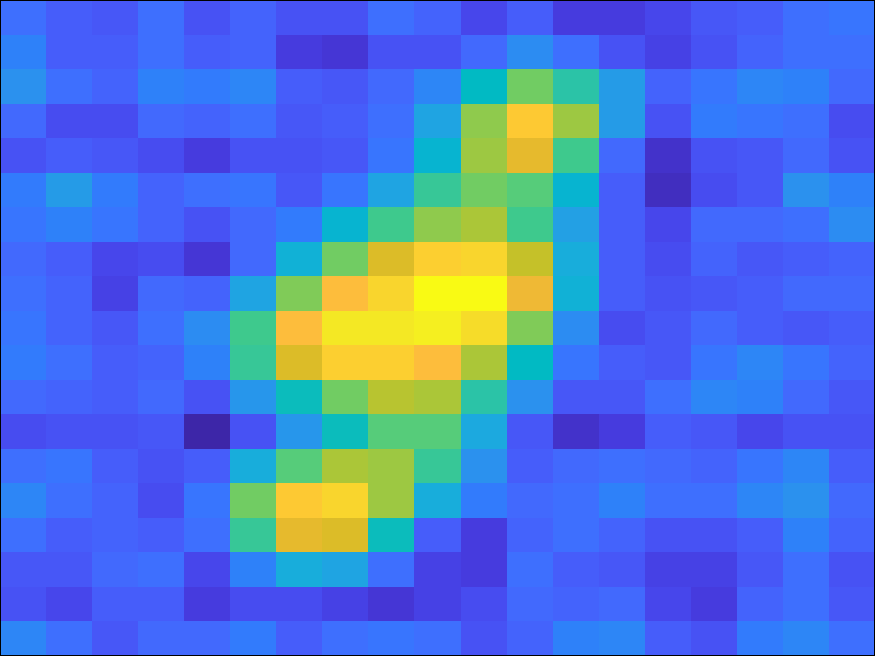}}
			\centerline{{\scriptsize (a) Ours\_Sep}}
		\end{minipage}
		\hspace{0.65cm}
		\begin{minipage}{0.05\textwidth}
			\centerline{\includegraphics[width=0.65in,height=0.65in]{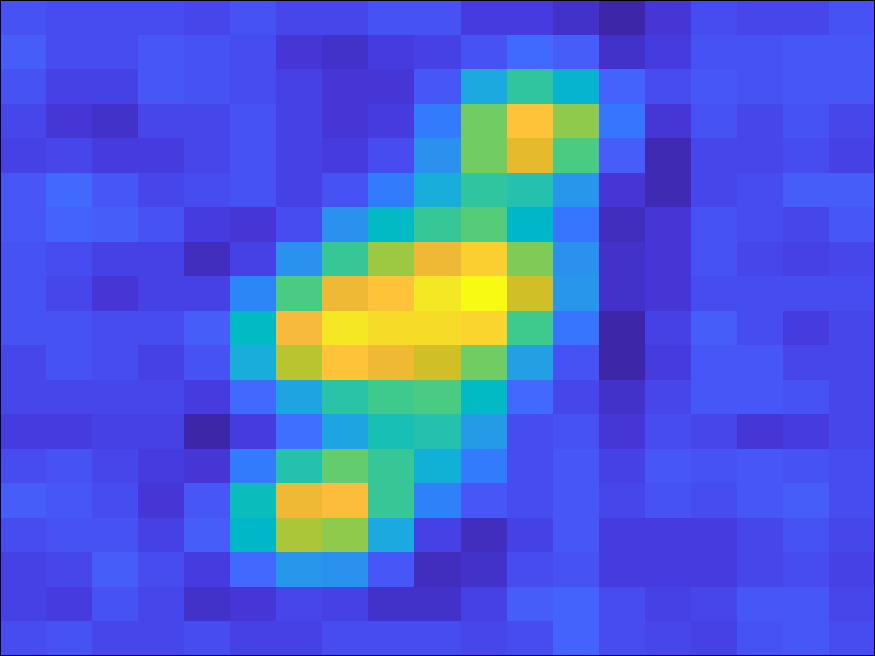}}
			\centerline{{\scriptsize (b) Ours\_Joint}}
		\end{minipage}
		\hspace{0.65cm}
		\begin{minipage}{0.05\textwidth}
			\centerline{\includegraphics[width=0.65in,height=0.65in]{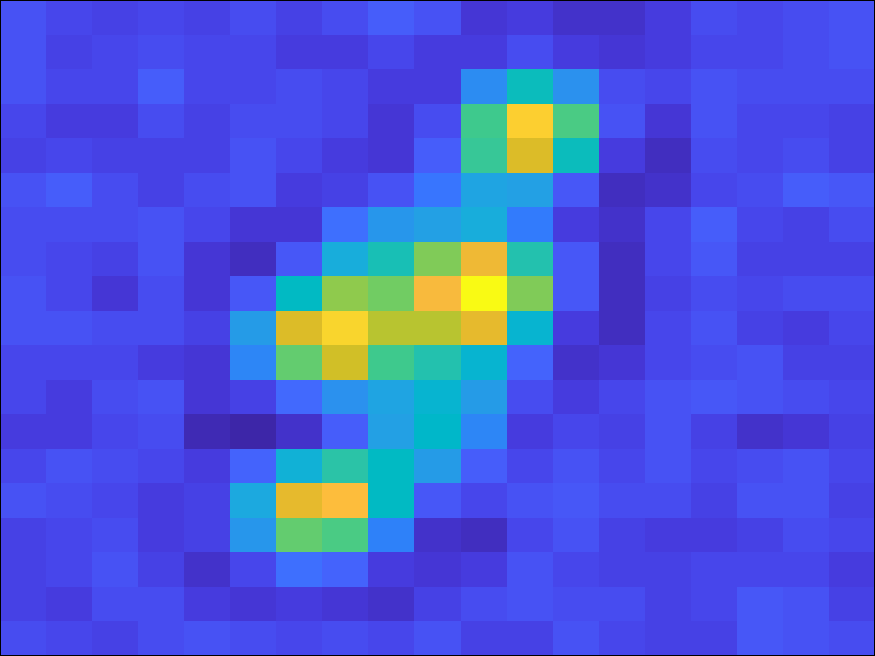}}
			\centerline{{\scriptsize (c) Ours\_Alter}}
		\end{minipage}
		\hspace{0.65cm}
		\begin{minipage}{0.05\textwidth}
			\centerline{\includegraphics[width=0.65in,height=0.65in]{./Picture_Result/KernelEstimation/K4/cloth_GT.png}}
			\centerline{{\scriptsize (d) GT}}
		\end{minipage}
		\vfill
		\vspace{0.1cm}
		\caption{The estimated four kind of kernels by different optimization manners, where the last column are Ground Truth (GT) and the first three columns are the results from Ours\_Alter, Ours\_Joint and Ours\_Sep, respectively.}
		\label{Fig_K_Comparison}
	\end{figure}

\begin{figure}[!htbp]
		\centering
		\hspace{-1.3cm}
		\begin{minipage}{0.05\textwidth}
			\centerline{\rotatebox{90}{ {\footnotesize Ours\_Sep}}}
		\end{minipage}
		\hspace{0.6cm}
		\begin{minipage}{0.05\textwidth}
			\centerline{\includegraphics[width=1.2in,height=0.8in]{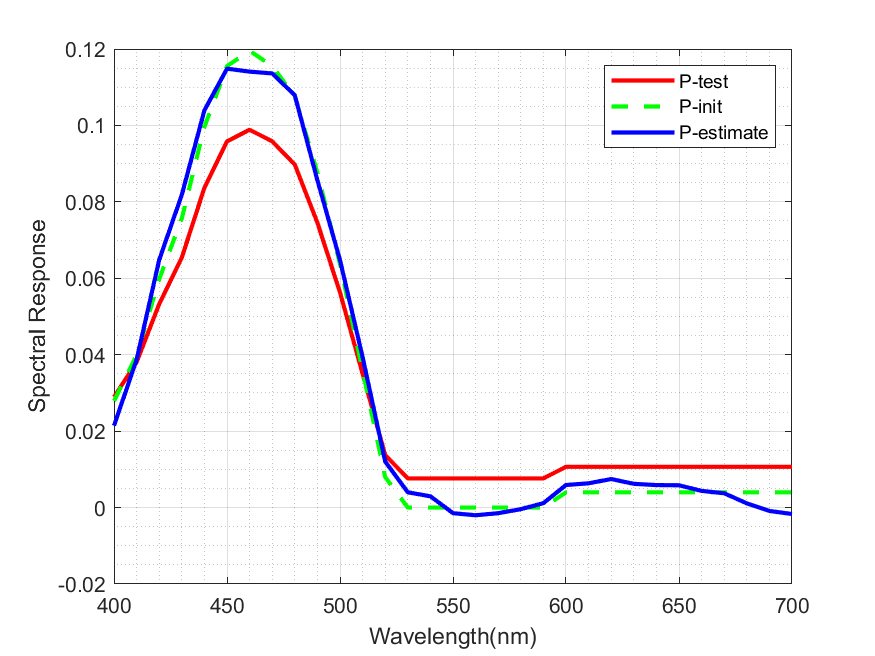}}
		\end{minipage}
		\hspace{1.75cm}
		\begin{minipage}{0.05\textwidth}
			\centerline{\includegraphics[width=1.2in,height=0.8in]{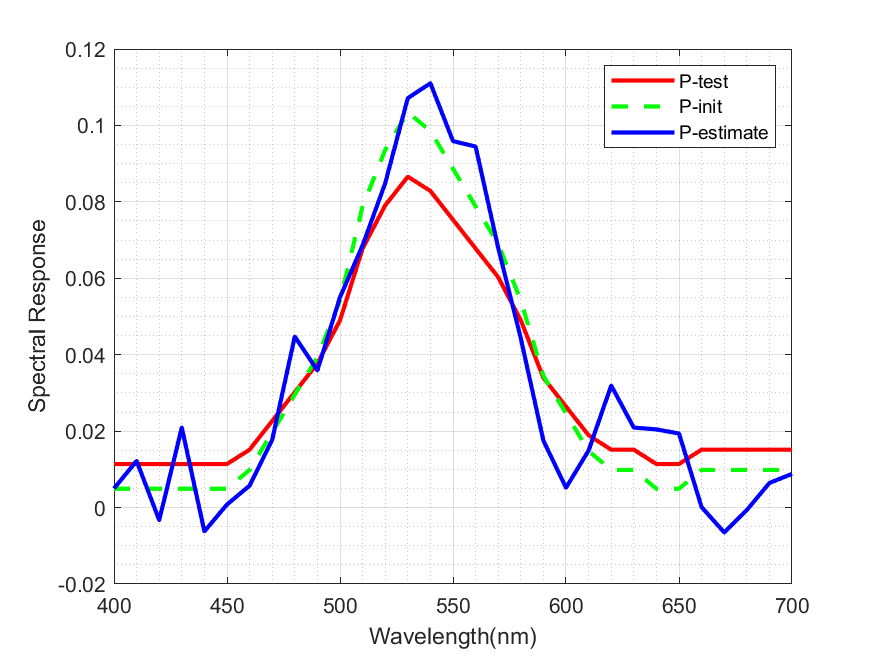}}
		\end{minipage}
		\hspace{1.75cm}
		\begin{minipage}{0.05\textwidth}
			\centerline{\includegraphics[width=1.2in,height=0.8in]{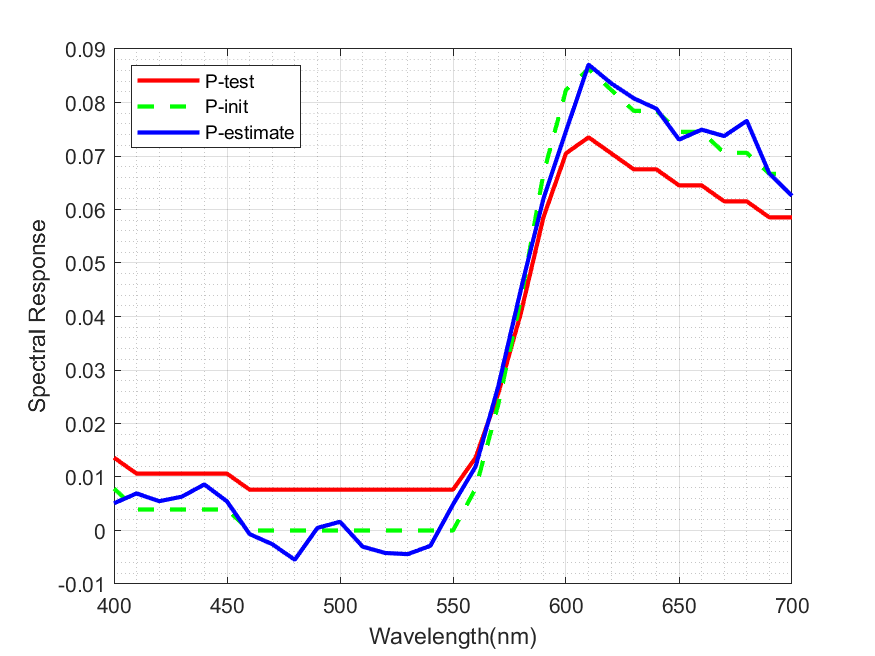}}
		\end{minipage}
		\vfill
		\hspace{-1.3cm}
		\begin{minipage}{0.05\textwidth}
			\centerline{\rotatebox{90}{ {\footnotesize Ours\_Joint}}}
		\end{minipage}
		\hspace{0.6cm}
		\begin{minipage}{0.05\textwidth}
			\centerline{\includegraphics[width=1.2in,height=0.8in]{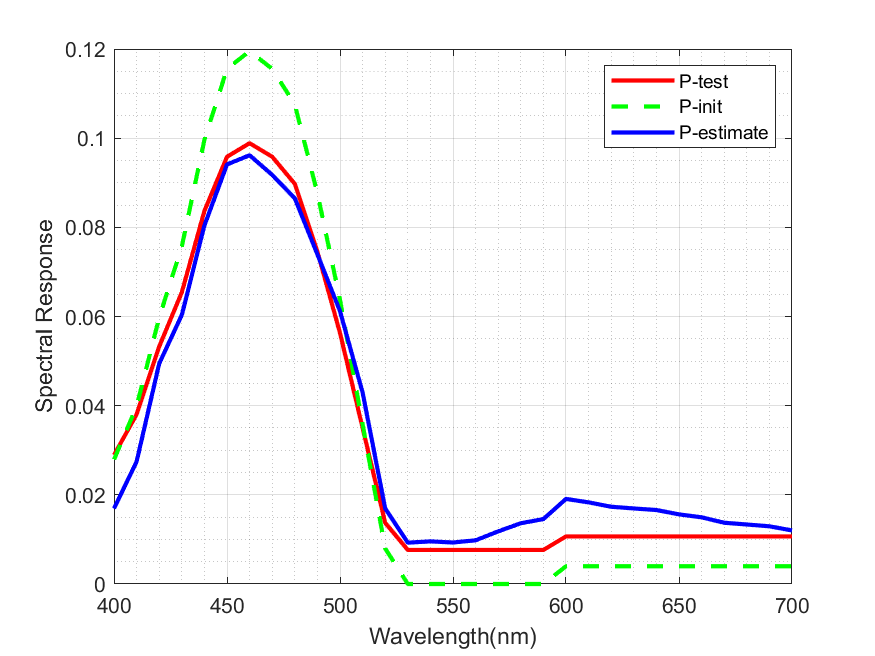}}
		\end{minipage}
		\hspace{1.75cm}
		\begin{minipage}{0.05\textwidth}
			\centerline{\includegraphics[width=1.2in,height=0.8in]{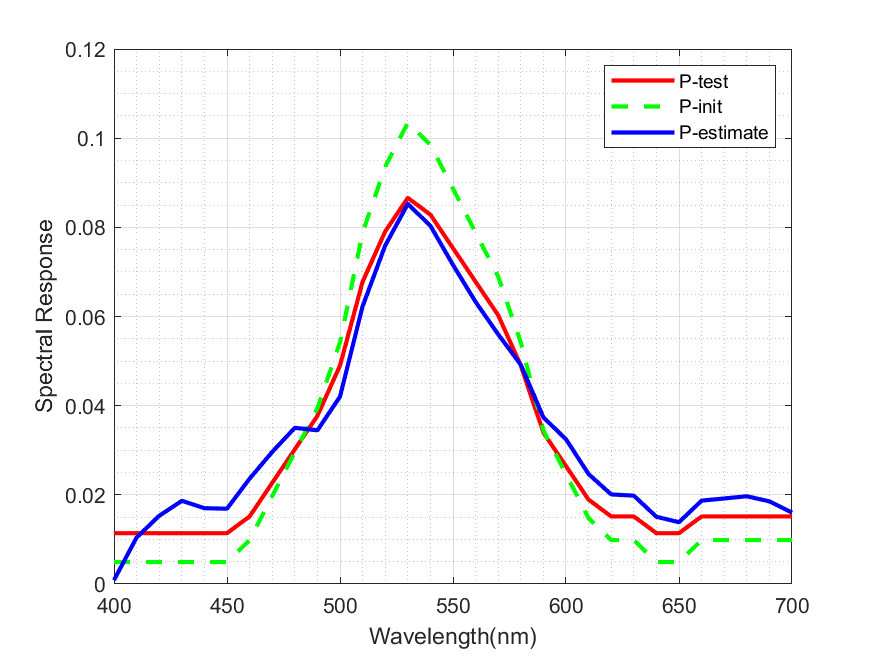}}
		\end{minipage}
		\hspace{1.75cm}
		\begin{minipage}{0.05\textwidth}
			\centerline{\includegraphics[width=1.2in,height=0.8in]{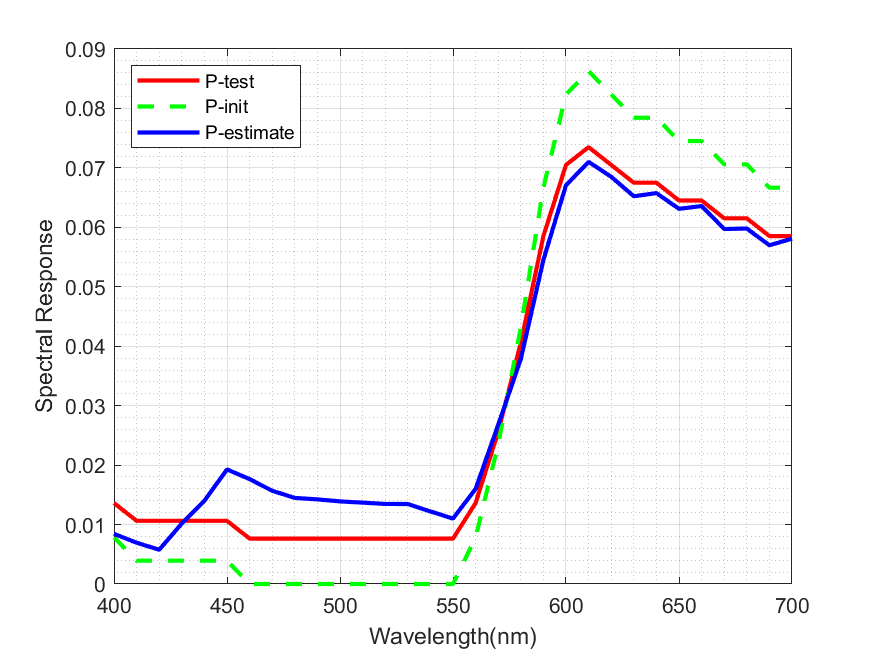}}
		\end{minipage}
		\vfill
		\hspace{-1.3cm}
		\begin{minipage}{0.05\textwidth}
			\centerline{\rotatebox{90}{ {\footnotesize Ours\_Alter}}}
		\end{minipage}
		\hspace{0.6cm}
		\begin{minipage}{0.05\textwidth}
			\centerline{\includegraphics[width=1.2in,height=0.8in]{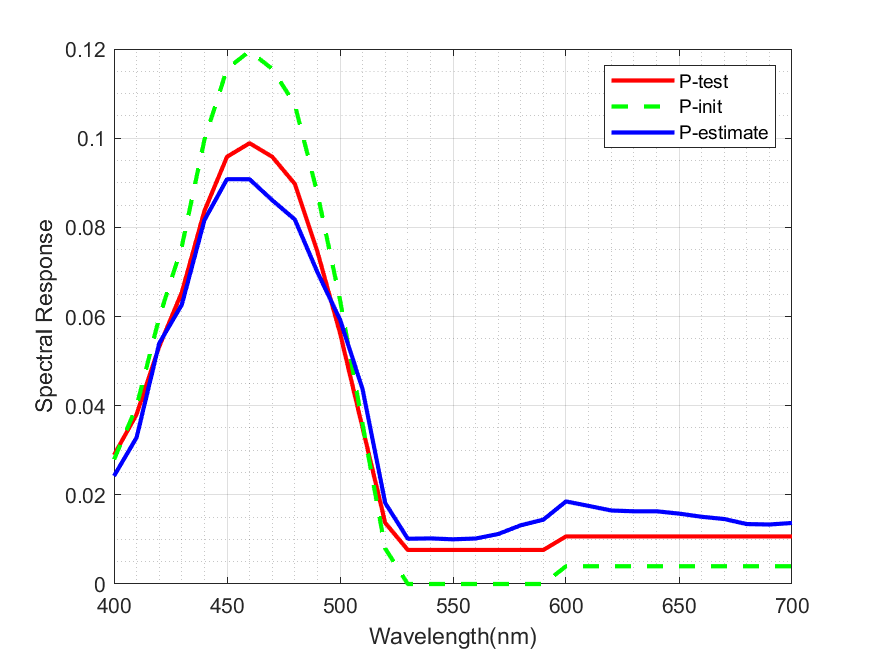}}
			\centerline{{\scriptsize (a) Blue band}}
		\end{minipage}
		\hspace{1.75cm}
		\begin{minipage}{0.05\textwidth}
			\centerline{\includegraphics[width=1.2in,height=0.8in]{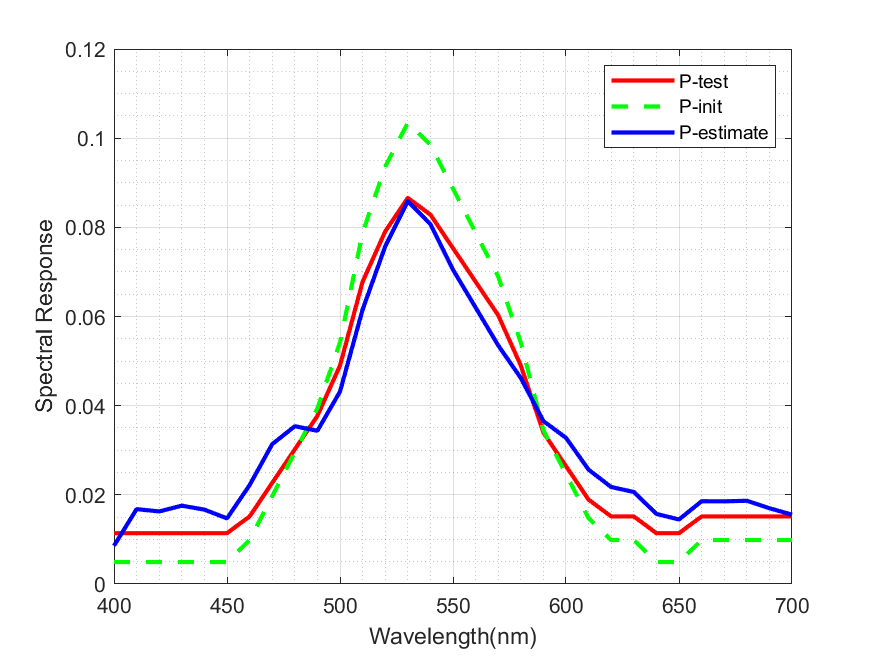}}
			\centerline{{\scriptsize (b) Green band}}
		\end{minipage}
		\hspace{1.75cm}
		\begin{minipage}{0.05\textwidth}
			\centerline{\includegraphics[width=1.2in,height=0.8in]{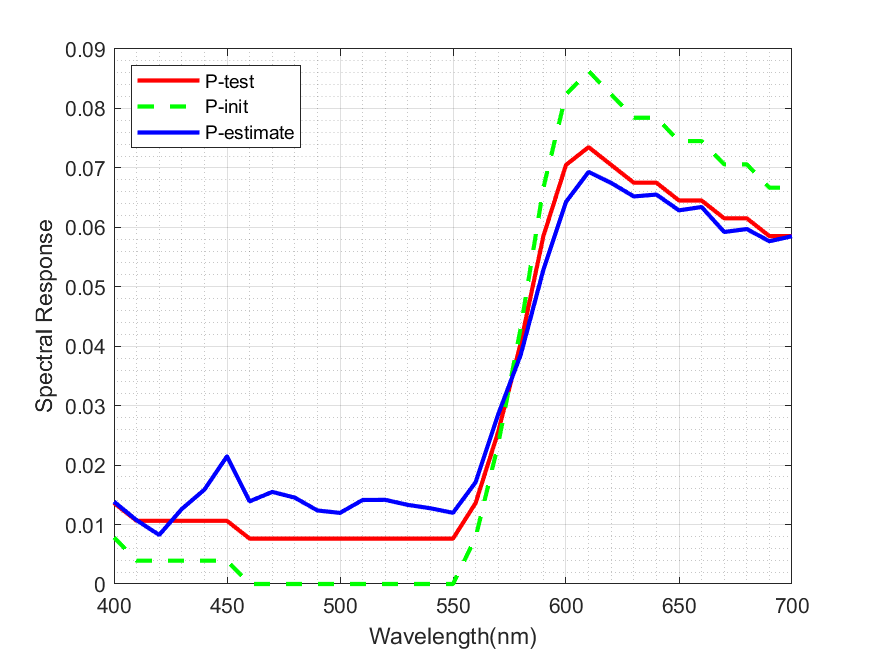}}
			\centerline{{\scriptsize (c) Red band}}
		\end{minipage}
		\vfill
		\caption{The estimated spectral response matrices by different optimization manners, from the top row to the bottom are Ours\_Sep, Ours\_Joint and Ours\_Alter, respectively.}
		\label{Fig_P_Estimation_Sup}
	\end{figure}

	\noindent \textbf{Effect of the optimization manner}\quad In the main manuscript, we have provided partial results to demonstrate the superiority of the alternating optimization framework. In this Subsection, we supplement an ablation study to further compare the performance of different optimization manners (\textit{e.g.}~Our\_Sep, Our\_Joint and Our\_Alter). Specifically, we testify these three methods using all four motion kernels with $\mathbf{P}_{0.02}$ as the test spectral response matrix. SNRs of both the observed LR HSI and HR MSI are 40dB and the SR scale is 8. The detailed numerical results are reported in the Table~\ref{Tab_Diff_Opti}. It can be seen the performance of alternating optimization based method is obvious superior to the other two kind of optimization manners, which is consistent with the conclusion in the main manuscript. In addition, we further plot the estimated blur kernels (\textit{e.g.}~Figure~\ref{Fig_K_Comparison}) and the spectral response matrices (\textit{e.g.}~Figure~\ref{Fig_P_Estimation_Sup}) of these three kinds of methods. 
	It can be seen that though all these three kind of methods can estimate the kernel with a similar shape as the test one, the estimated kernels of Ours\_Joint and Ours\_Sep are much fuzzier than that estimated by Ours\_Alter. The estimated spectral response matrices shown in the Figure~\ref{Fig_P_Estimation_Sup} also demonstrate the Ours\_Sep inferiors to the other two kinds of methods. Furthermore, the Ours\_Alter can well estimate the degeneration models in both spectral and spatial domains, thus presents a better performance than Ours\_Joint on blind HSI fusion SR.
	
\begin{figure}[!tbp]
		\centerline{\includegraphics[width=3in,height=2in]{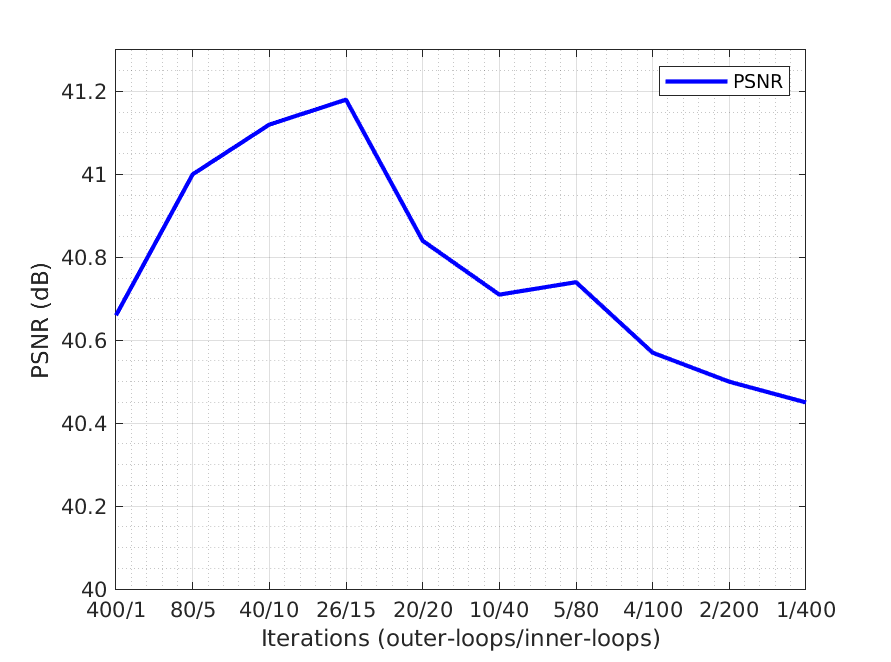}}
		\centering
		\caption{The variations of PSNR when changing the iteration numbers of outer-loop and inner-loop.}
		\label{VariationWithIteration}
	\end{figure}

	\noindent \textbf{Effect of different iteration numbers}\quad
	The proposed method is implemented by iteratively alternating optimization strategy, which is consisted by an outer-loop iteration and an inner-loop iteration. $I_{out}$ denotes the number of conducting alternating optimization (outer-loop) and $I_{in}$ indicates the independent optimization for the estimation and reconstruction networks (inner-loop). In this experiment, we fix the total iteration number $I$ as 400 ($I = I_{out} \times I_{in}$). Thus, the increase of  $I_{out}$ will decrease the correspondingly $I_{in}$, and vice versa. Therefore, in our experiment, we only verify the influence of changing $I_{out}$, which is directly related with alternating optimization. In experiment, we conduct the experiments on the CAVE dataset with different $I_{out}$. SNRs of both the observed LR HSI and HR MSI are 40dB and the SR scale is 8. The blur kernels and the spectral response matrix for the test data is $\mathbf{k}_1$ with $\mathbf{P}_{0.01}$, respectively. The experimental results are plotted in Figure~\ref{VariationWithIteration}. It can be seen a large $I_{out}$ often presents a better PSNR result, which demonstrates that the increase of time on alternation optimization is benefit to improve the performance of blind HSI fusion SR. Thus, in all experiments, we empirically set the number of $I_{out}$ is 40.     
	
	\begin{figure}[!htbp]
		\centering
		\begin{minipage}{0.05\textwidth}
			\centerline{\includegraphics[width=0.65in,height=0.65in]{./Picture_Result/KernelEstimation/K1/cloth_GT.png}}
		\end{minipage}
		\hspace{0.65cm}
		\begin{minipage}{0.05\textwidth}
			\centerline{\includegraphics[width=0.65in,height=0.65in]{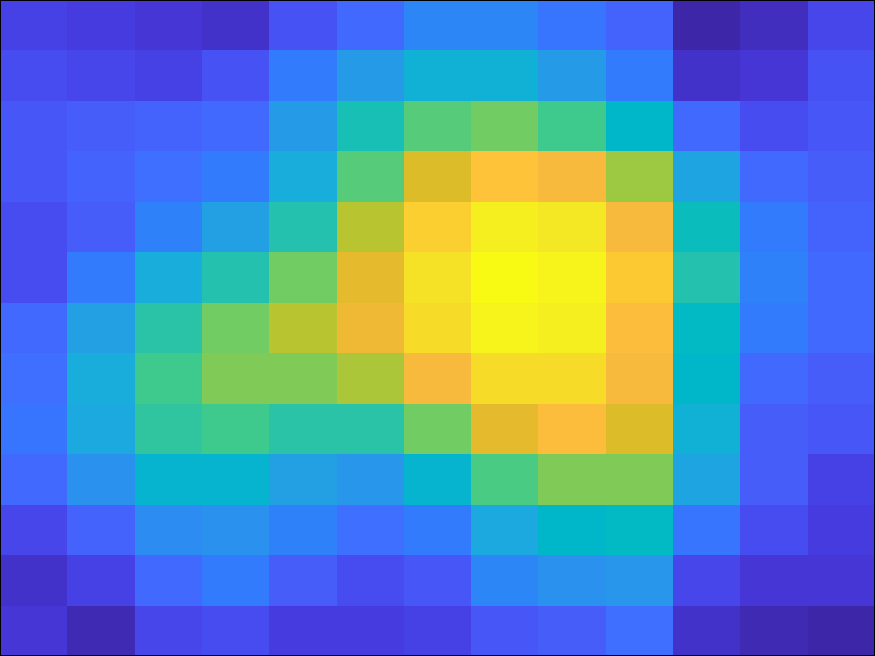}}
		\end{minipage}
		\hspace{0.65cm}
		\begin{minipage}{0.05\textwidth}
			\centerline{\includegraphics[width=0.65in,height=0.65in]{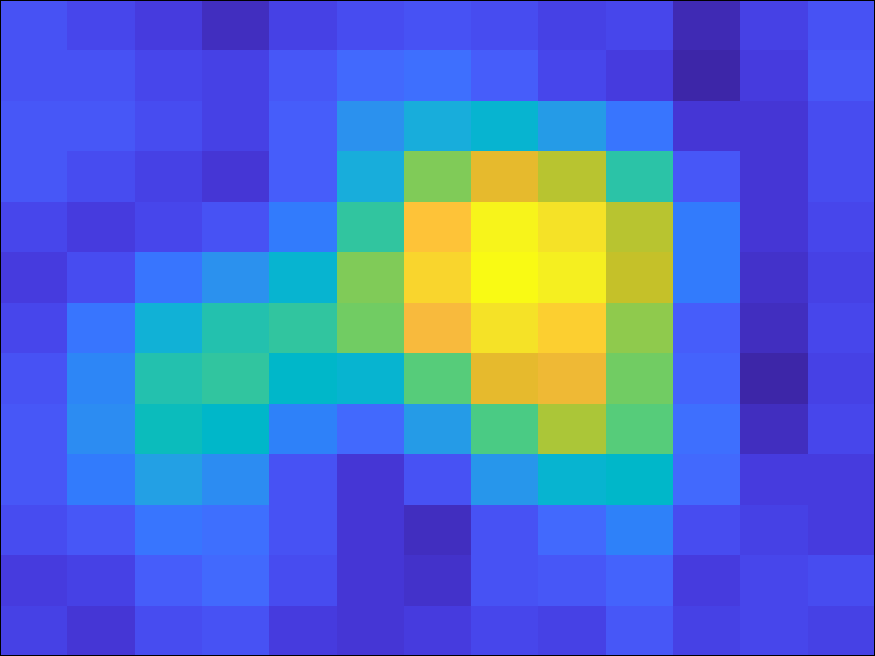}}
		\end{minipage}
		\hspace{0.65cm}
		\begin{minipage}{0.05\textwidth}
			\centerline{\includegraphics[width=0.65in,height=0.65in]{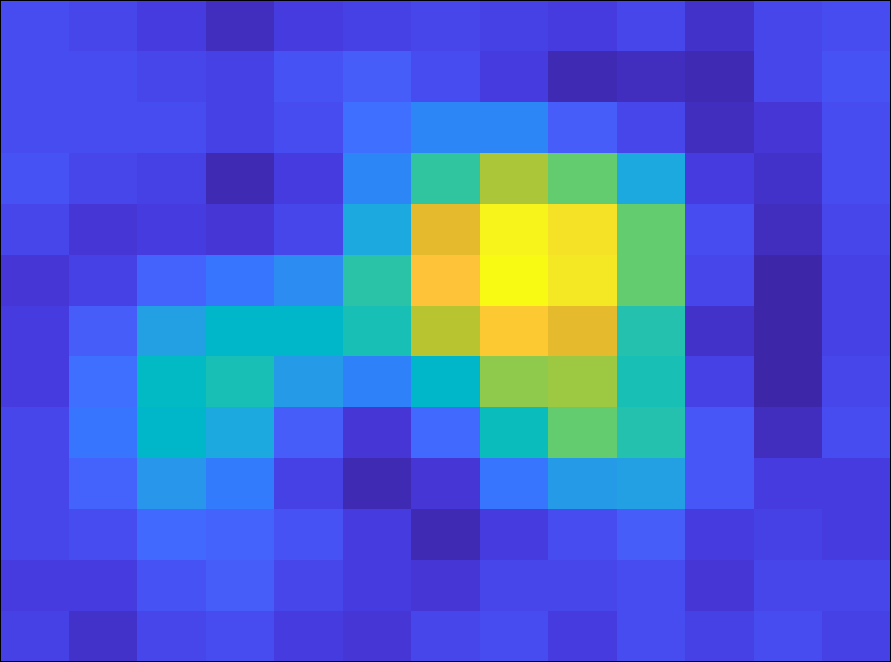}}
		\end{minipage}
		\hspace{0.65cm}
		\begin{minipage}{0.05\textwidth}
			\centerline{\includegraphics[width=0.65in,height=0.65in]{./Picture_Result/KernelEstimation/K1/cloth_39.png}}
		\end{minipage}
		\vfill
		\begin{minipage}{0.05\textwidth}
			\centerline{\includegraphics[width=0.65in,height=0.65in]{./Picture_Result/KernelEstimation/K2/cloth_GT.png}}
		\end{minipage}
		\hspace{0.65cm}
		\begin{minipage}{0.05\textwidth}
			\centerline{\includegraphics[width=0.65in,height=0.65in]{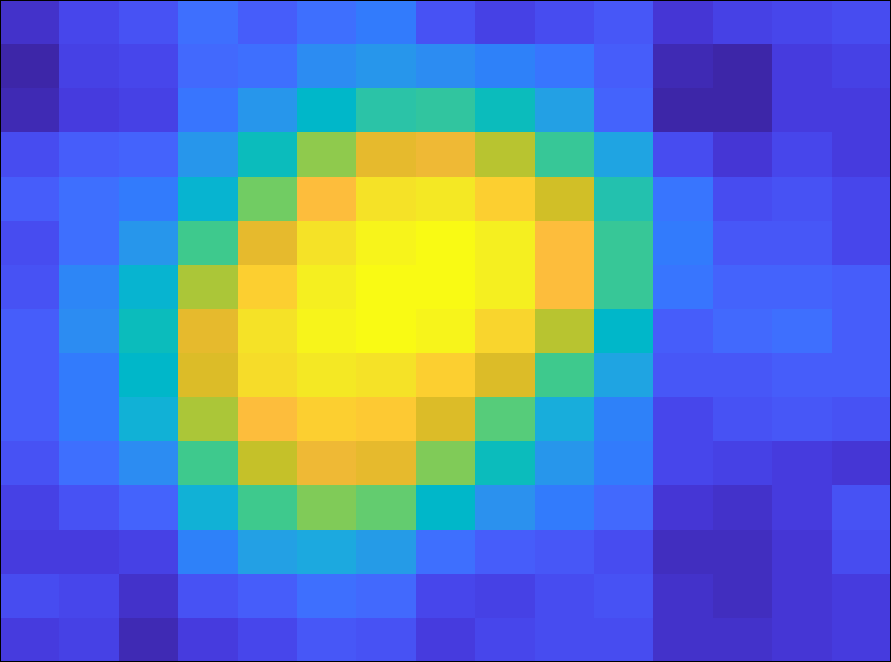}}
		\end{minipage}
		\hspace{0.65cm}
		\begin{minipage}{0.05\textwidth}
			\centerline{\includegraphics[width=0.65in,height=0.65in]{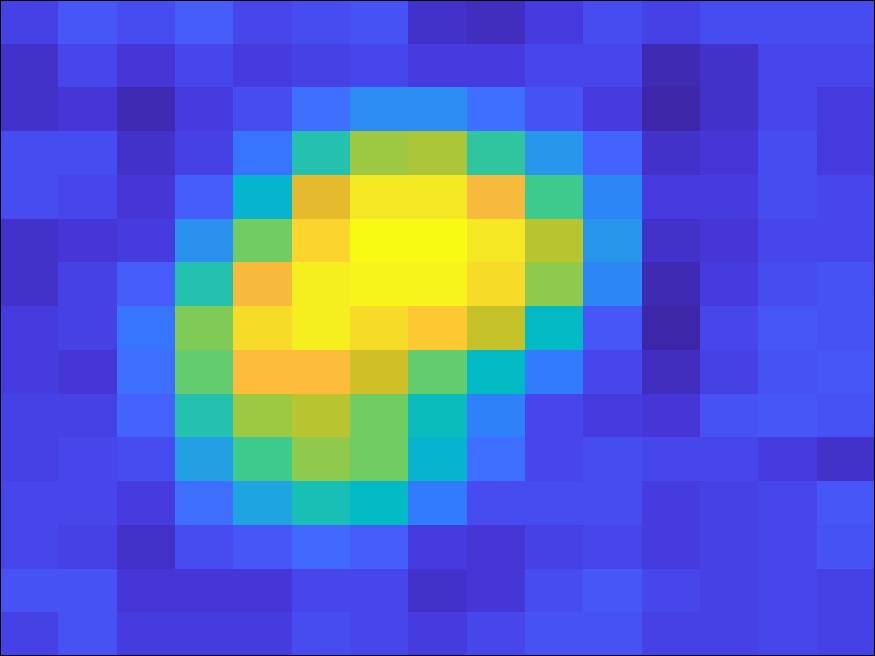}}
		\end{minipage}
		\hspace{0.65cm}
		\begin{minipage}{0.05\textwidth}
			\centerline{\includegraphics[width=0.65in,height=0.65in]{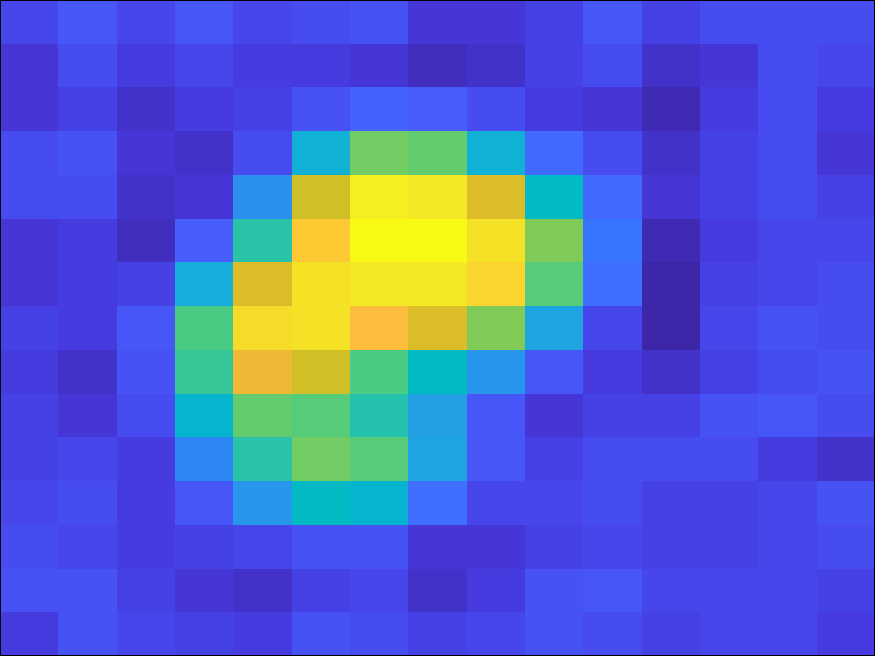}}
		\end{minipage}
		\hspace{0.65cm}
		\begin{minipage}{0.05\textwidth}
			\centerline{\includegraphics[width=0.65in,height=0.65in]{./Picture_Result/KernelEstimation/K2/cloth_39.png}}
		\end{minipage}
		\vfill
		\begin{minipage}{0.05\textwidth}
			\centerline{\includegraphics[width=0.65in,height=0.65in]{./Picture_Result/KernelEstimation/K3/cloth_GT.png}}
		\end{minipage}
		\hspace{0.65cm}
		\begin{minipage}{0.05\textwidth}
			\centerline{\includegraphics[width=0.65in,height=0.65in]{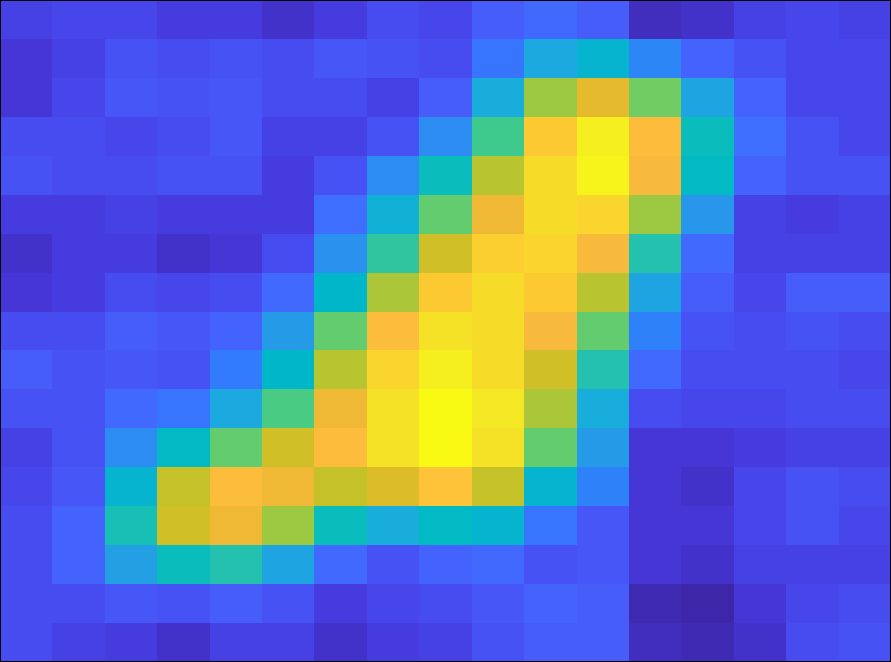}}
		\end{minipage}
		\hspace{0.65cm}
		\begin{minipage}{0.05\textwidth}
			\centerline{\includegraphics[width=0.65in,height=0.65in]{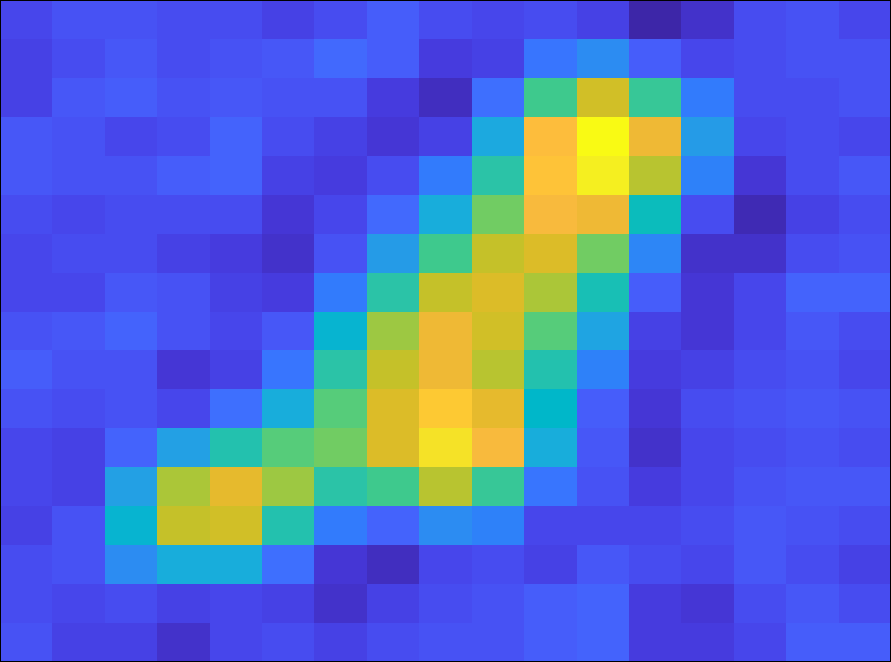}}
		\end{minipage}
		\hspace{0.65cm}
		\begin{minipage}{0.05\textwidth}
			\centerline{\includegraphics[width=0.65in,height=0.65in]{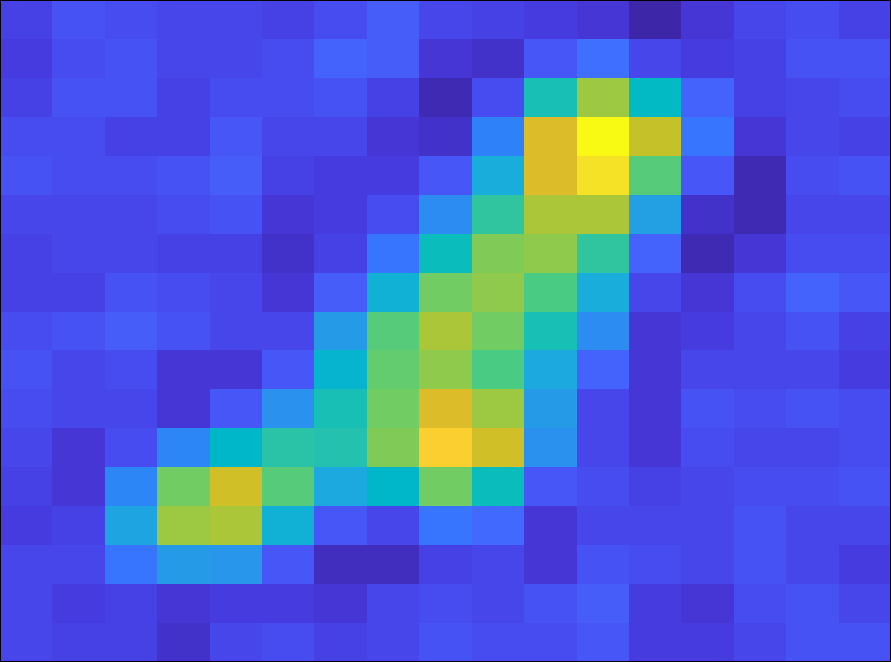}}
		\end{minipage}
		\hspace{0.65cm}
		\begin{minipage}{0.05\textwidth}
			\centerline{\includegraphics[width=0.65in,height=0.65in]{./Picture_Result/KernelEstimation/K3/cloth_39.png}}
		\end{minipage}
		\vfill
		\begin{minipage}{0.05\textwidth}
			\centerline{\includegraphics[width=0.65in,height=0.65in]{./Picture_Result/KernelEstimation/K4/cloth_GT.png}}
			\centerline{{\scriptsize (a) GT}}
		\end{minipage}
		\hspace{0.65cm}
		\begin{minipage}{0.05\textwidth}
			\centerline{\includegraphics[width=0.65in,height=0.65in]{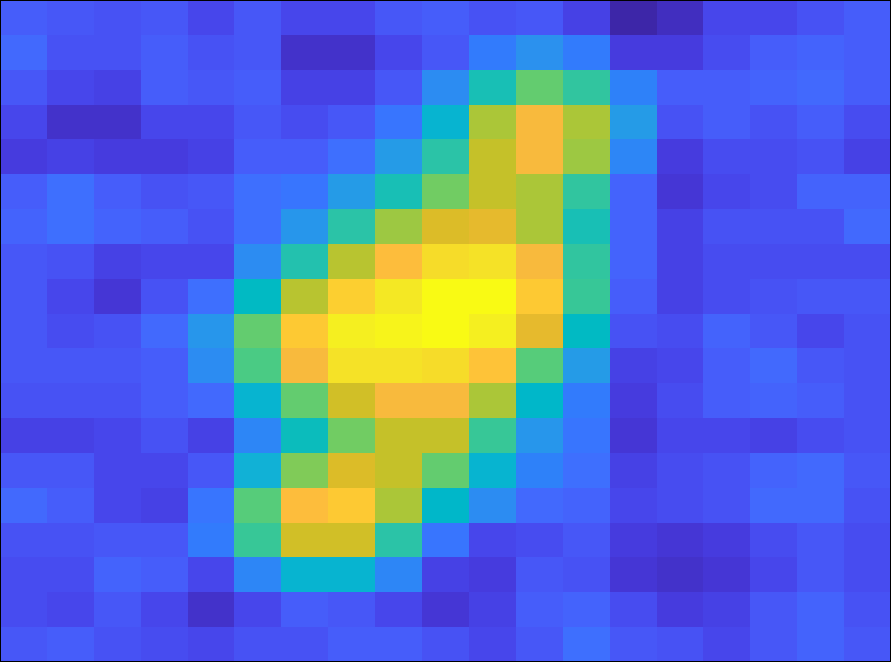}}
			\centerline{{\scriptsize (b) $I_{out}$ = 10}}
		\end{minipage}
		\hspace{0.65cm}
		\begin{minipage}{0.05\textwidth}
			\centerline{\includegraphics[width=0.65in,height=0.65in]{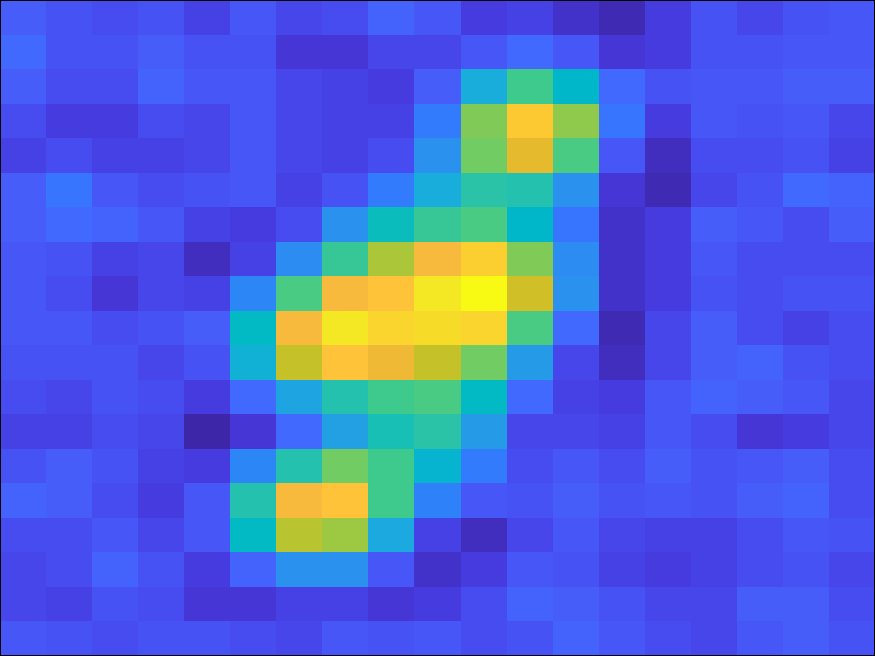}}
			\centerline{{\scriptsize (c) $I_{out}$ = 20}}
		\end{minipage}
		\hspace{0.65cm}
		\begin{minipage}{0.05\textwidth}
			\centerline{\includegraphics[width=0.65in,height=0.65in]{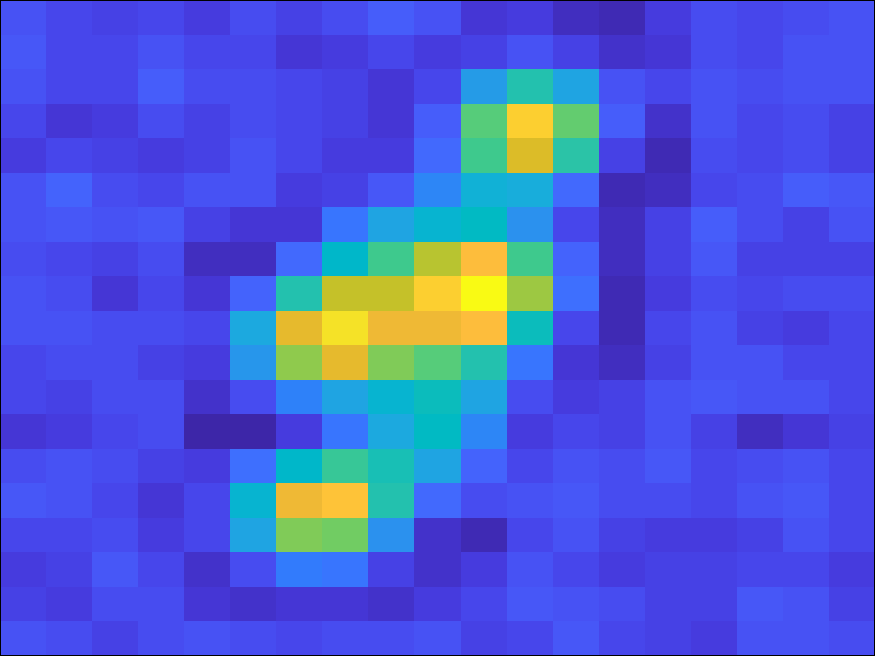}}
			\centerline{{\scriptsize (d) $I_{out}$ = 30}}
		\end{minipage}
		\hspace{0.65cm}
		\begin{minipage}{0.05\textwidth}
			\centerline{\includegraphics[width=0.65in,height=0.65in]{./Picture_Result/KernelEstimation/K4/cloth_39.png}}
			\centerline{{\scriptsize (e) $I_{out}$ = 40}}
		\end{minipage}
		\vfill
		\vspace{0.1cm}
		\caption{The estimated kernels by the proposed method with different iteration numbers.}
		\label{Fig_K_Estimation_Sup}
	\end{figure}
	
	In the experiments corresponding to Figure~\ref{Fig_K_Comparison} and Figure~\ref{Fig_P_Estimation_Sup}, we have demonstrated the proposed method can well estimate the degeneration models. In this experiment, we further visualize the intermediate results (the estimated kernels in the intermediate stages) of the reconstructed blur kernel with different iteration number of alternating optimization. We conduct the experiments on different four kind of motion kernels with $\mathbf{P}_{0.01}$ as the test spectral response matrix. SNRs of both the observed LR HSI and HR MSI are 40dB, and the SR scale is 8. The estimated kernels with different iteration numbers are shown in Figure~\ref{Fig_K_Estimation_Sup}. It can be seen that the estimation of blur kernels can be gradually improved with more iteration numbers, which also results in better performance of reconstruction network accordingly.
	
	\subsection{More experimental results on the benchmark datasets}
	\begin{table}[!htbp]\small
		\centering
		\caption{The performance of each method on the CAVE dataset with input images, which suffers from noise with different intensities. 
			The best results are in bold.
		}
		\setlength{\tabcolsep}{1.3mm}{\begin{tabular}{c|c|c|c|c|c}
				\hline\hline
				Noise intensity&Method&RMSE&PSNR&SAM&SSIM\\
				\hline
				\multirow{6}*{\shortstack{HSI 40dB\\MSI 40dB }}&NSSR~\cite{7438864}&8.11&30.53&18.85&0.8533\\	
				\cline{2-6}&MHFnet~\cite{8953470}&4.53&35.67&14.01&0.9409\\
				\cline{2-6}&Yong~\cite{8019510}&4.46&35.72&11.29&0.9491\\
				\cline{2-6}&UAL~\cite{Ours_CVPR2020}&3.93&37.07&7.57&0.9724\\
				\cline{2-6}&DBSR~\cite{9136736}&4.01&36.66&\textbf{5.53}&0.9739\\
				\cline{2-6}&Ours&\textbf{2.40}&\textbf{41.12}&6.53&\textbf{0.9854}\\
				\hline
				\hline
				\multirow{6}*{\shortstack{HSI 35dB\\MSI 35dB }}&NSSR~\cite{7438864}&8.95&29.52&26.27&0.7462\\	
				\cline{2-6}&MHFnet~\cite{8953470}&4.85&34.91&18.40&0.9040\\
				\cline{2-6}&Yong~\cite{8019510}&4.58&35.48&12.70&0.9415\\
				\cline{2-6}&UAL~\cite{Ours_CVPR2020}&4.09&36.57&9.70&0.9652\\
				\cline{2-6}&DBSR~\cite{9136736}&4.61&35.56&\textbf{7.74}&0.9624\\
				\cline{2-6}&Ours&\textbf{2.67}&\textbf{40.21}&8.74&\textbf{0.9797}\\
				\hline
				\hline
				\multirow{6}*{\shortstack{HSI 30dB\\MSI 30dB }}&NSSR~\cite{7438864}&11.16&27.40&35.34&0.5623\\	
				\cline{2-6}&MHFnet~\cite{8953470}&5.69&33.30&24.54&0.8279\\
				\cline{2-6}&Yong~\cite{8019510}&4.87&34.82&14.92&0.9244\\
				\cline{2-6}&UAL~\cite{Ours_CVPR2020}&4.45&35.70&13.17&0.9495\\
				\cline{2-6}&DBSR~\cite{9136736}&6.15&32.97&12.49&0.9243\\
				\cline{2-6}&Ours&\textbf{3.20}&\textbf{38.36}&\textbf{12.45}&\textbf{0.9632}\\
				\hline\hline
		\end{tabular}}
		\label{Tab_CAVE_Result_Diff_Noise}
	\end{table}

	\begin{figure*}[!ht]
		\centering
		\vspace{-0.4cm}
		\hspace{0.55cm}
		\begin{minipage}{0.05\textwidth}
			\centerline{\includegraphics[width=0.94in,height=0.94in]{./Picture_Result/CAVE_K1_S8_4040-0.01/chart_and_stuffed_toy_RGB/GT.pdf}}
		\end{minipage}
		\hspace{1.4cm}
		\begin{minipage}{0.05\textwidth}
			\centerline{\includegraphics[width=0.94in,height=0.94in]{./Picture_Result/CAVE_K1_S8_4040-0.01/chart_and_stuffed_toy_RGB/NSSR.pdf}}
		\end{minipage}
		\hspace{1.4cm}
		\begin{minipage}{0.05\textwidth}
			\centerline{\includegraphics[width=0.94in,height=0.94in]{./Picture_Result/CAVE_K1_S8_4040-0.01/chart_and_stuffed_toy_RGB/MHFnet.pdf}}
		\end{minipage}
		\hspace{1.4cm}
		\begin{minipage}{0.05\textwidth}
			\centerline{\includegraphics[width=0.94in,height=0.94in]{./Picture_Result/CAVE_K1_S8_4040-0.01/chart_and_stuffed_toy_RGB/Yong.pdf}}
		\end{minipage}
		\hspace{1.4cm}
		\begin{minipage}{0.05\textwidth}
			\centerline{\includegraphics[width=0.94in,height=0.94in]{./Picture_Result/CAVE_K1_S8_4040-0.01/chart_and_stuffed_toy_RGB/Ours_2020.pdf}}
		\end{minipage}
		\hspace{1.4cm}
		\begin{minipage}{0.05\textwidth}
			\centerline{\includegraphics[width=0.94in,height=0.94in]{./Picture_Result/CAVE_K1_S8_4040-0.01/chart_and_stuffed_toy_RGB/DBSR.pdf}}
		\end{minipage}
		\hspace{1.4cm}
		\begin{minipage}{0.05\textwidth}
			\centerline{\includegraphics[width=0.94in,height=0.94in]{./Picture_Result/CAVE_K1_S8_4040-0.01/chart_and_stuffed_toy_RGB/Ours_2021.pdf}}
		\end{minipage}
		\hspace{0.8cm}
		\begin{minipage}{0.005\textwidth}
			\centerline{\includegraphics[width=0.1in,height=0.94in]{./Picture_Result/CAVE_K1_S8_4040-0.01/chart_and_stuffed_toy_RGB/colorbar_placeholder.png}}
		\end{minipage}
		\vfill
		\hspace{0.55cm}
		\begin{minipage}{0.05\textwidth}
			\centerline{\includegraphics[width=0.94in,height=0.94in]{./Picture_Result/CAVE_K1_S8_4040-0.01/chart_and_stuffed_toy_RGB/GT_mse.pdf}}
		\end{minipage}
		\hspace{1.4cm}
		\begin{minipage}{0.05\textwidth}
			\centerline{\includegraphics[width=0.94in,height=0.94in]{./Picture_Result/CAVE_K1_S8_4040-0.01/chart_and_stuffed_toy_RGB/NSSR_mse.pdf}}
		\end{minipage}
		\hspace{1.4cm}
		\begin{minipage}{0.05\textwidth}
			\centerline{\includegraphics[width=0.94in,height=0.94in]{./Picture_Result/CAVE_K1_S8_4040-0.01/chart_and_stuffed_toy_RGB/MHFnet_mse.pdf}}
		\end{minipage}
		\hspace{1.4cm}
		\begin{minipage}{0.05\textwidth}
			\centerline{\includegraphics[width=0.94in,height=0.94in]{./Picture_Result/CAVE_K1_S8_4040-0.01/chart_and_stuffed_toy_RGB/Yong_mse.pdf}}
		\end{minipage}
		\hspace{1.4cm}
		\begin{minipage}{0.05\textwidth}
			\centerline{\includegraphics[width=0.94in,height=0.94in]{./Picture_Result/CAVE_K1_S8_4040-0.01/chart_and_stuffed_toy_RGB/Ours_2020_mse.pdf}}
		\end{minipage}
		\hspace{1.4cm}
		\begin{minipage}{0.05\textwidth}
			\centerline{\includegraphics[width=0.94in,height=0.94in]{./Picture_Result/CAVE_K1_S8_4040-0.01/chart_and_stuffed_toy_RGB/DBSR_mse.pdf}}
		\end{minipage}
		\hspace{1.4cm}
		\begin{minipage}{0.05\textwidth}
			\centerline{\includegraphics[width=0.94in,height=0.94in]{./Picture_Result/CAVE_K1_S8_4040-0.01/chart_and_stuffed_toy_RGB/Ours_2021_mse.pdf}}
		\end{minipage}
		\hspace{0.8cm}
		\begin{minipage}{0.005\textwidth}
			\centerline{\includegraphics[width=0.15in,height=0.94in]{./Picture_Result/CAVE_K1_S8_4040-0.01/chart_and_stuffed_toy_RGB/CB_30_R.pdf}}
		\end{minipage}
		\vfill
		\hspace{0.55cm}
		\begin{minipage}{0.05\textwidth}
			\centerline{\includegraphics[width=0.94in,height=0.94in]{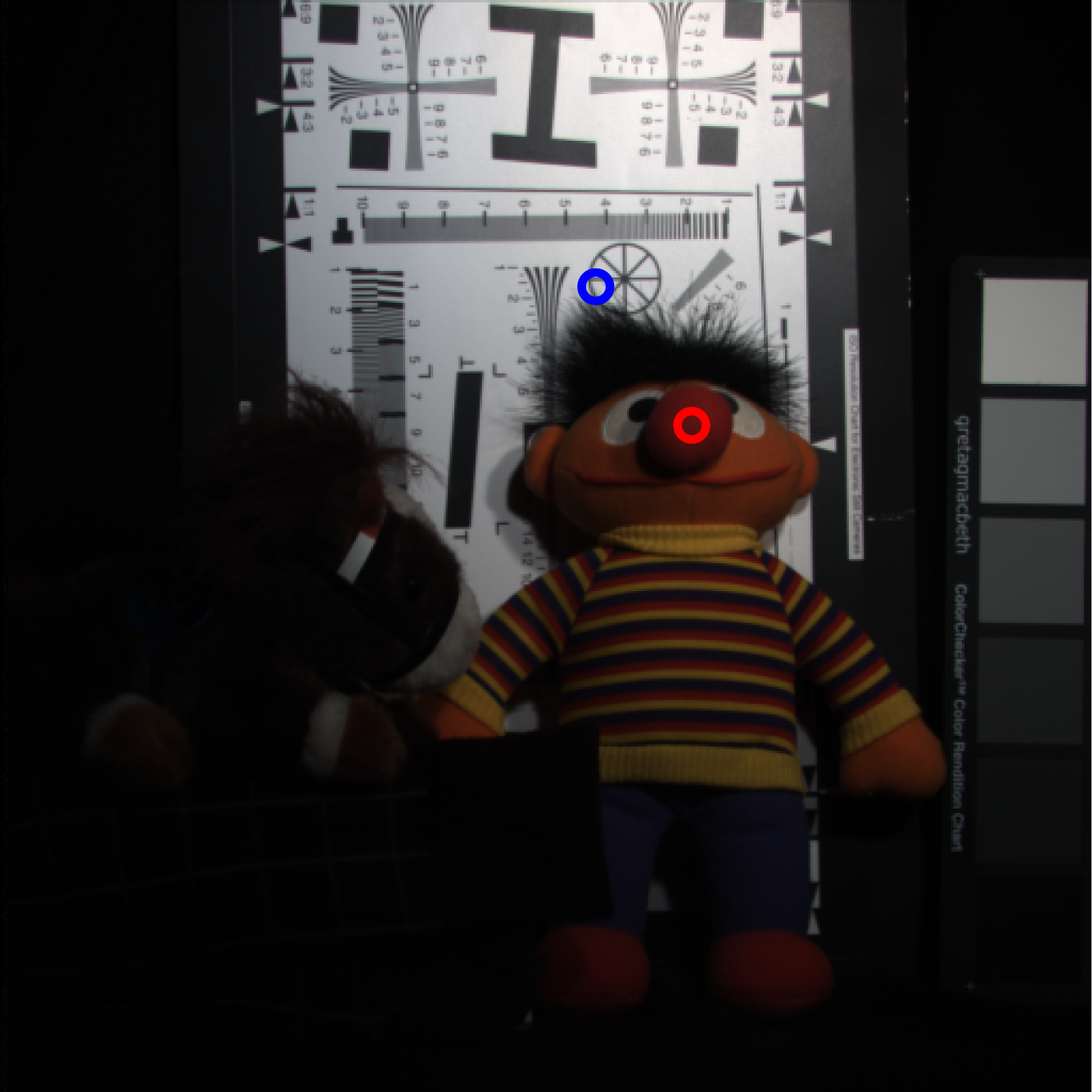}}
		\end{minipage}
		\hspace{1.4cm}
		\begin{minipage}{0.05\textwidth}
			\centerline{\includegraphics[width=0.94in,height=0.8in]{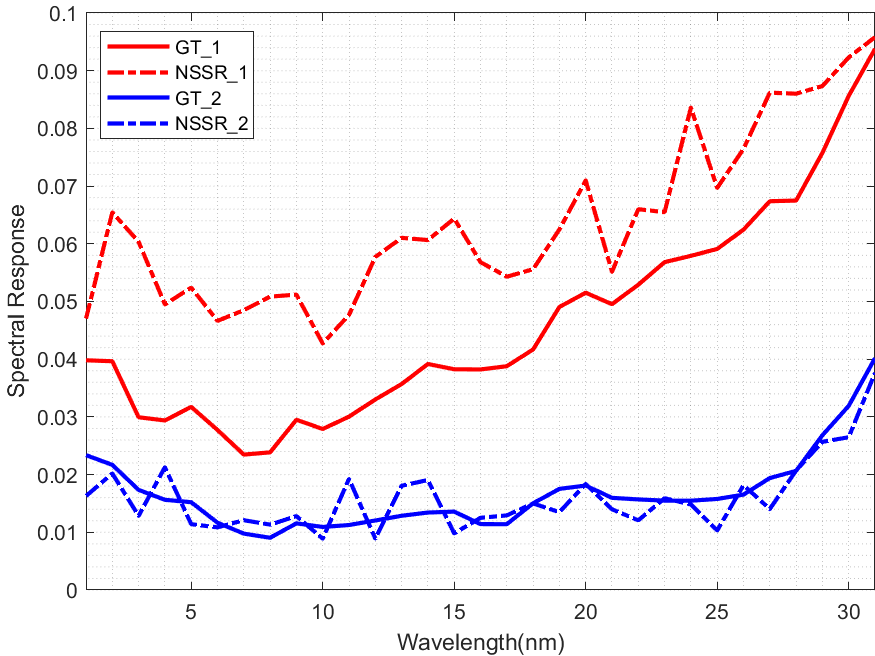}}
		\end{minipage}
		\hspace{1.4cm}
		\begin{minipage}{0.05\textwidth}
			\centerline{\includegraphics[width=0.94in,height=0.8in]{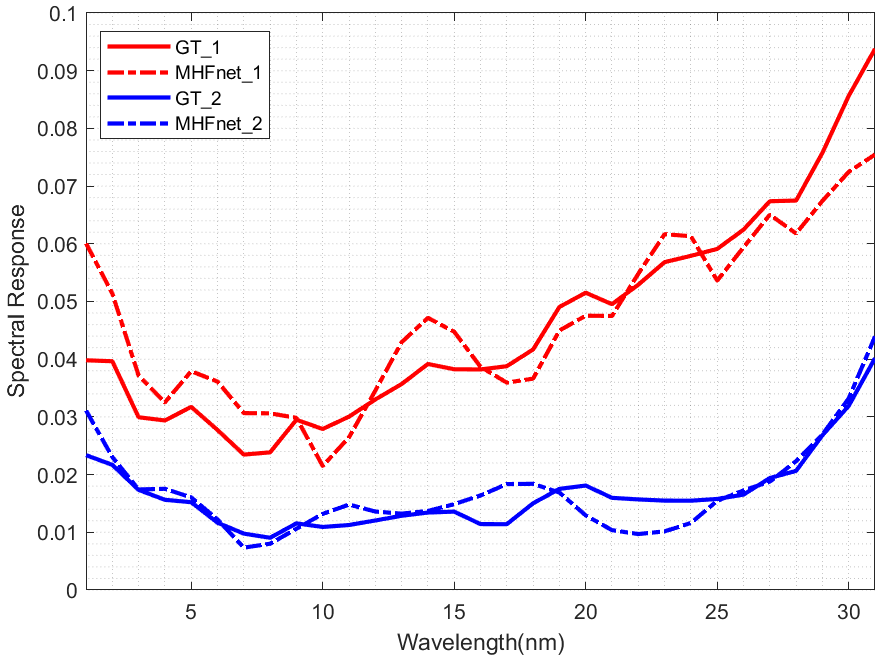}}
		\end{minipage}
		\hspace{1.4cm}
		\begin{minipage}{0.05\textwidth}
			\centerline{\includegraphics[width=0.94in,height=0.8in]{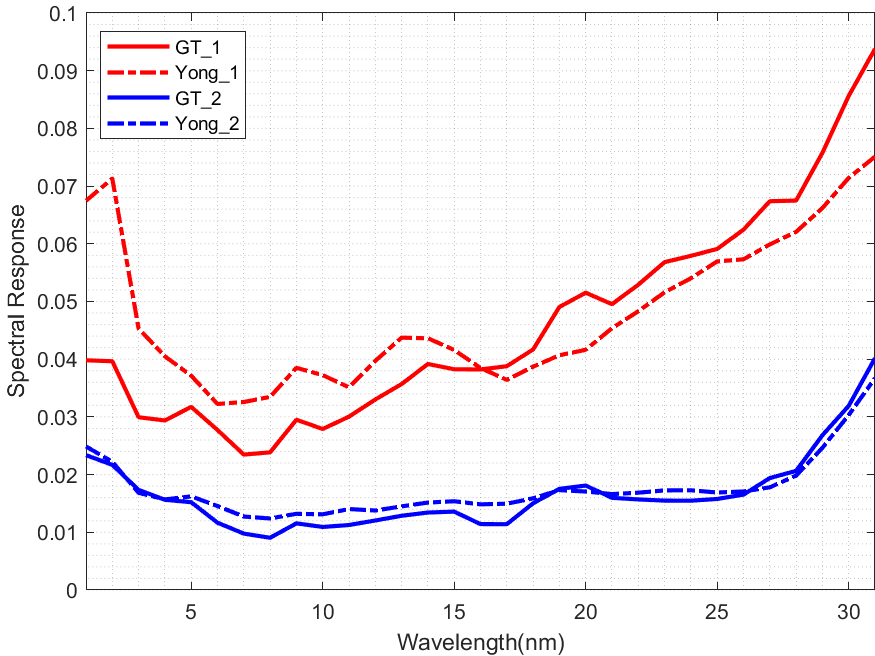}}
		\end{minipage}
		\hspace{1.4cm}
		\begin{minipage}{0.05\textwidth}
			\centerline{\includegraphics[width=0.94in,height=0.8in]{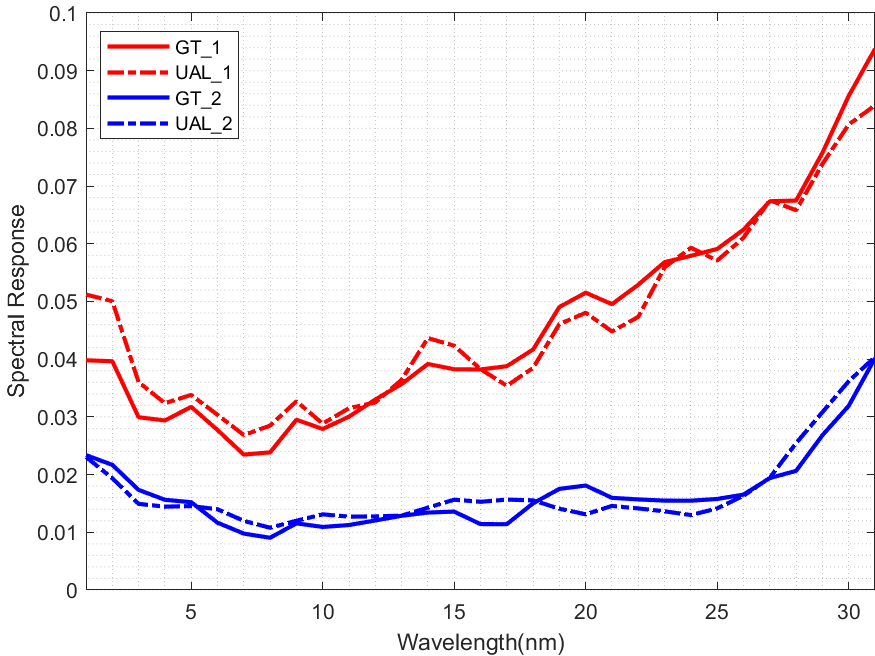}}
		\end{minipage}
		\hspace{1.4cm}
		\begin{minipage}{0.05\textwidth}
			\centerline{\includegraphics[width=0.94in,height=0.8in]{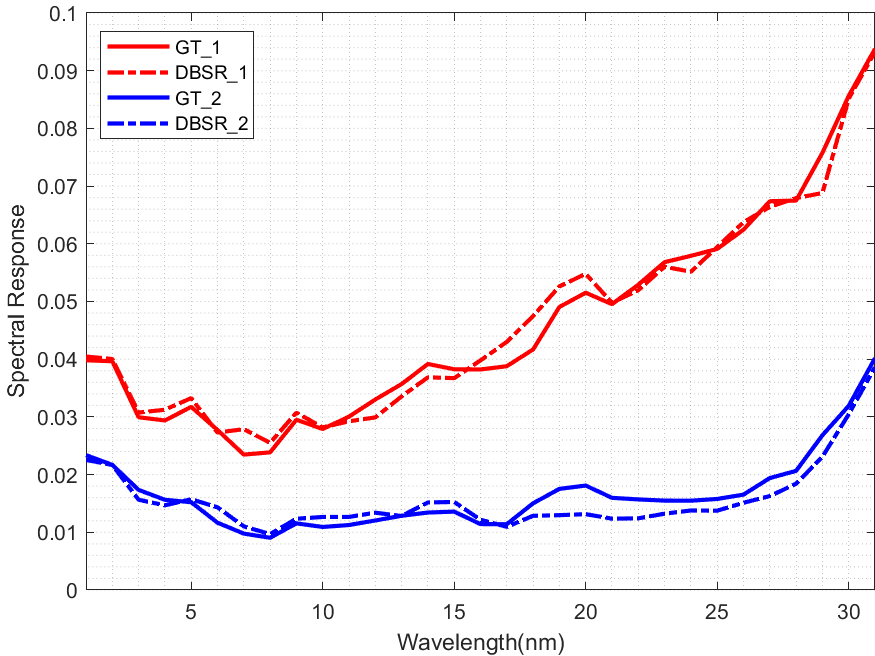}}
		\end{minipage}
		\hspace{1.4cm}
		\begin{minipage}{0.05\textwidth}
			\centerline{\includegraphics[width=0.94in,height=0.8in]{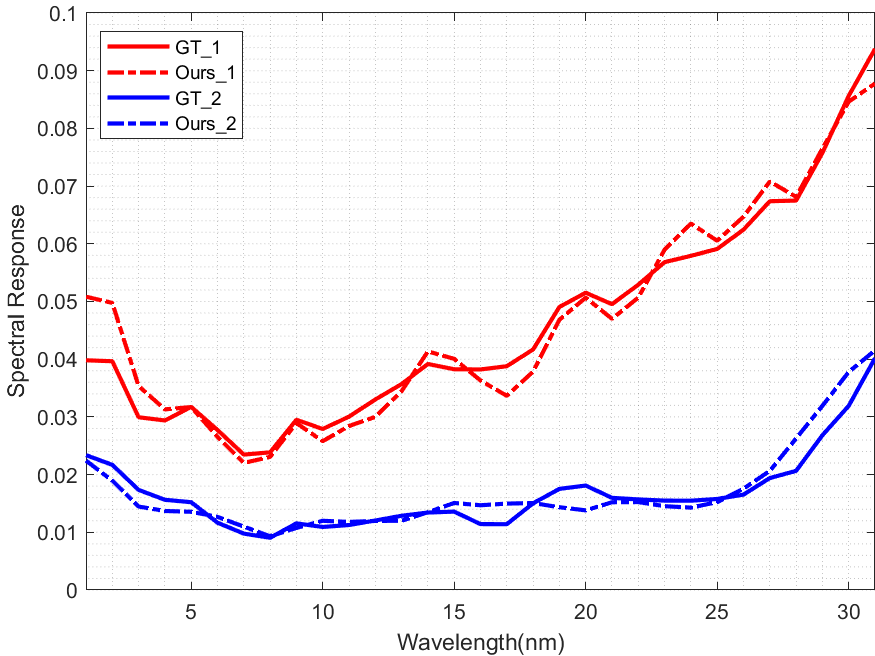}}
		\end{minipage}
		\hspace{0.8cm}
		\begin{minipage}{0.005\textwidth}
			\centerline{\includegraphics[width=0.15in,height=0.8in]{./Picture_Result/CAVE_K1_S8_4040-0.01/chart_and_stuffed_toy_RGB/colorbar_placeholder.png}}
		\end{minipage}
		\vfill
		\hspace{0.55cm}
		\begin{minipage}{0.05\textwidth}
			\centerline{{\scriptsize (a) GT}}
		\end{minipage}
		\hspace{1.4cm}
		\begin{minipage}{0.05\textwidth}
			\centerline{{\scriptsize (b) NSSR~\cite{7438864}}}
		\end{minipage}
		\hspace{1.4cm}
		\begin{minipage}{0.05\textwidth}
			\centerline{{\scriptsize (c) MHFnet~\cite{8953470}}}
		\end{minipage}
		\hspace{1.4cm}
		\begin{minipage}{0.05\textwidth}
			\centerline{{\scriptsize (d) Yong~\cite{8019510}}}
		\end{minipage}
		\hspace{1.4cm}
		\begin{minipage}{0.05\textwidth}
			\centerline{{\scriptsize (e) UAL~\cite{Ours_CVPR2020}}}
		\end{minipage}
		\hspace{1.4cm}
		\begin{minipage}{0.05\textwidth}
			\centerline{{\scriptsize (f) DBSR~\cite{9136736}}}
		\end{minipage}
		\hspace{1.4cm}
		\begin{minipage}{0.05\textwidth}
			\centerline{{\scriptsize (g) Ours}}
		\end{minipage}
		\hspace{0.8cm}
		\begin{minipage}{0.005\textwidth}
			\centerline{}
		\end{minipage}
		\caption{The visual SR results of all methods on the CAVE dataset. The observed LR HSI and HR MSI generated by $\mathbf{k}_1$ and $\mathbf{P}_{0.01}$, respectively. SNRs of both two observed images are 40dB, and the SR scale is 8.}
		\label{Fig_CAVE_Visible_sup}
	\end{figure*}

	\begin{figure*}[!htbp]
		\centering
		\hspace{0.55cm}
		\begin{minipage}{0.05\textwidth}
			\centerline{\includegraphics[width=0.94in,height=0.94in]{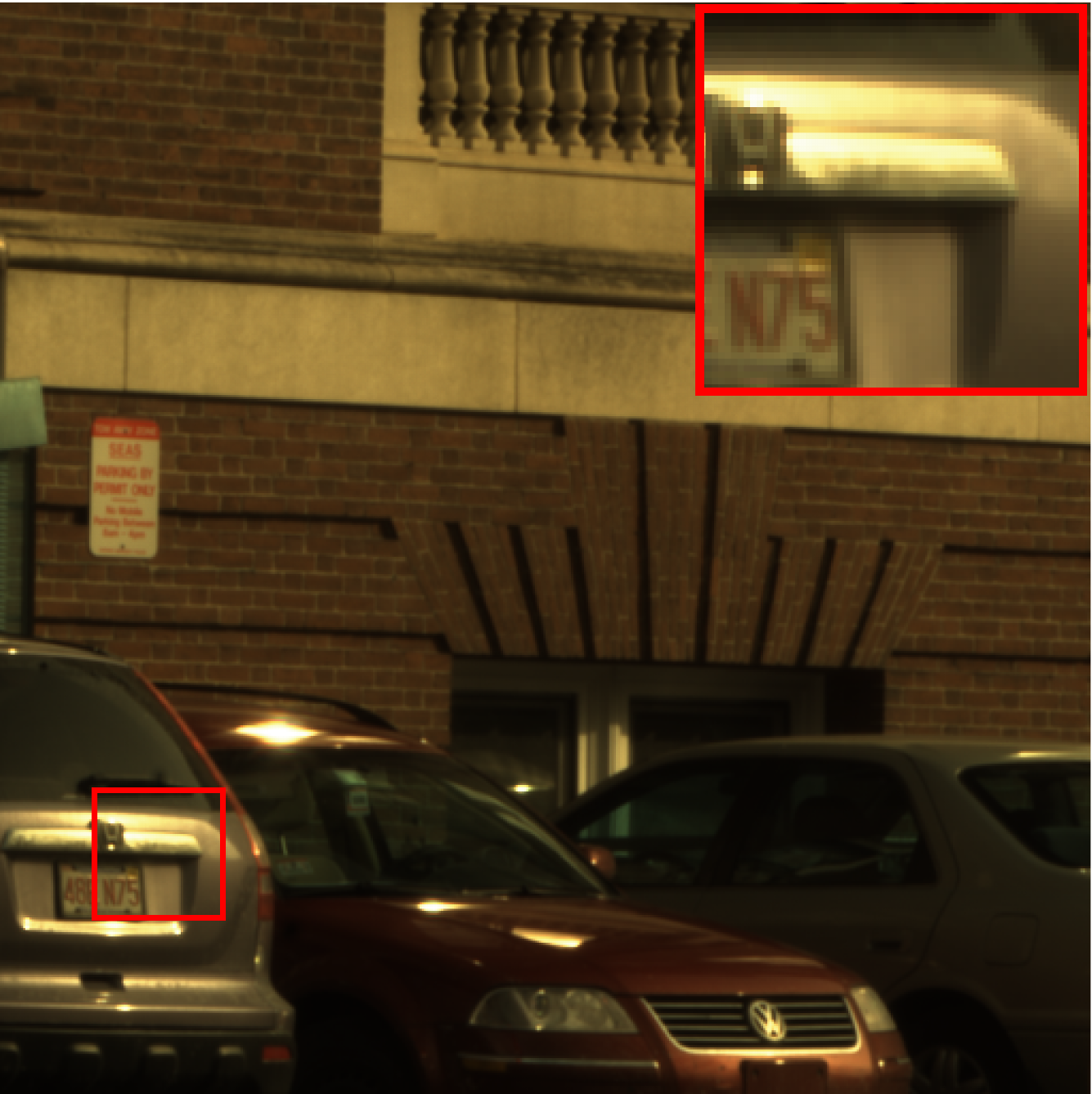}}
		\end{minipage}
		\hspace{1.4cm}
		\begin{minipage}{0.05\textwidth}
			\centerline{\includegraphics[width=0.94in,height=0.94in]{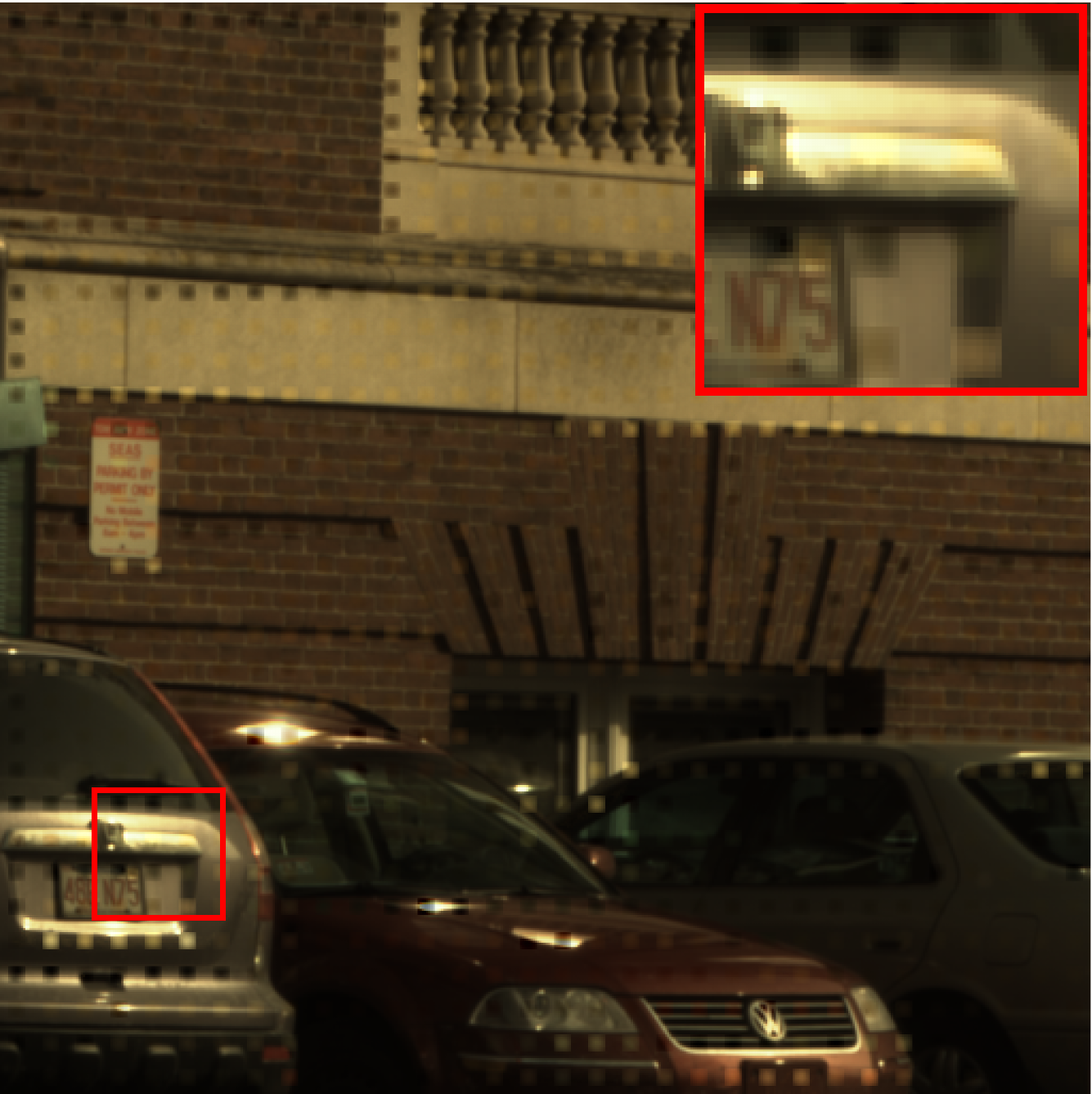}}
		\end{minipage}
		\hspace{1.4cm}
		\begin{minipage}{0.05\textwidth}
			\centerline{\includegraphics[width=0.94in,height=0.94in]{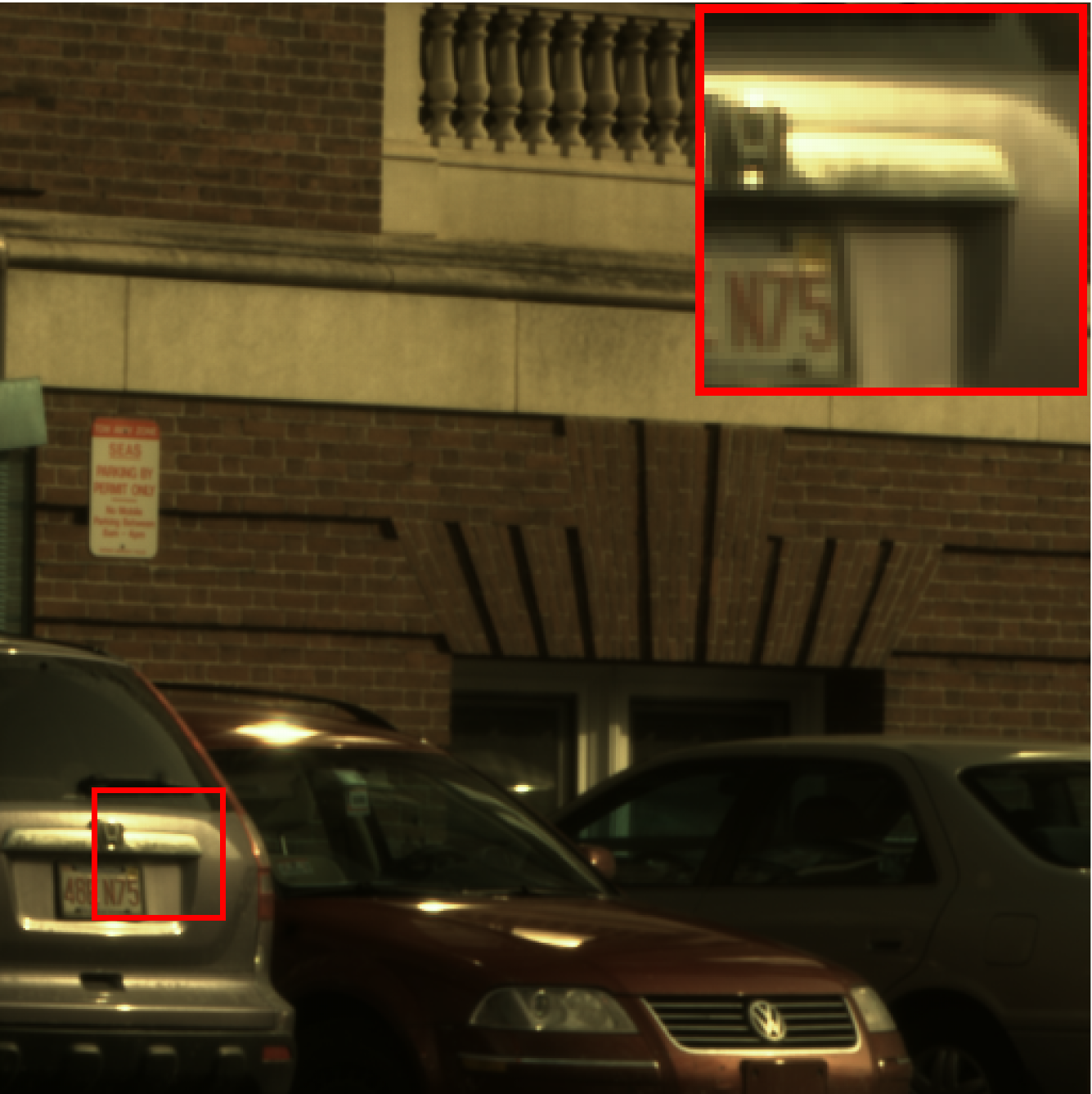}}
		\end{minipage}
		\hspace{1.4cm}
		\begin{minipage}{0.05\textwidth}
			\centerline{\includegraphics[width=0.94in,height=0.94in]{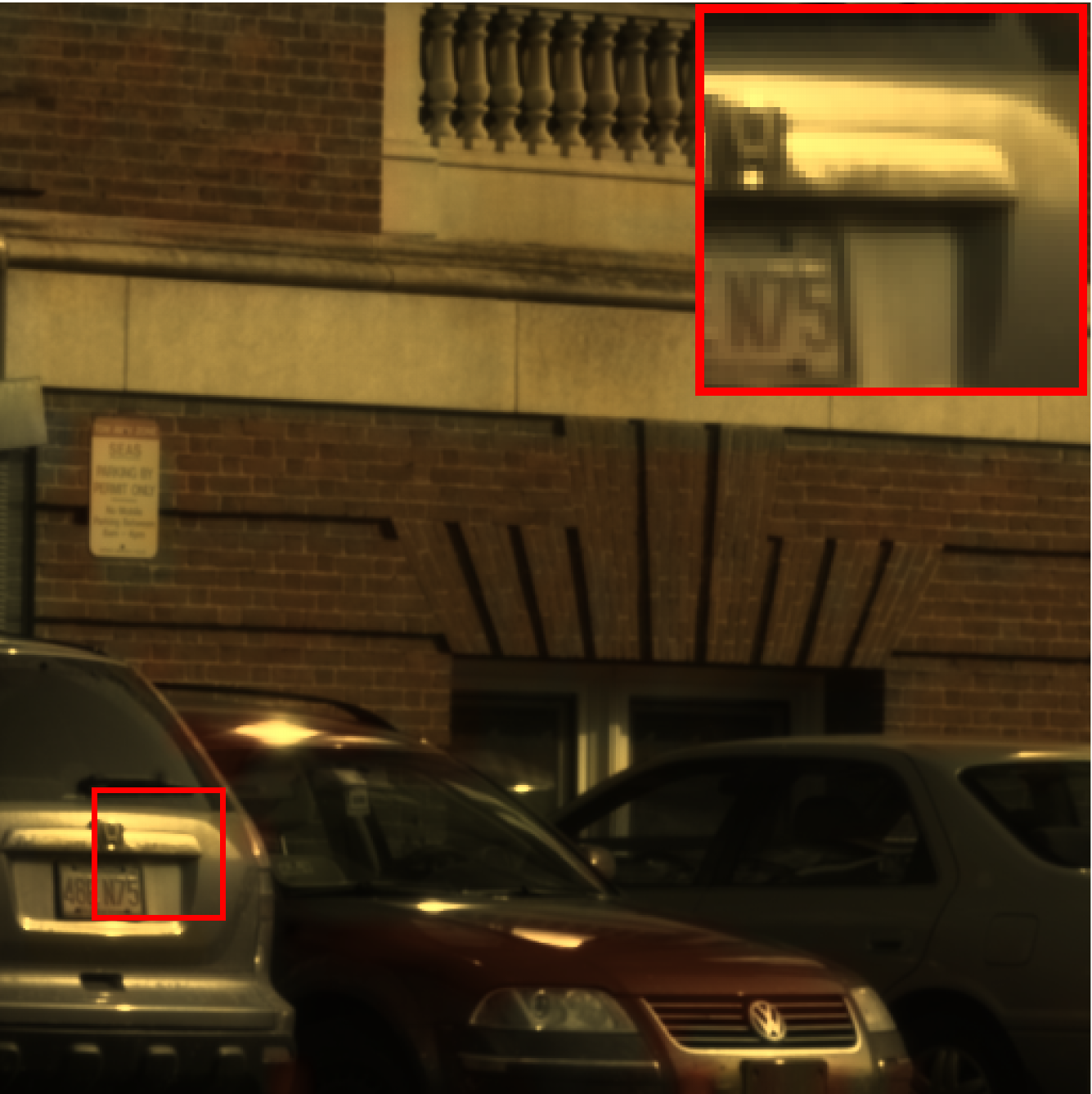}}
		\end{minipage}
		\hspace{1.4cm}
		\begin{minipage}{0.05\textwidth}
			\centerline{\includegraphics[width=0.94in,height=0.94in]{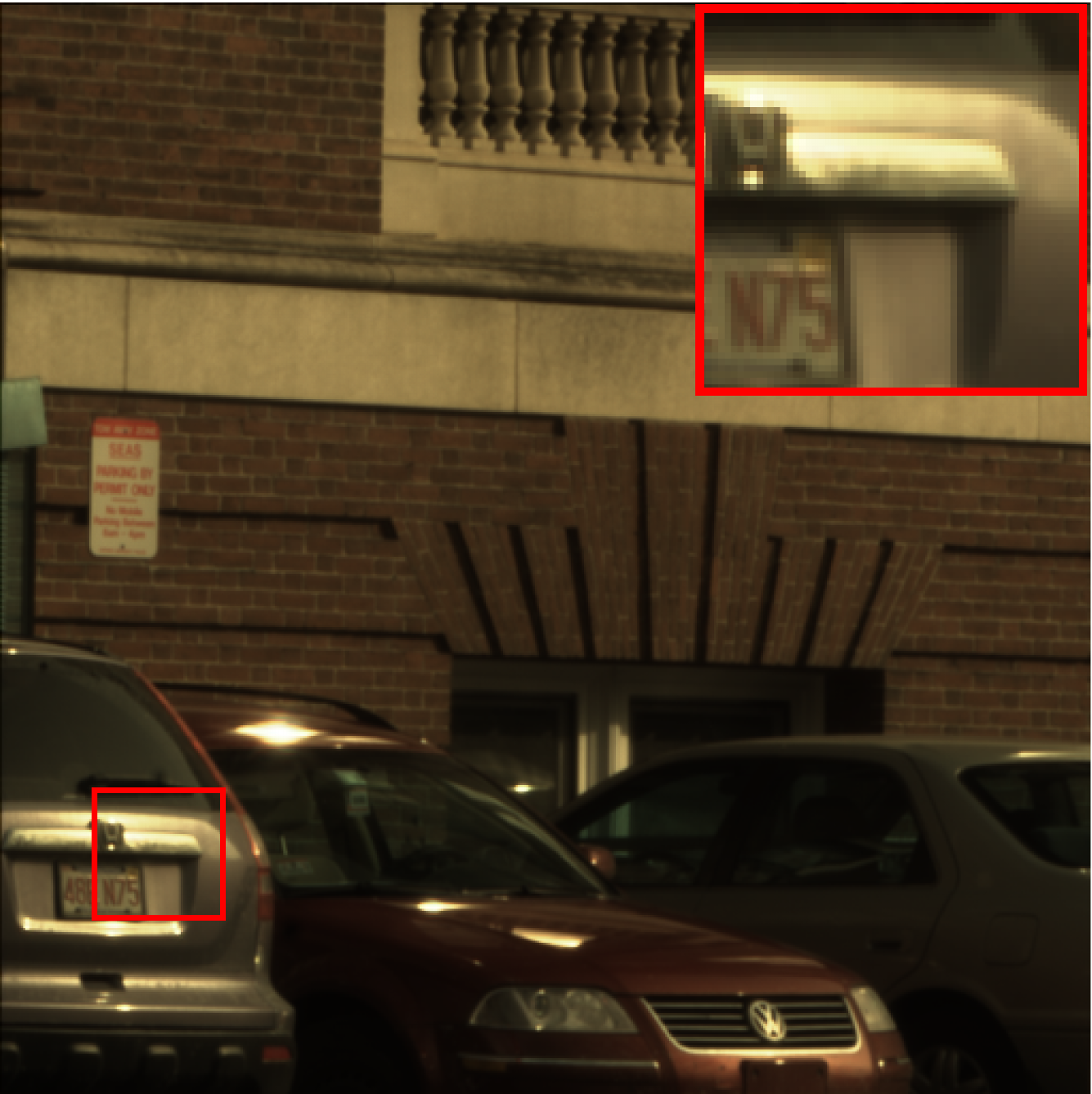}}
		\end{minipage}
		\hspace{1.4cm}
		\begin{minipage}{0.05\textwidth}
			\centerline{\includegraphics[width=0.94in,height=0.94in]{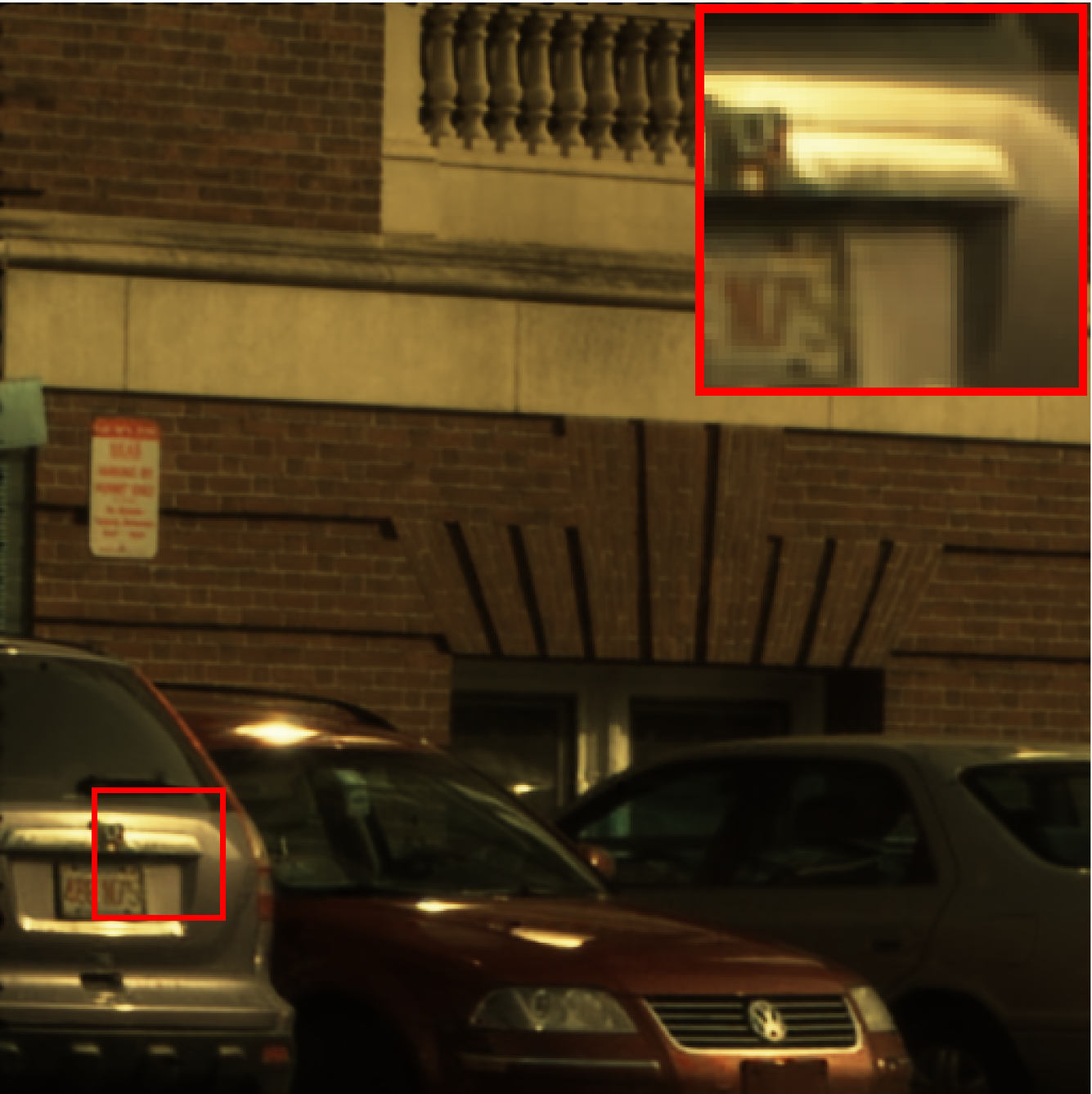}}
		\end{minipage}
		\hspace{1.4cm}
		\begin{minipage}{0.05\textwidth}
			\centerline{\includegraphics[width=0.94in,height=0.94in]{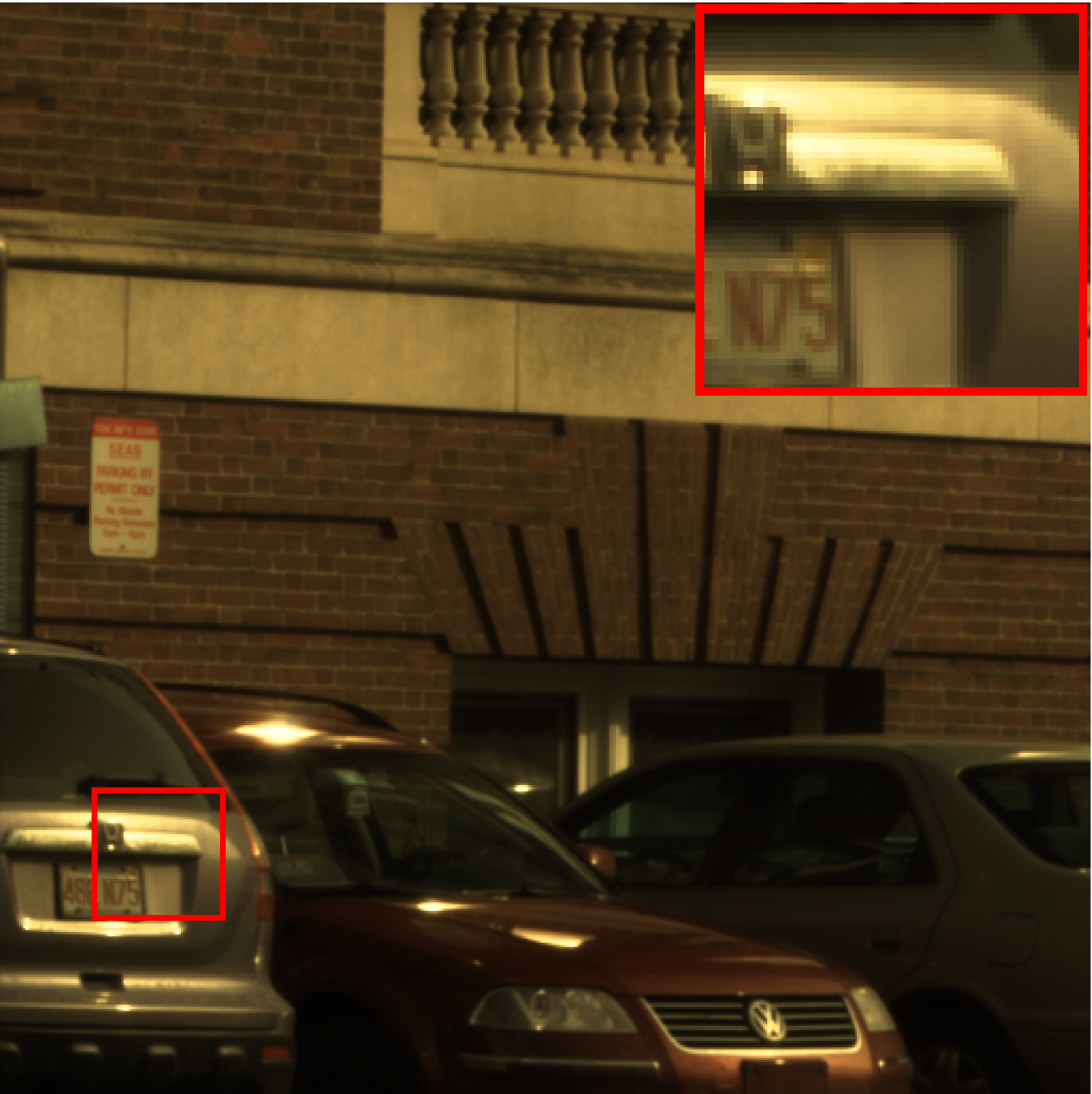}}
		\end{minipage}
		\hspace{0.8cm}
		\begin{minipage}{0.005\textwidth}
			\centerline{\includegraphics[width=0.1in,height=0.94in]{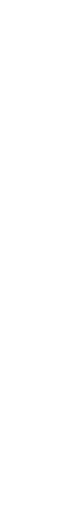}}
		\end{minipage}
		\vfill
		\hspace{0.55cm}
		\begin{minipage}{0.05\textwidth}
			\centerline{\includegraphics[width=0.94in,height=0.94in]{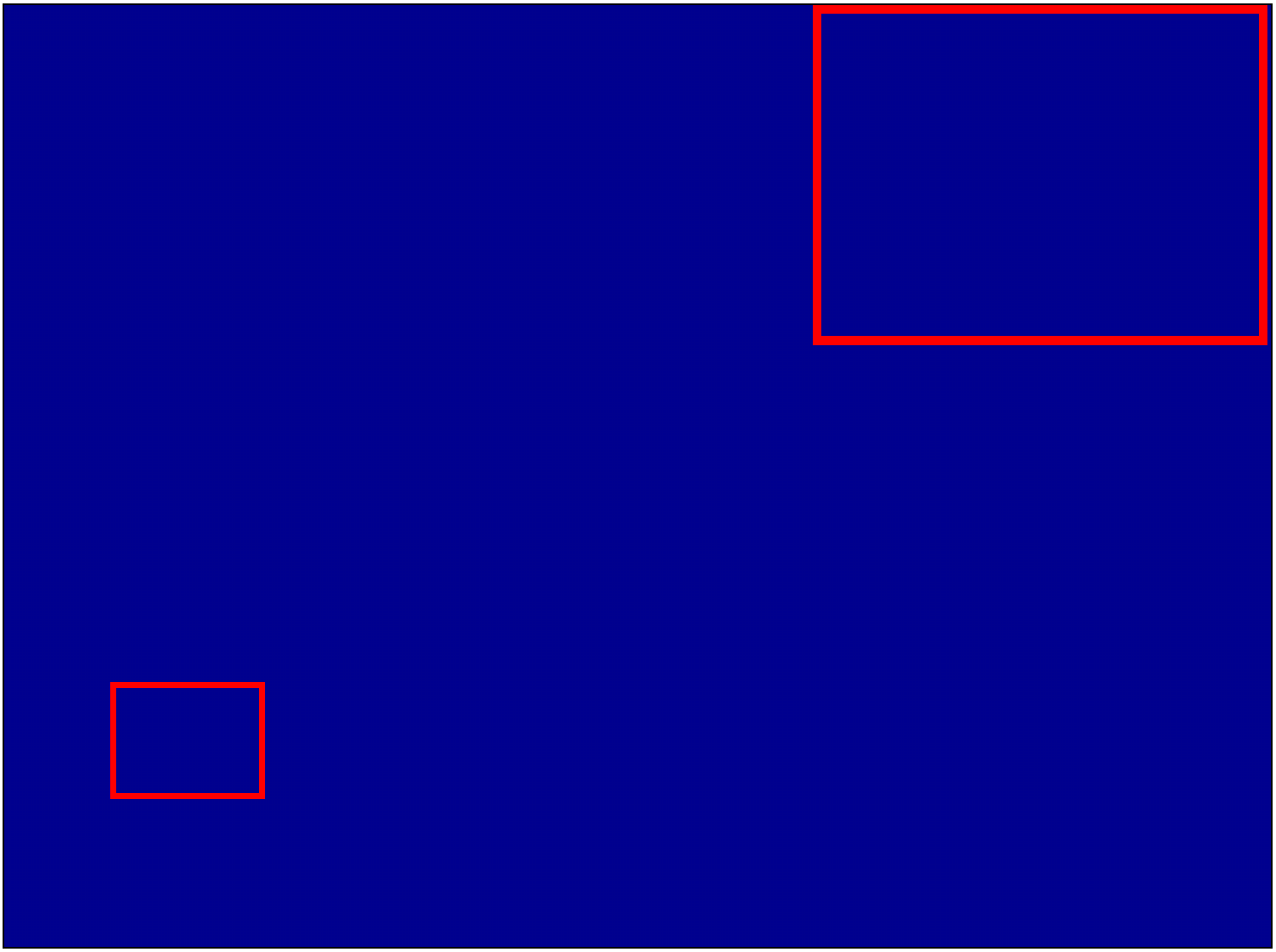}}
		\end{minipage}
		\hspace{1.4cm}
		\begin{minipage}{0.05\textwidth}
			\centerline{\includegraphics[width=0.94in,height=0.94in]{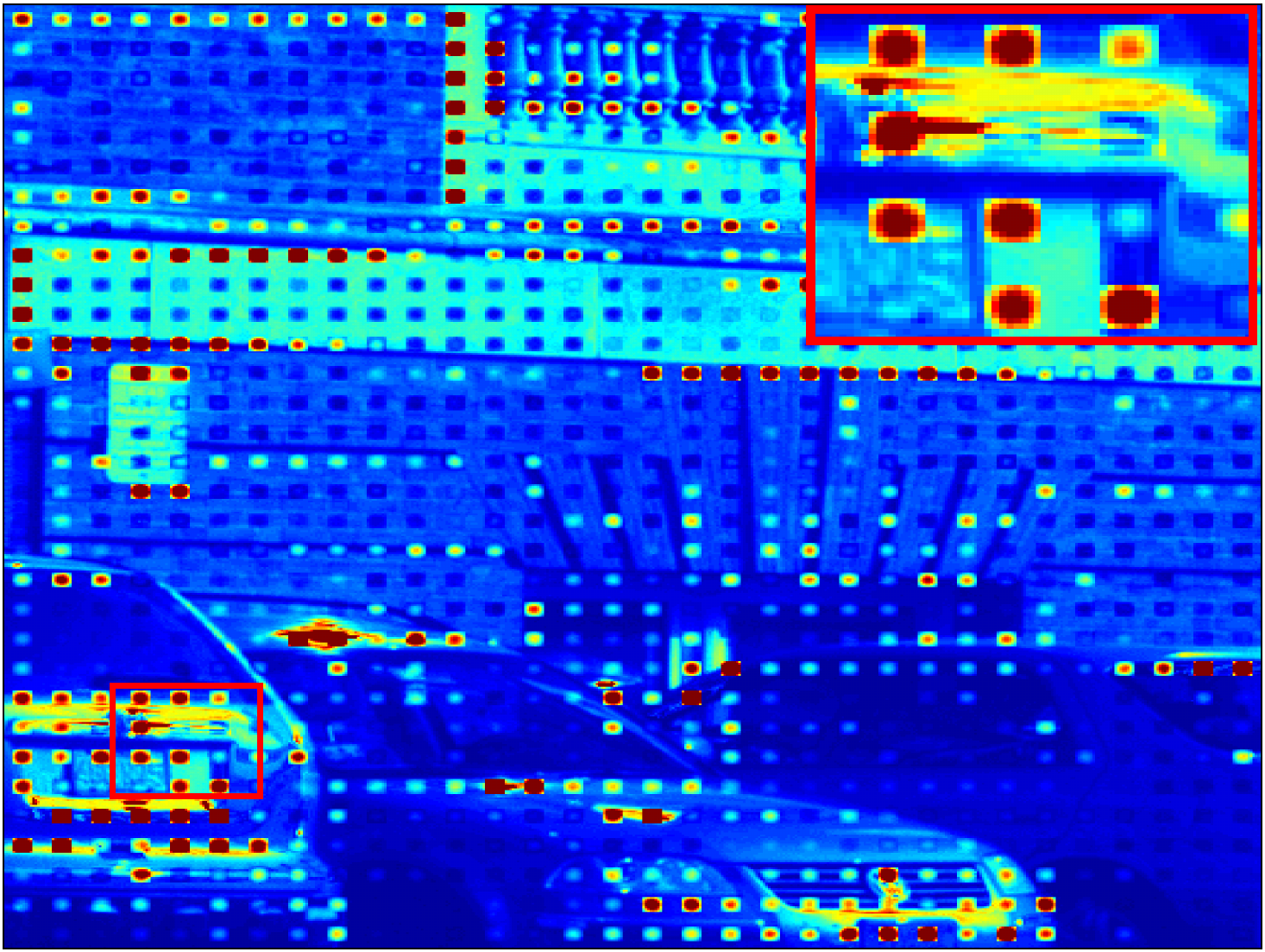}}
		\end{minipage}
		\hspace{1.4cm}
		\begin{minipage}{0.05\textwidth}
			\centerline{\includegraphics[width=0.94in,height=0.94in]{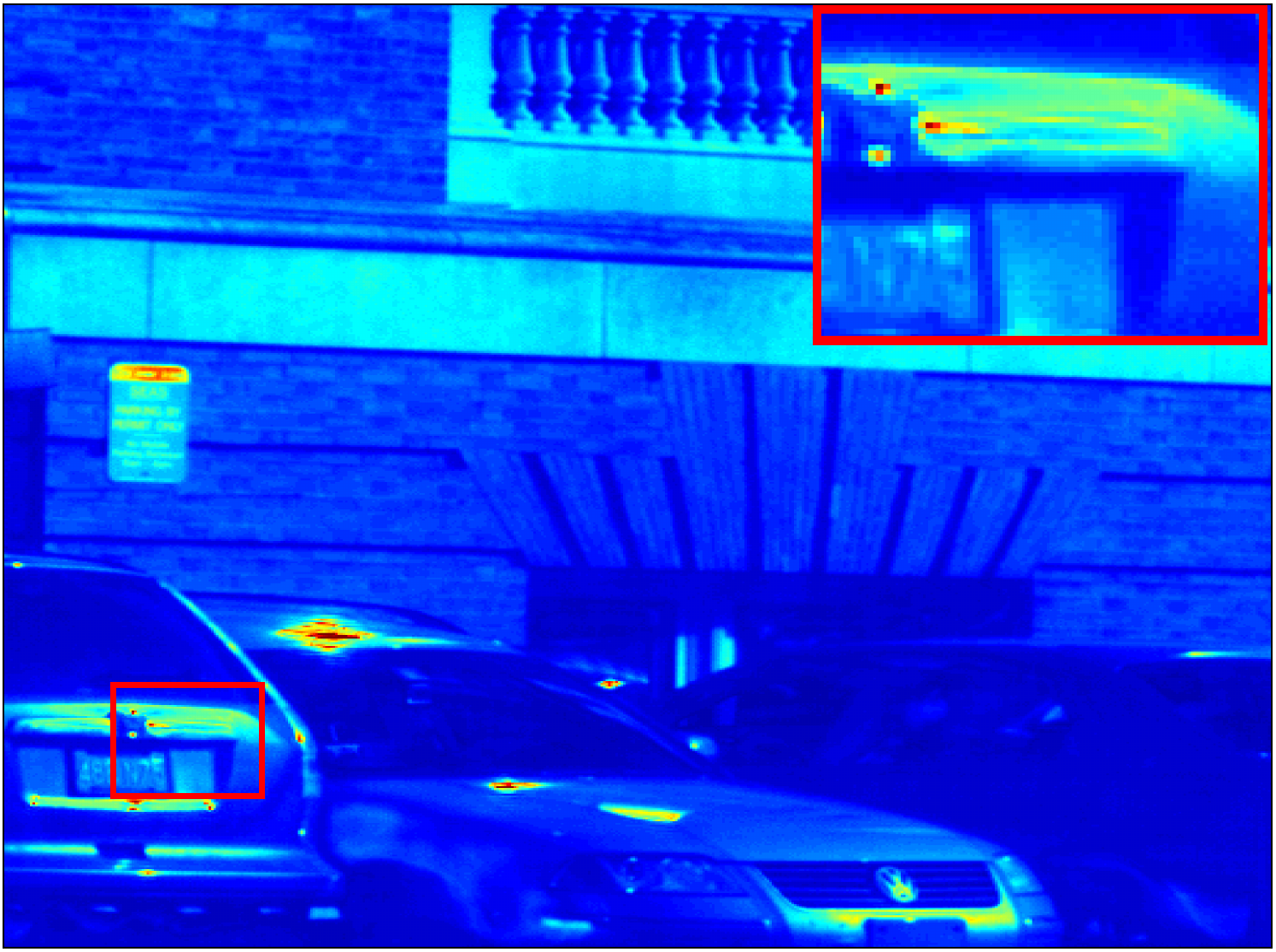}}
		\end{minipage}
		\hspace{1.4cm}
		\begin{minipage}{0.05\textwidth}
			\centerline{\includegraphics[width=0.94in,height=0.94in]{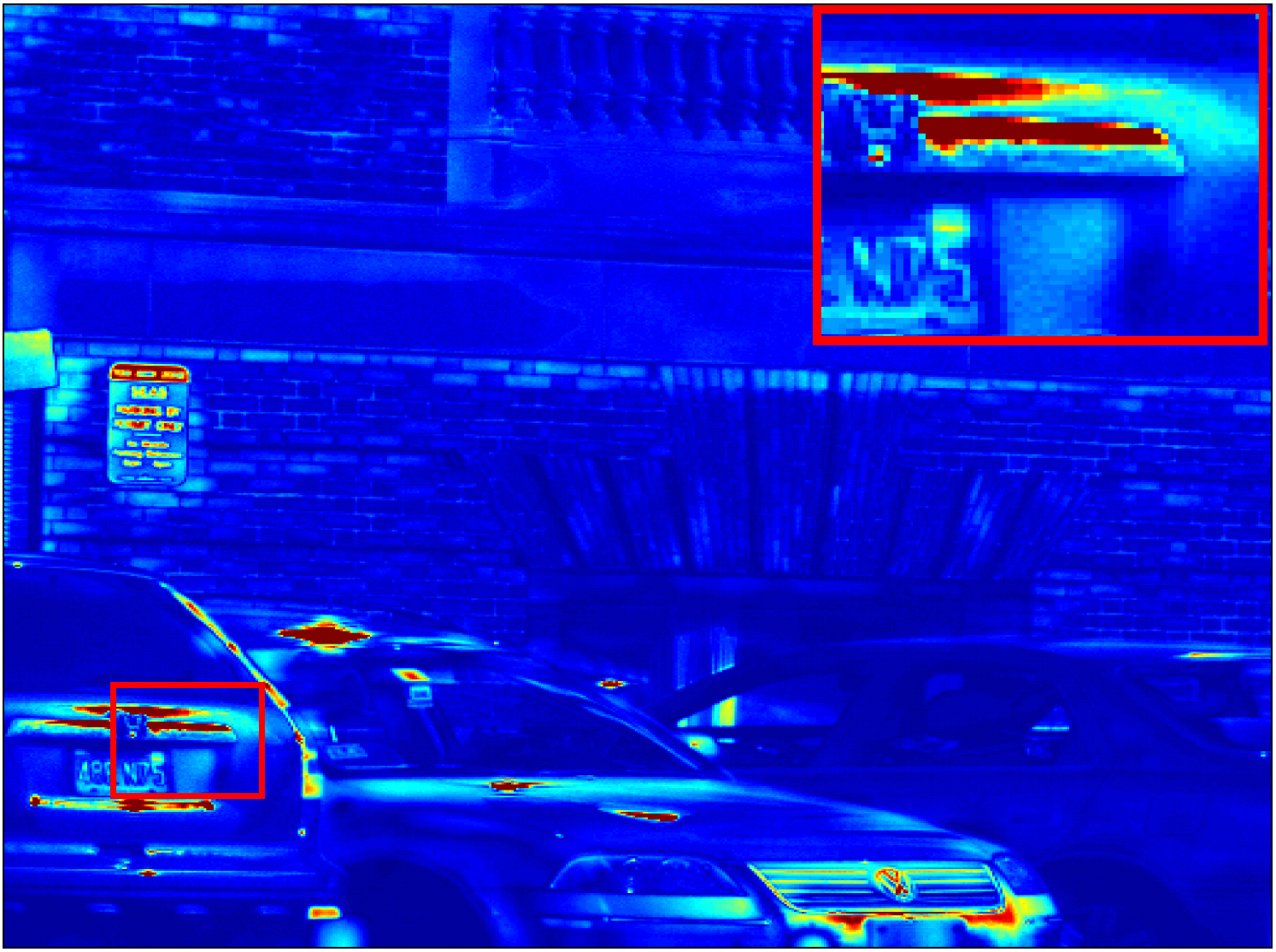}}
		\end{minipage}
		\hspace{1.4cm}
		\begin{minipage}{0.05\textwidth}
			\centerline{\includegraphics[width=0.94in,height=0.94in]{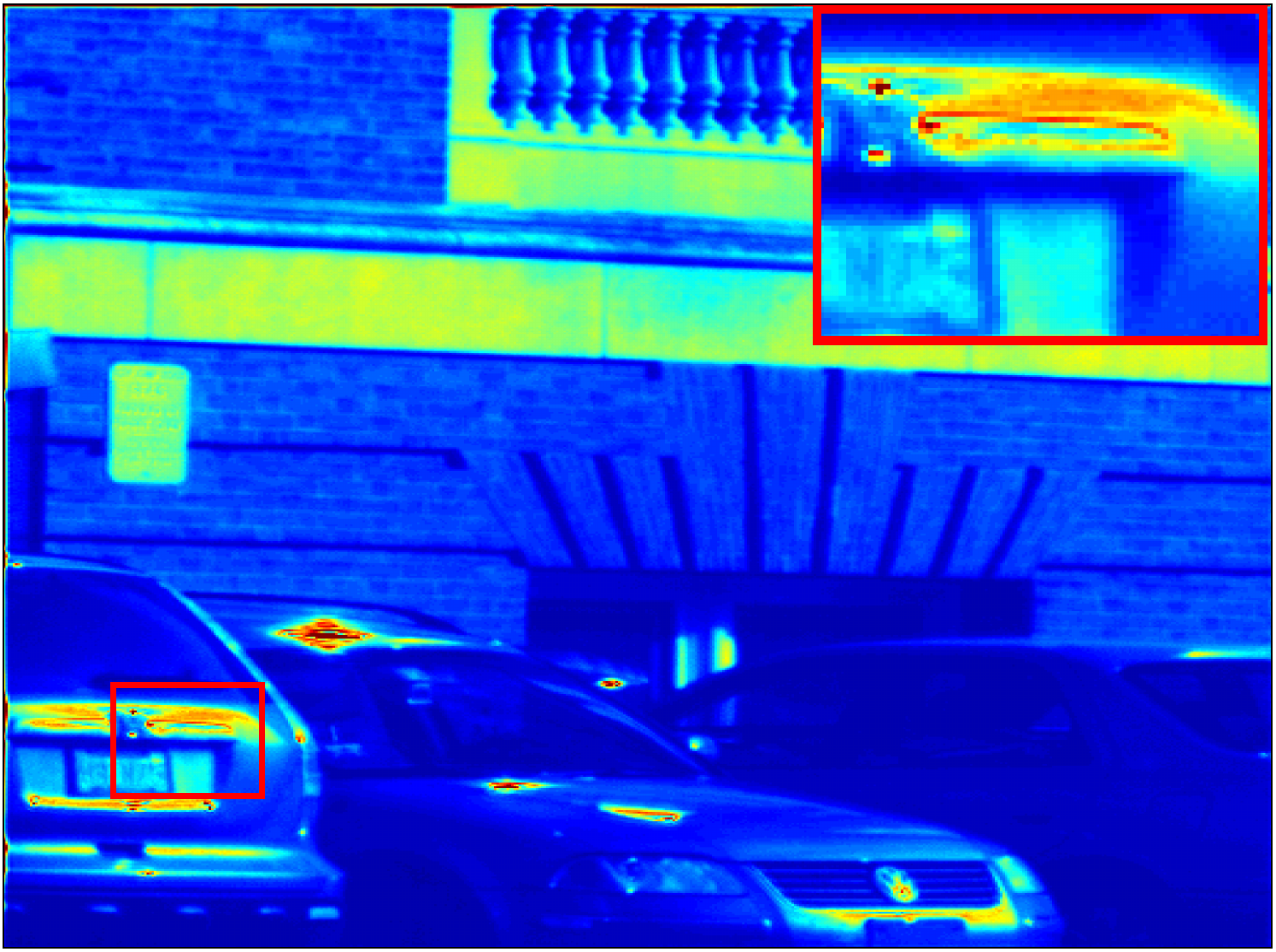}}
		\end{minipage}
		\hspace{1.4cm}
		\begin{minipage}{0.05\textwidth}
			\centerline{\includegraphics[width=0.94in,height=0.94in]{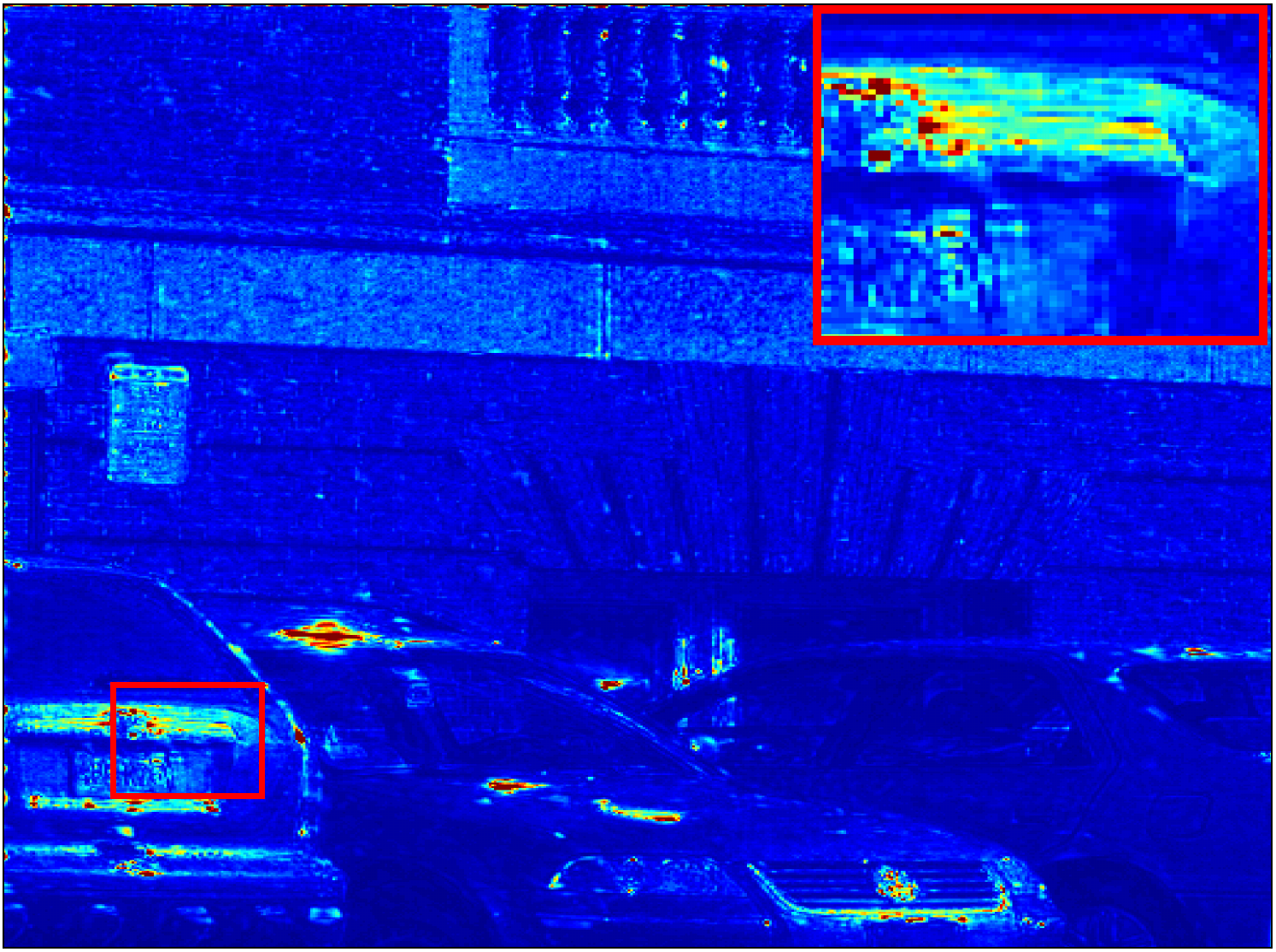}}
		\end{minipage}
		\hspace{1.4cm}
		\begin{minipage}{0.05\textwidth}
			\centerline{\includegraphics[width=0.94in,height=0.94in]{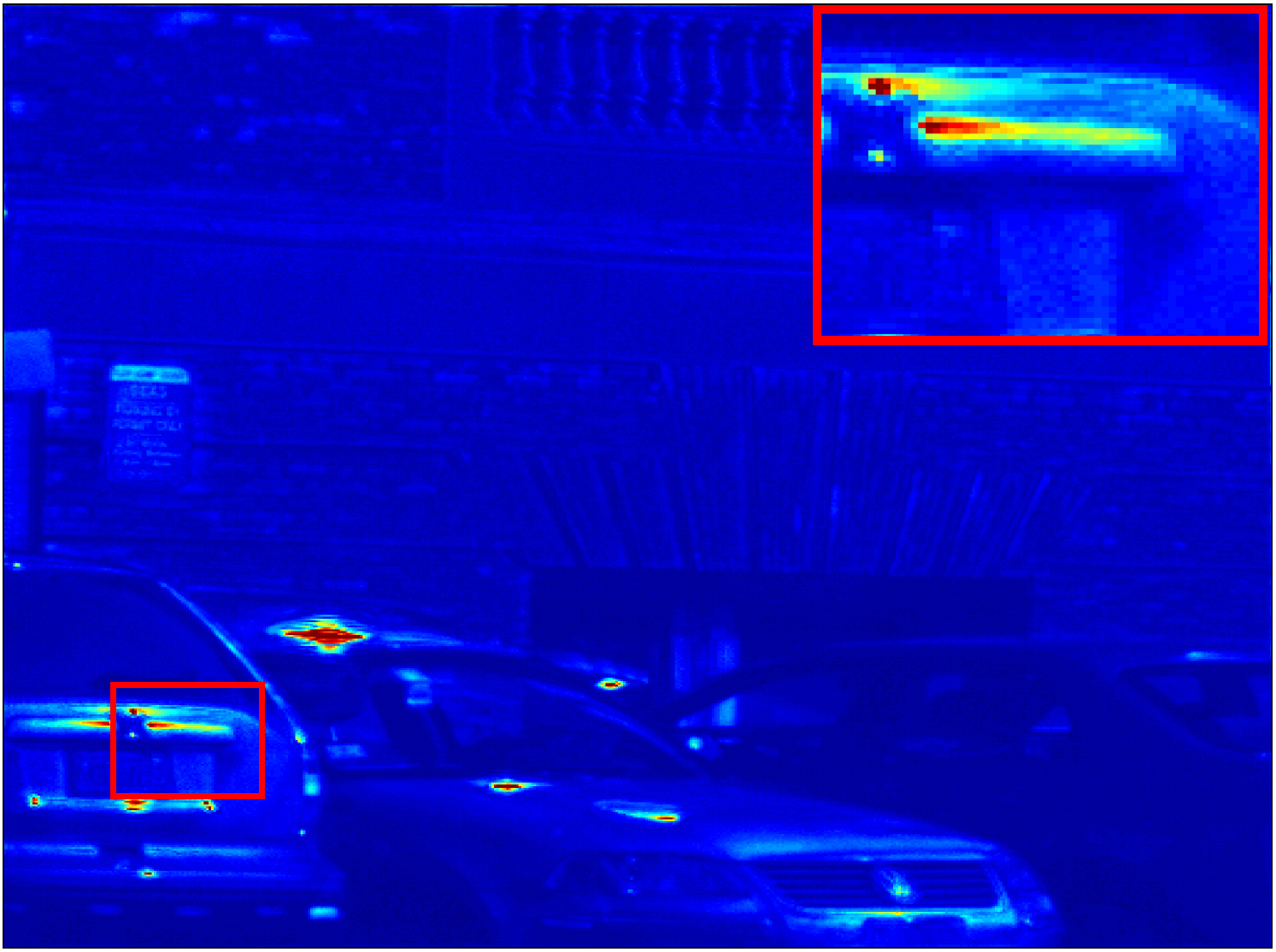}}
		\end{minipage}
		\hspace{0.8cm}
		\begin{minipage}{0.005\textwidth}
			\centerline{\includegraphics[width=0.15in,height=0.94in]{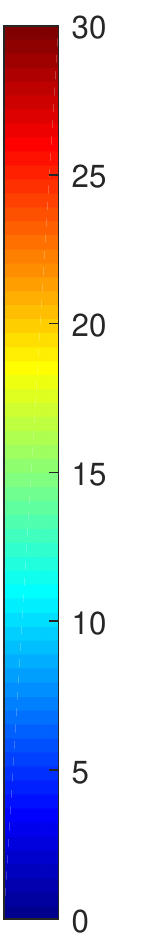}}
		\end{minipage}
		\vfill
		\hspace{0.55cm}
		\begin{minipage}{0.05\textwidth}
			\centerline{\includegraphics[width=0.94in,height=0.94in]{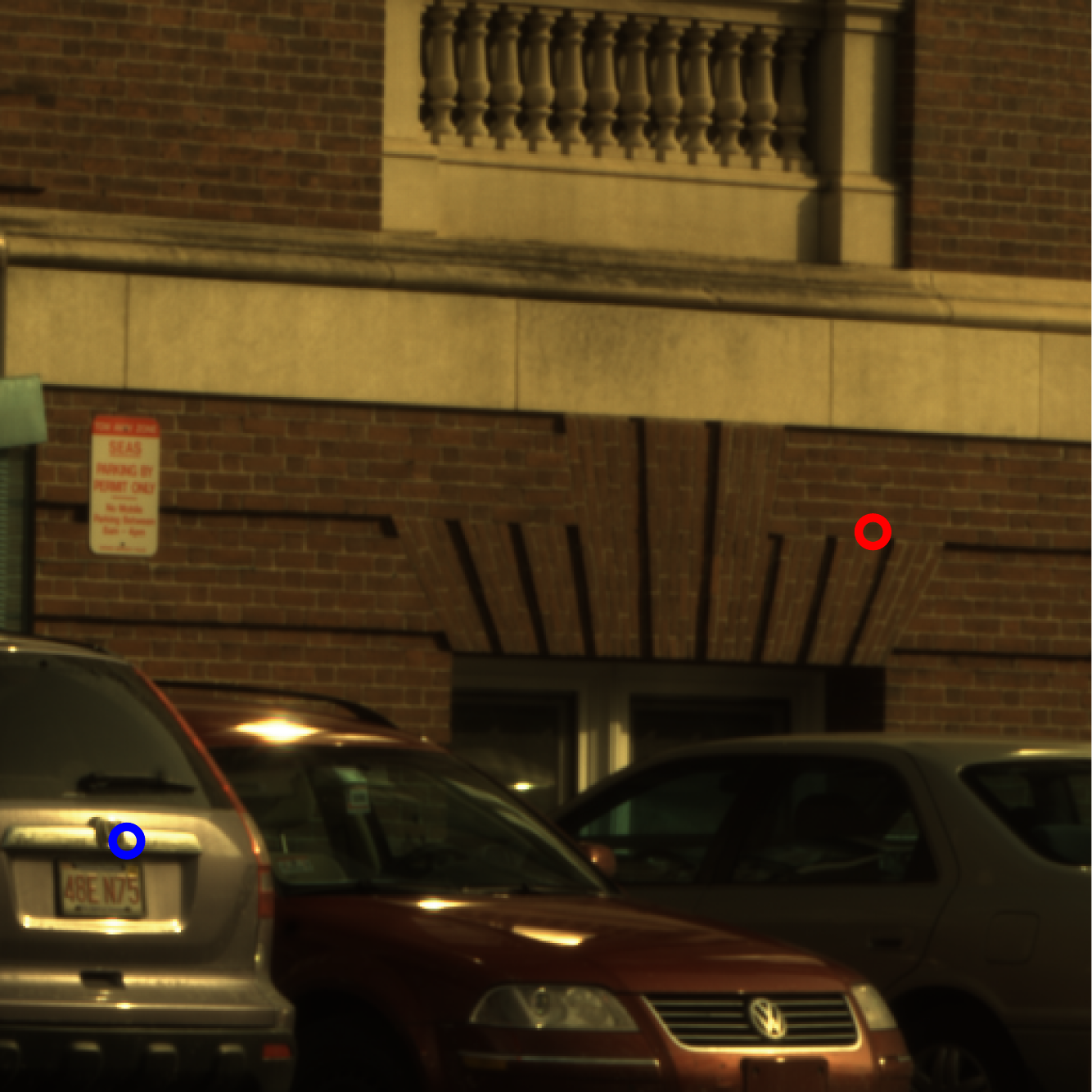}}
		\end{minipage}
		\hspace{1.4cm}
		\begin{minipage}{0.05\textwidth}
			\centerline{\includegraphics[width=0.94in,height=0.8in]{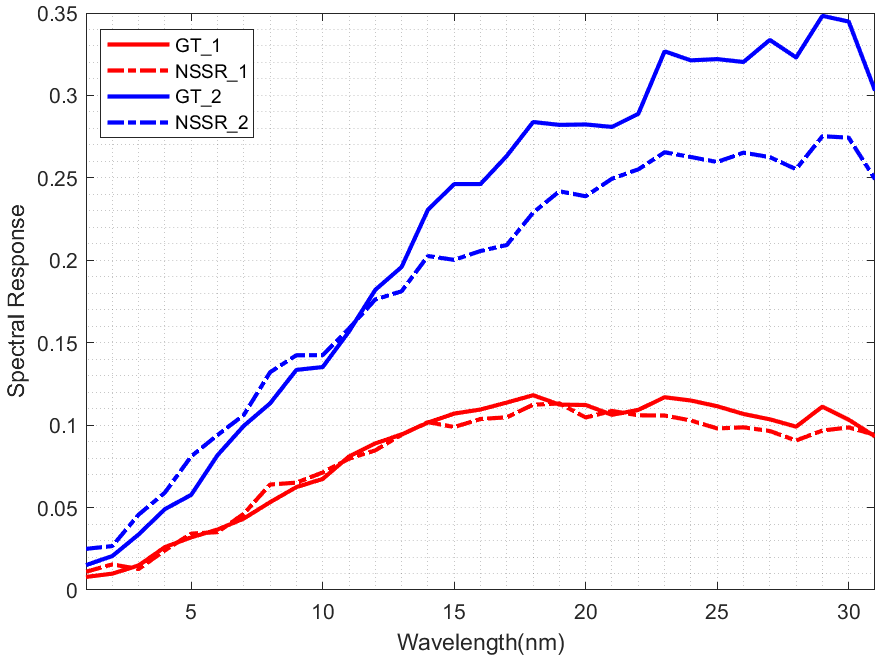}}
		\end{minipage}
		\hspace{1.4cm}
		\begin{minipage}{0.05\textwidth}
			\centerline{\includegraphics[width=0.94in,height=0.8in]{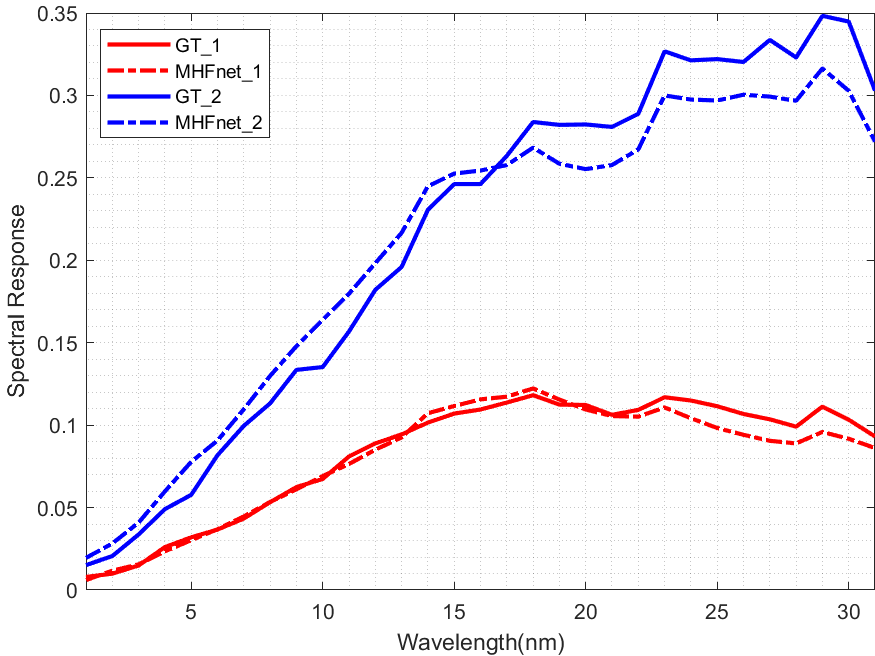}}
		\end{minipage}
		\hspace{1.4cm}
		\begin{minipage}{0.05\textwidth}
			\centerline{\includegraphics[width=0.94in,height=0.8in]{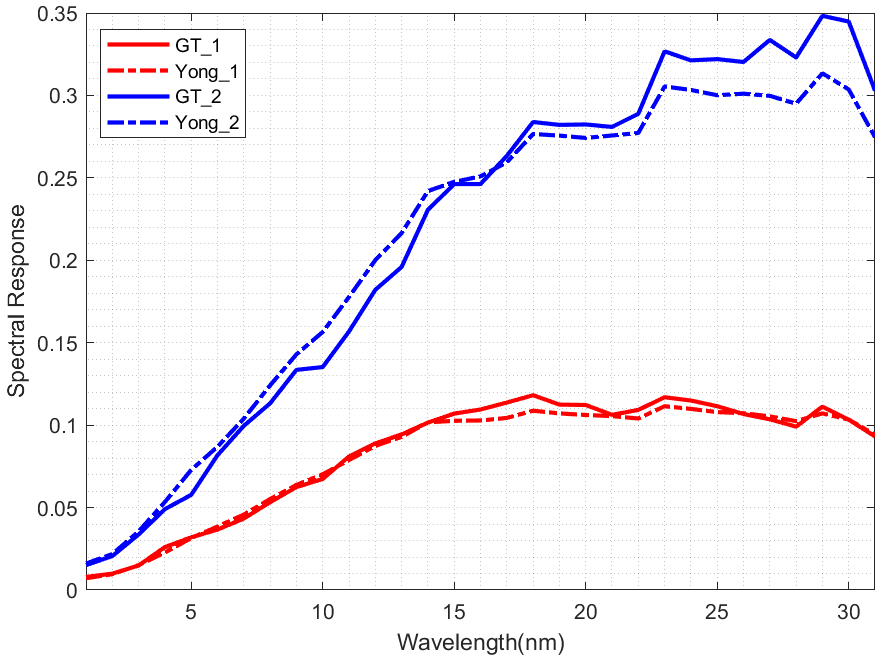}}
		\end{minipage}
		\hspace{1.4cm}
		\begin{minipage}{0.05\textwidth}
			\centerline{\includegraphics[width=0.94in,height=0.8in]{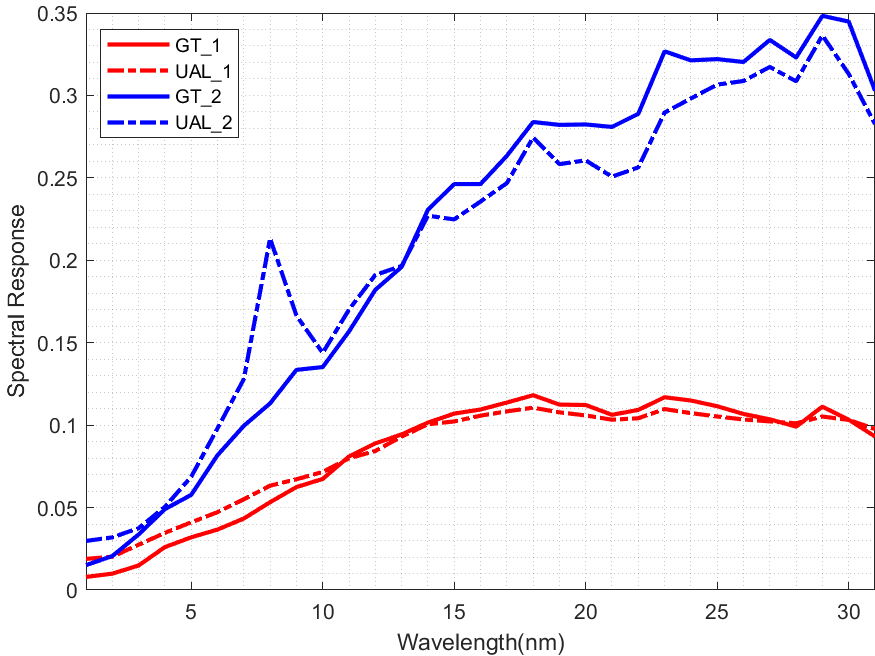}}
		\end{minipage}
		\hspace{1.4cm}
		\begin{minipage}{0.05\textwidth}
			\centerline{\includegraphics[width=0.94in,height=0.8in]{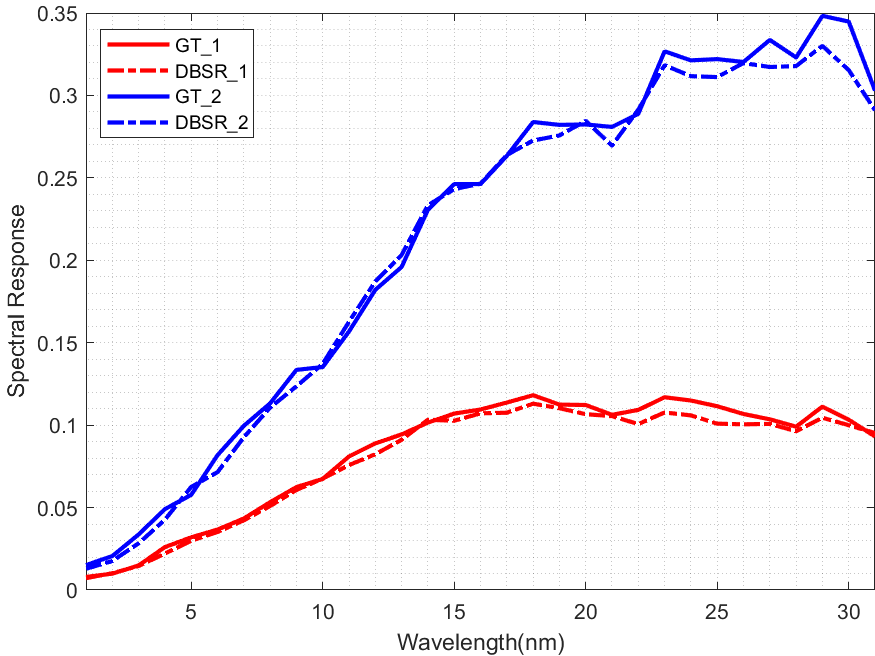}}
		\end{minipage}
		\hspace{1.4cm}
		\begin{minipage}{0.05\textwidth}
			\centerline{\includegraphics[width=0.94in,height=0.8in]{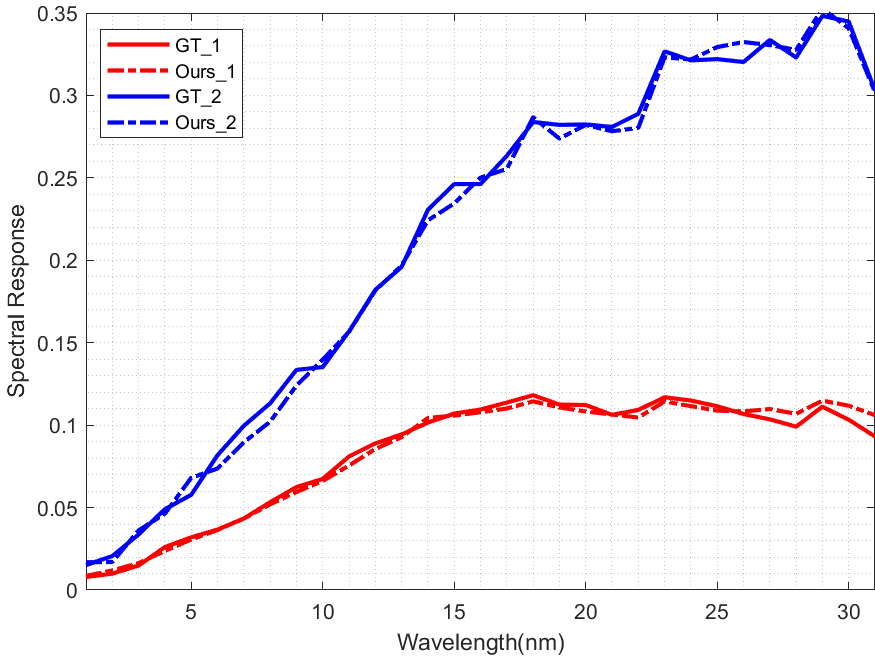}}
		\end{minipage}
		\hspace{0.8cm}
		\begin{minipage}{0.005\textwidth}
			\centerline{\includegraphics[width=0.15in,height=0.8in]{./Picture_Result/Harvard_K1_S16_4040-0.01/imgb2_RGB/colorbar_placeholder.png}}
		\end{minipage}
		\vfill
		\hspace{0.55cm}
		\begin{minipage}{0.05\textwidth}
			\centerline{{\scriptsize (a) GT}}
		\end{minipage}
		\hspace{1.4cm}
		\begin{minipage}{0.05\textwidth}
			\centerline{{\scriptsize (b) NSSR~\cite{7438864}}}
		\end{minipage}
		\hspace{1.4cm}
		\begin{minipage}{0.05\textwidth}
			\centerline{{\scriptsize (c) MHFnet~\cite{8953470}}}
		\end{minipage}
		\hspace{1.4cm}
		\begin{minipage}{0.05\textwidth}
			\centerline{{\scriptsize (d) Yong~\cite{8019510}}}
		\end{minipage}
		\hspace{1.4cm}
		\begin{minipage}{0.05\textwidth}
			\centerline{{\scriptsize (e) UAL~\cite{Ours_CVPR2020}}}
		\end{minipage}
		\hspace{1.4cm}
		\begin{minipage}{0.05\textwidth}
			\centerline{{\scriptsize (f) DBSR~\cite{9136736}}}
		\end{minipage}
		\hspace{1.4cm}
		\begin{minipage}{0.05\textwidth}
			\centerline{{\scriptsize (g) Ours}}
		\end{minipage}
		\hspace{0.8cm}
		\begin{minipage}{0.005\textwidth}
			\centerline{}
		\end{minipage}
		\caption{The visual SR results of all methods on the Harvard dataset. The observed LR HSI and HR MSI generated by $\mathbf{k}_1$ and $\mathbf{P}_{0.01}$, respectively. SNRs of both two observed images are 40dB, and the SR scale is 16. }
		\label{Fig_Harvard_Visible}
	\end{figure*}
	
	\noindent \textbf{The experiment results on the CAVE dataset}\quad In the main manuscript, we have demonstrated several experimental results on the CAVE dataset with different settings. In this experiment, considering the observed images are always contaminated by noise, we further verify the performance of the proposed method influenced by the observed images with noise. For this purpose, we add the noise with different intensities into the original noisy-free images, which results in the observed images different SNRs. In experiment, the test LR HSI and HR MSI are generated from $\mathbf{k}_1$ and $\mathbf{P}_{0.01}$, respectively, and the SR scale is 8. We summarize the corresponding numerical results in Table~\ref{Tab_CAVE_Result_Diff_Noise}. It can be seen that the performance of all methods degrade with larger noise intensity. For example, the SAM of NSSR~\cite{7438864} decreases to 35 when the SNR is 30dB. Nevertheless, the results of Ours still obtain obvious advantages over the competing four evaluation metrics.

	In addition, to further clarify the superiority of the proposed method, we select two representative pixels (marked by different colors) in the displayed HSI to plot the reconstructed spectral curve. The corresponding spectral curves are provided in the third row of Figure~\ref{Fig_CAVE_Visible_sup}, in which the reconstructed spectrum is plotted with a dotted line and the ground truth spectrum is plotted with a solid line. It can be seen that the proposed method can accurately reconstructs the spectra, which demonstrates it has smaller spectral distortion compared than the competing methods.

	\noindent \textbf{The experiment results on the Harvard dataset}\quad
	In this experiment, we give the visual SR results and the reconstruction curves of spectra in Figure~\ref{Fig_Harvard_Visible}. It can be seen that the reconstructed results of the proposed method have more details and lower reconstruction errors than the competing methods. In addition, the reconstructed spectrum of the proposed method is much closer to the ground truth than the other methods. From above experimental results, we can conclude the proposed blind HSI SR method can well reconstruct the latent HSI in both spatial and spectral domain on the Harvard dataset.

	\begin{table*}[!htbp]\small
		\centering
		\caption{The performance of each method on the NTIRE dataset with different SR scale factors, while the input LR HSI and HR MSI generated by $\mathbf{k}_1$ and $\mathbf{P}_{0.01}$, respectively. The SNR of this two input images are both 40dB, and the best results are in bold.}
		\setlength{\tabcolsep}{2.4mm}{\begin{tabular}{c|c|c|c|c|c|c|c|c|c|c|c|c}
				\hline
				\multirow{2}*{Methods}& \multicolumn{4}{c|}{s = 8}&\multicolumn{4}{c|}{s = 16}& \multicolumn{4}{c}{s = 32}\\
				\cline{2-13}&RMSE&PSNR&SAM&SSIM&RMSE&PSNR&SAM&SSIM&RMSE&PSNR&SAM&SSIM\\
				\hline	
				NSSR~\cite{7438864}&7.11&31.43&1.55&0.9447&6.69&32.01&2.46&0.9454&5.91&33.09&2.81&0.9696\\
				MHFnet~\cite{8953470}&3.16&38.78&1.53&0.9928&3.44&38.07&1.69&0.9925&4.06&36.58&2.14&0.9920\\
				\hline
				Yong~\cite{8019510}&3.27&38.23&1.57&0.9804&4.19&36.23&2.11&0.9774&4.51&35.59&2.50&0.9801\\
				UAL~\cite{Ours_CVPR2020}&4.68&35.61&2.21&0.9924&5.61&33.64&2.27&0.9891&6.05&32.97&2.27&0.9864\\
				\hline
				DBSR~\cite{9136736}&7.08&33.10&\textbf{0.73}&0.9799&6.61&33.47&\textbf{0.95}&0.9788&11.43&27.85&1.54&0.9542\\
				\hline
				Ours&\textbf{2.19}&\textbf{41.75}&0.88&\textbf{0.9955}&\textbf{2.31}&\textbf{41.28}&1.09&\textbf{0.9951}&\textbf{3.11}&\textbf{38.69}&\textbf{1.46}&\textbf{0.9936}\\
				\hline
		\end{tabular}}
		\label{Tab_NTIRE_Result_Diff_Scale}
	\end{table*}
	
	\begin{figure*}[!htbp]
		\centering
		\hspace{0.55cm}
		\begin{minipage}{0.05\textwidth}
			\centerline{\includegraphics[width=0.94in,height=0.94in]{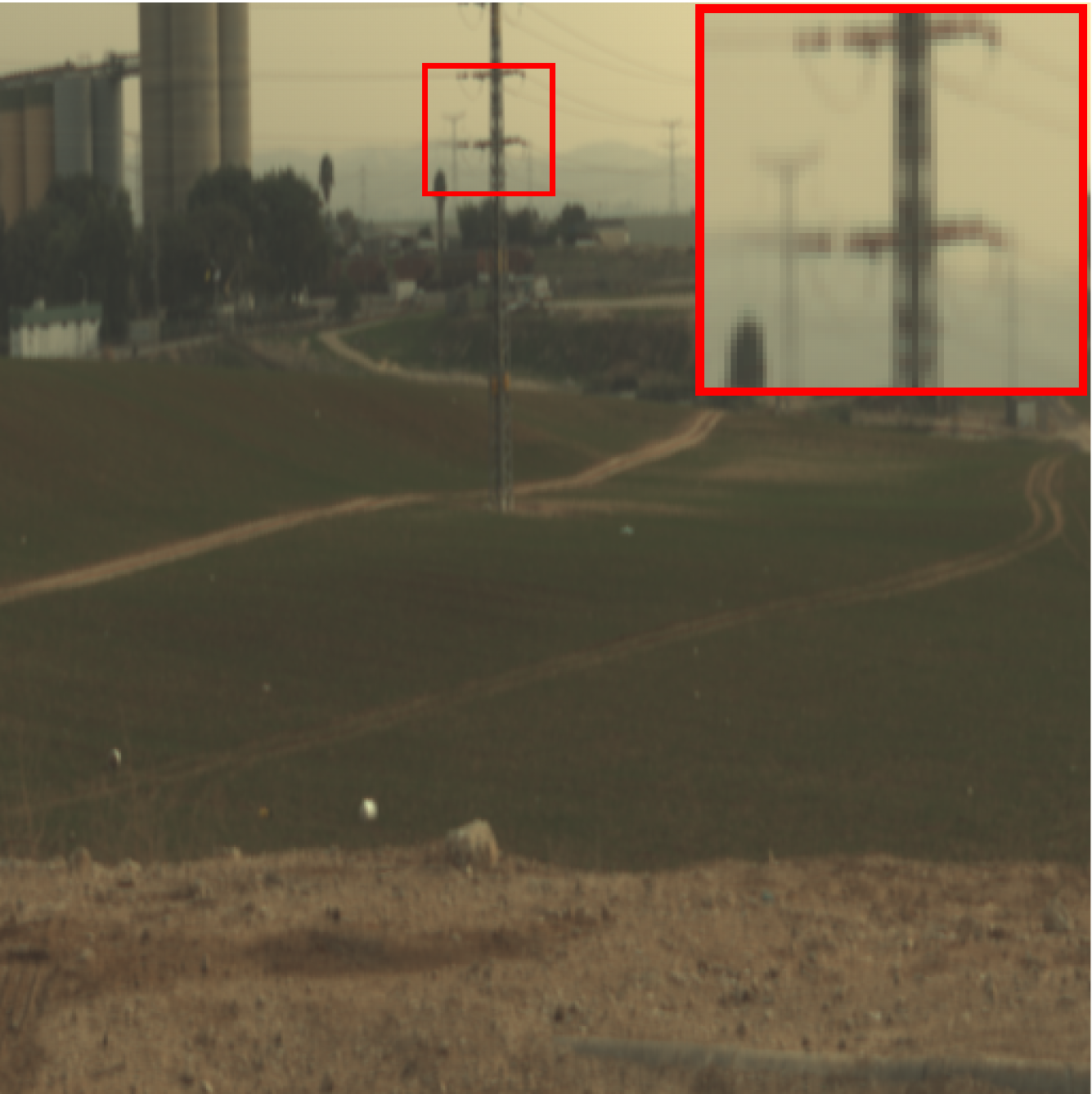}}
		\end{minipage}
		\hspace{1.4cm}
		\begin{minipage}{0.05\textwidth}
			\centerline{\includegraphics[width=0.94in,height=0.94in]{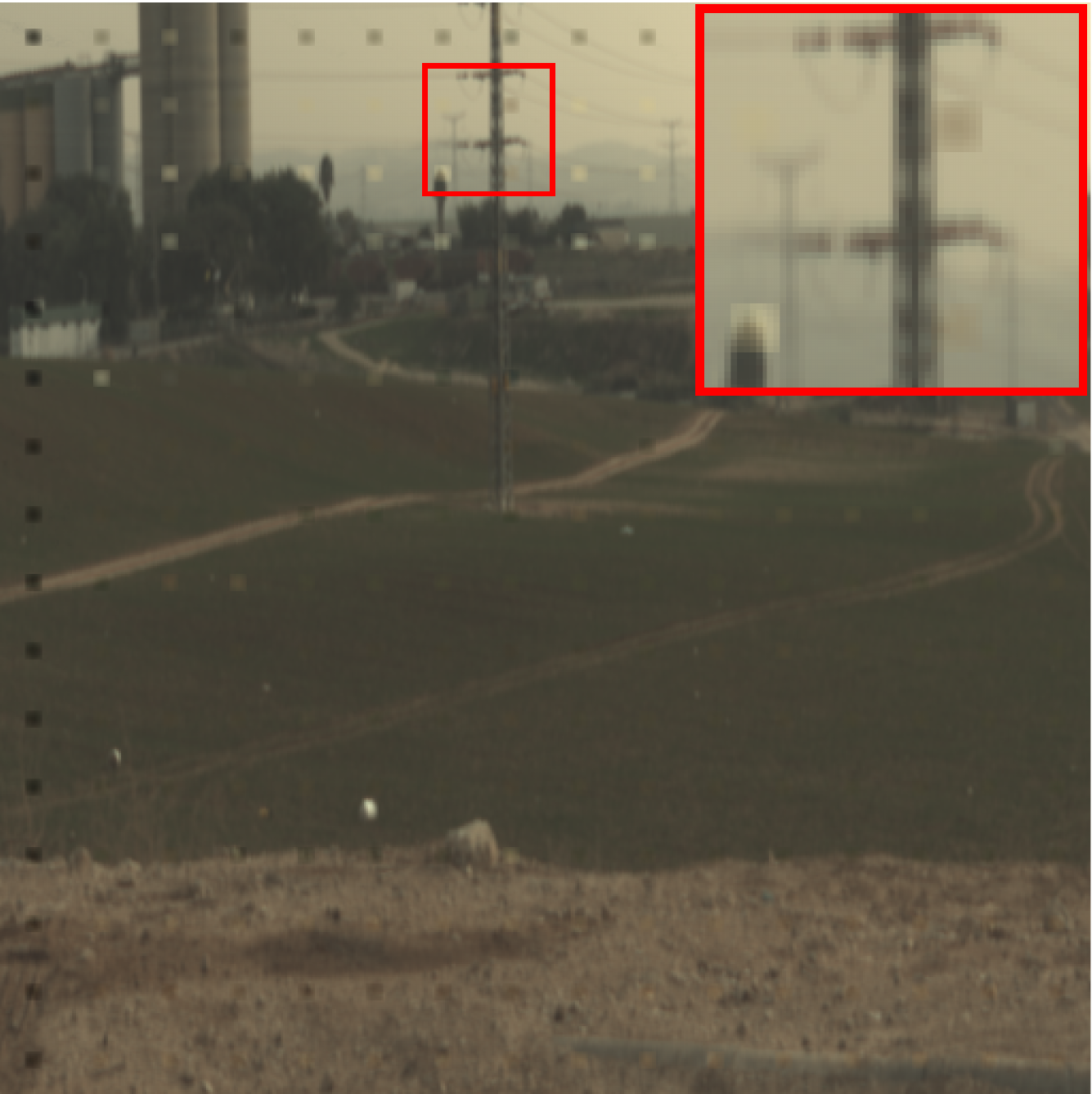}}
		\end{minipage}
		\hspace{1.4cm}
		\begin{minipage}{0.05\textwidth}
			\centerline{\includegraphics[width=0.94in,height=0.94in]{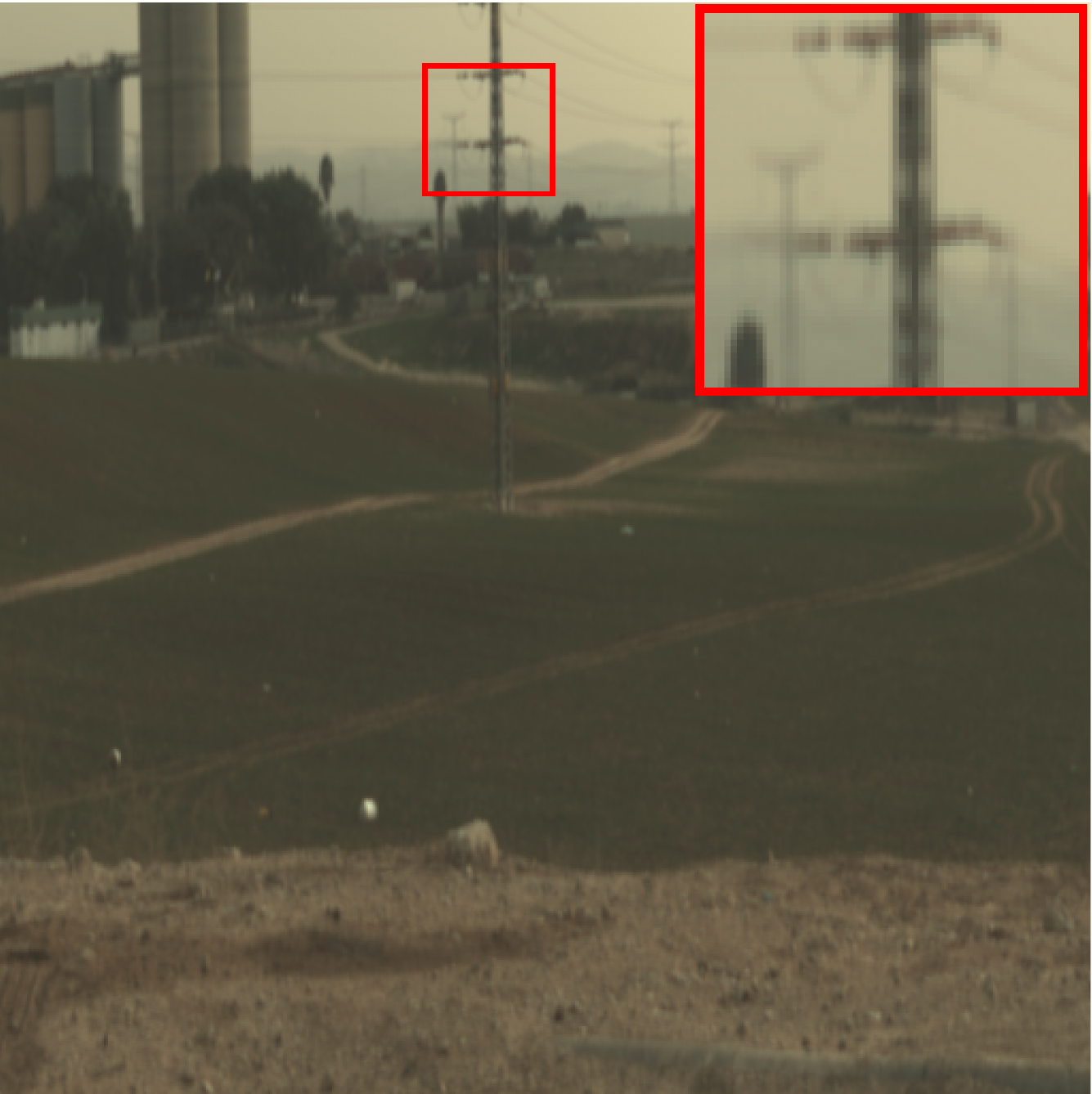}}
		\end{minipage}
		\hspace{1.4cm}
		\begin{minipage}{0.05\textwidth}
			\centerline{\includegraphics[width=0.94in,height=0.94in]{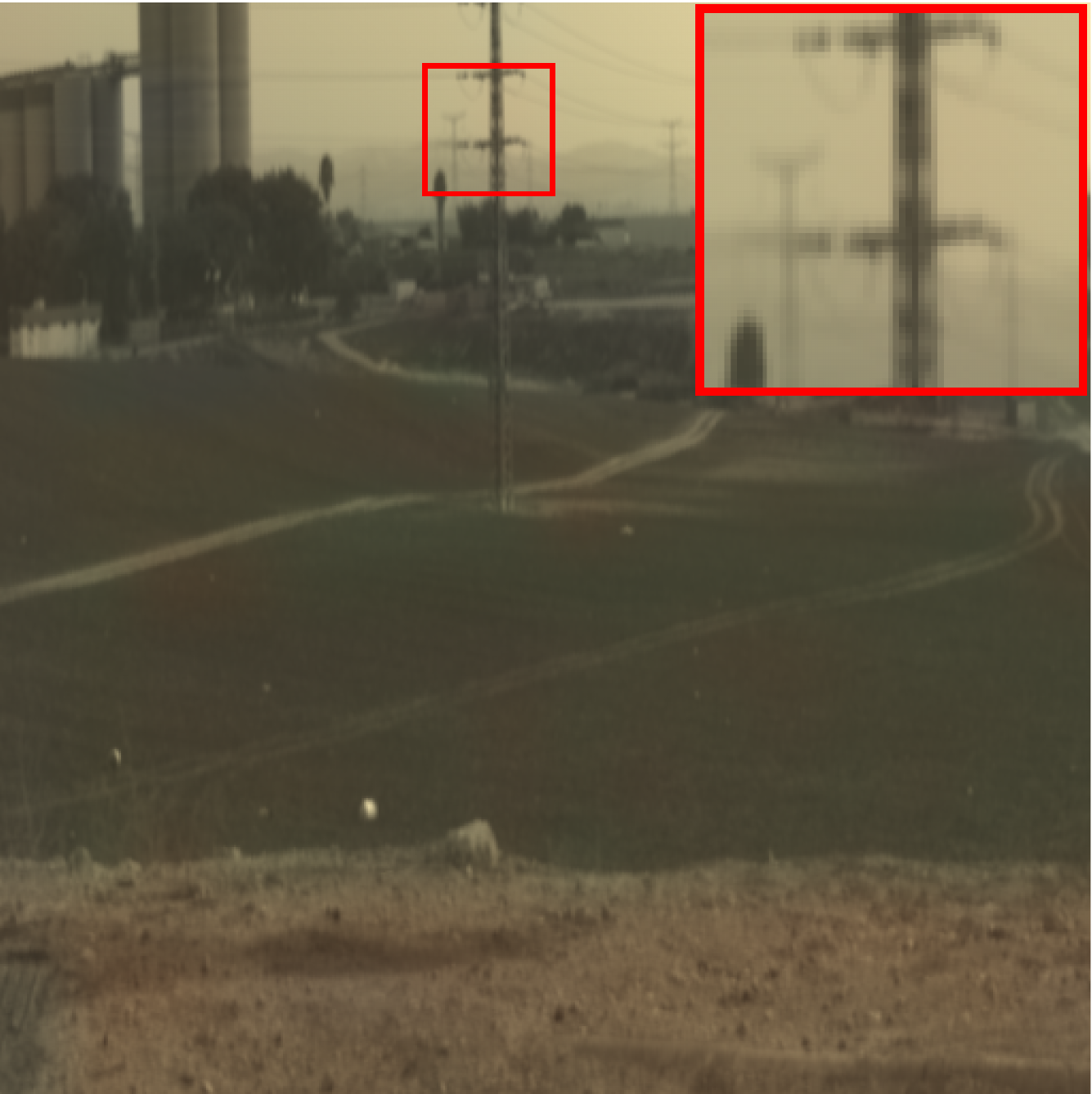}}
		\end{minipage}
		\hspace{1.4cm}
		\begin{minipage}{0.05\textwidth}
			\centerline{\includegraphics[width=0.94in,height=0.94in]{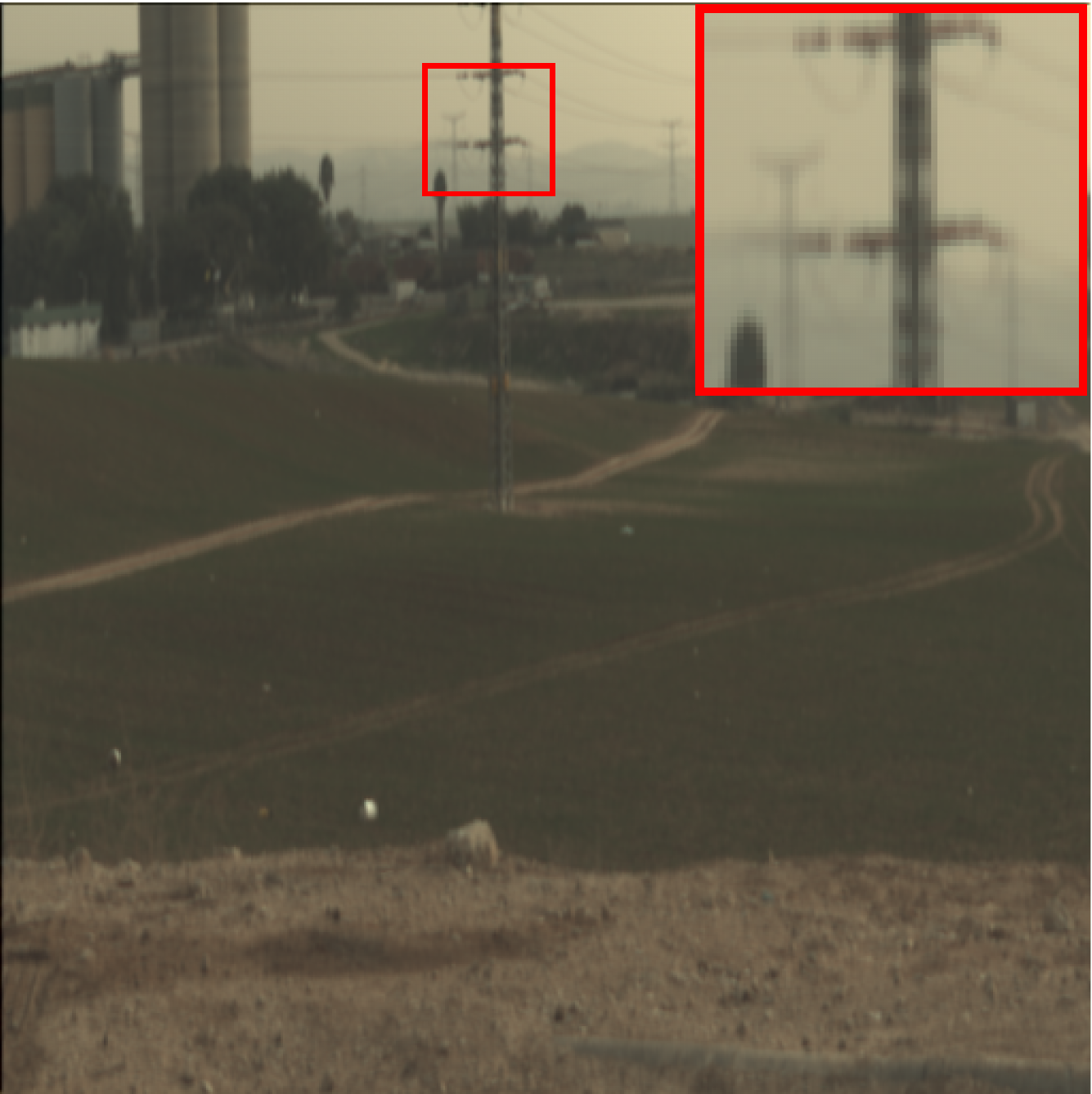}}
		\end{minipage}
		\hspace{1.4cm}
		\begin{minipage}{0.05\textwidth}
			\centerline{\includegraphics[width=0.94in,height=0.94in]{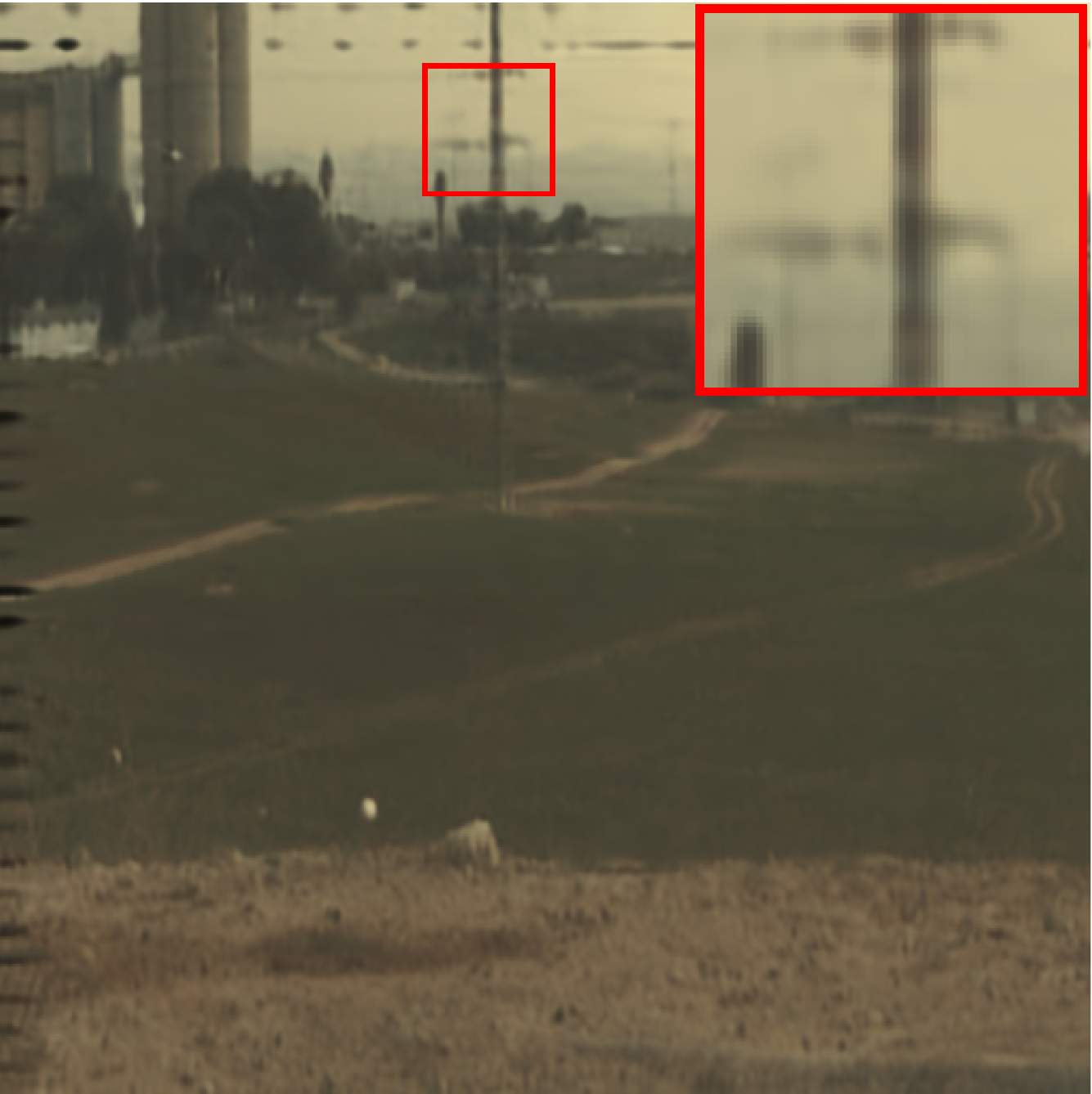}}
		\end{minipage}
		\hspace{1.4cm}
		\begin{minipage}{0.05\textwidth}
			\centerline{\includegraphics[width=0.94in,height=0.94in]{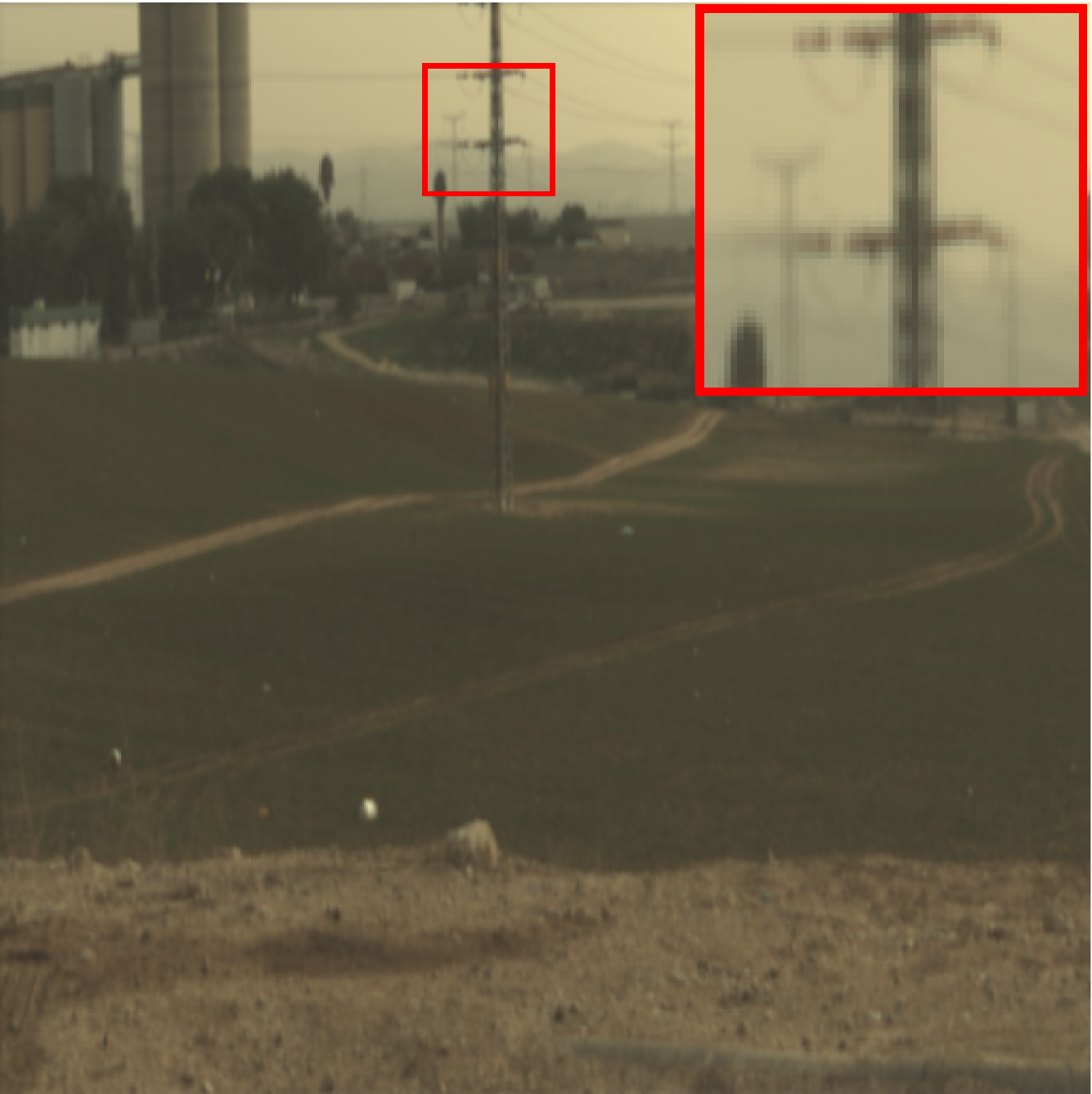}}
		\end{minipage}
		\hspace{0.8cm}
		\begin{minipage}{0.005\textwidth}
			\centerline{\includegraphics[width=0.1in,height=0.94in]{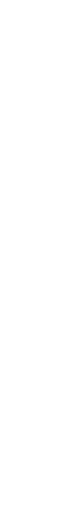}}
		\end{minipage}
		\vfill
		\hspace{0.55cm}
		\begin{minipage}{0.05\textwidth}
			\centerline{\includegraphics[width=0.94in,height=0.94in]{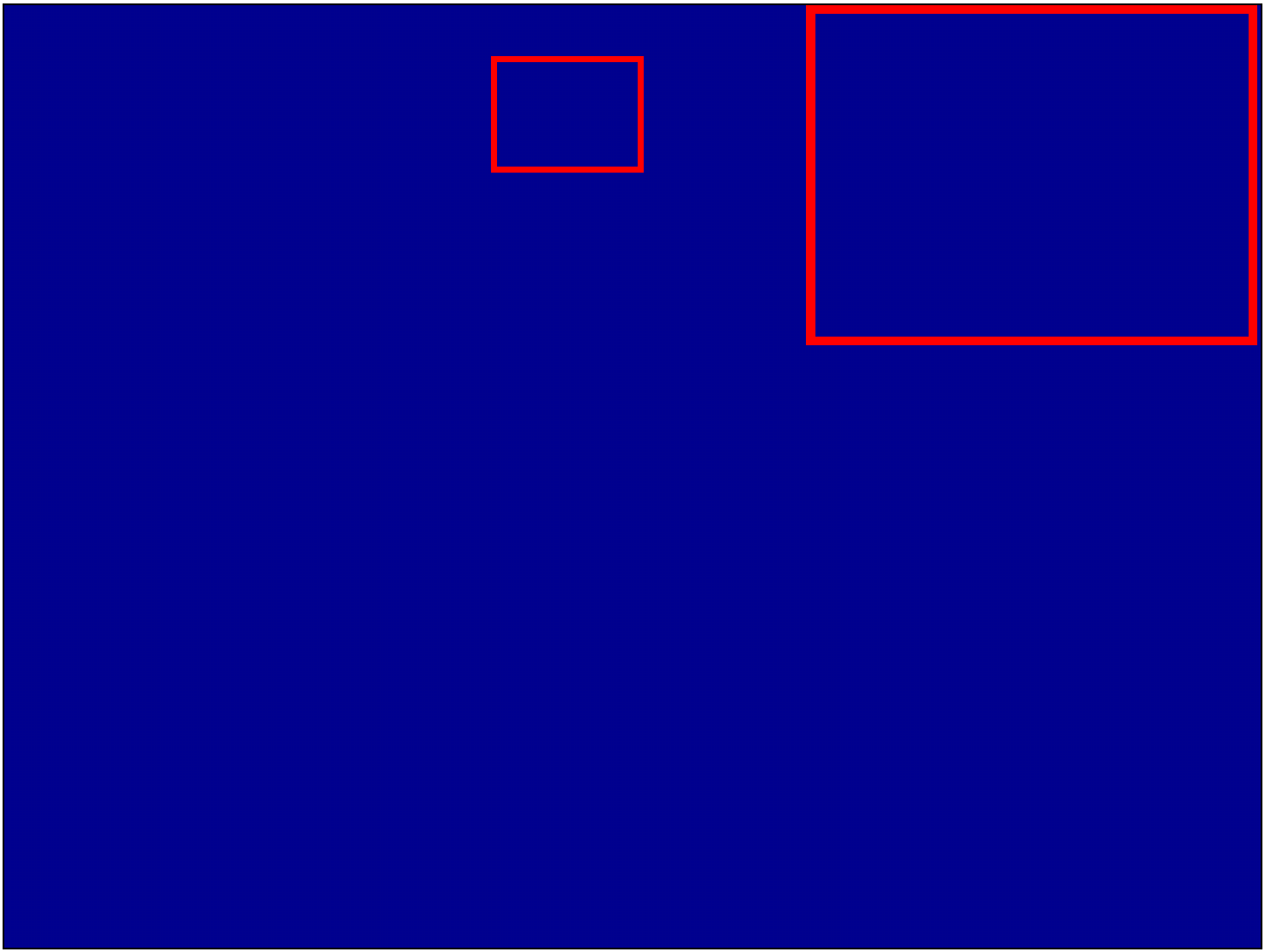}}
		\end{minipage}
		\hspace{1.4cm}
		\begin{minipage}{0.05\textwidth}
			\centerline{\includegraphics[width=0.94in,height=0.94in]{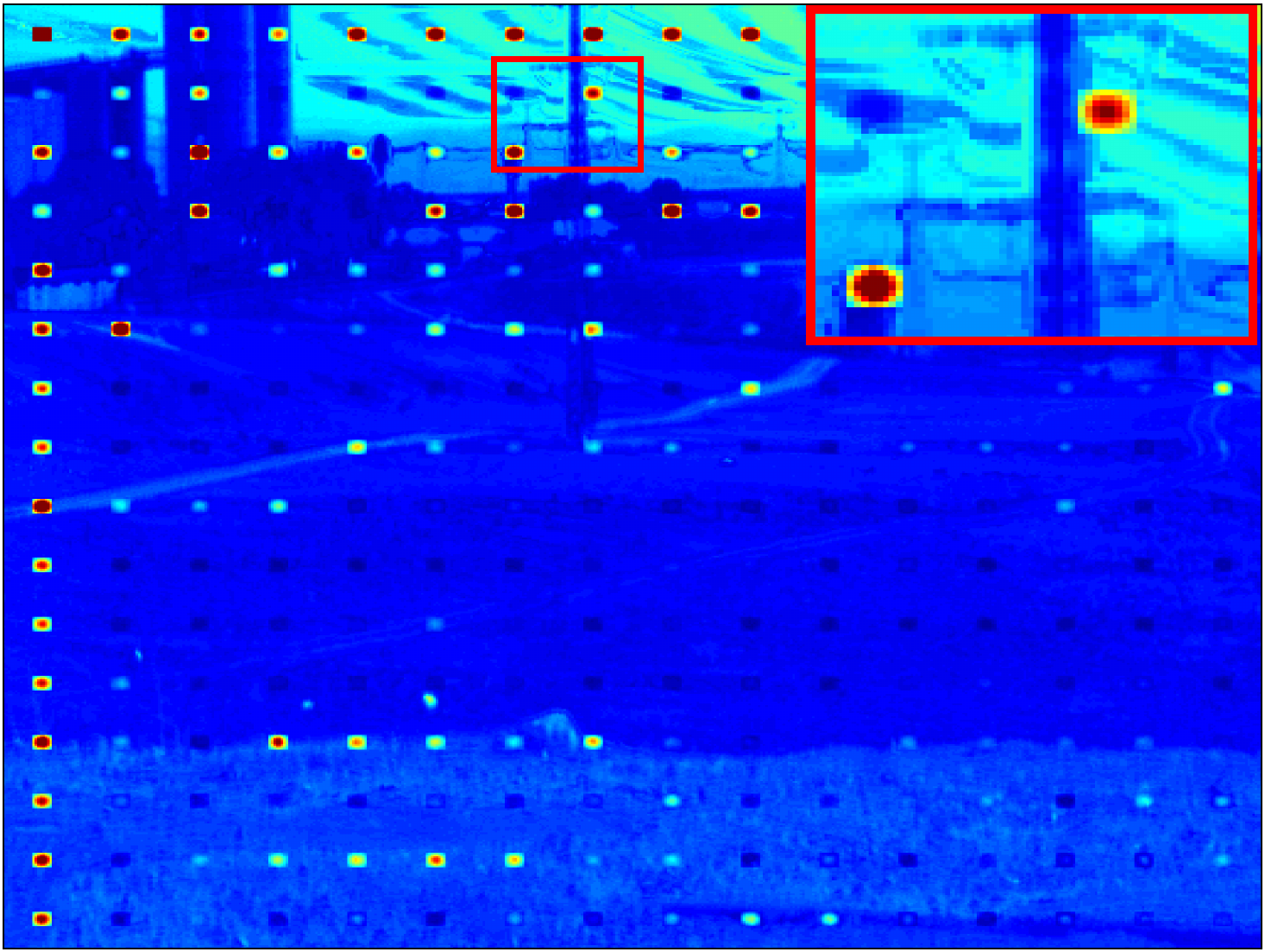}}
		\end{minipage}
		\hspace{1.4cm}
		\begin{minipage}{0.05\textwidth}
			\centerline{\includegraphics[width=0.94in,height=0.94in]{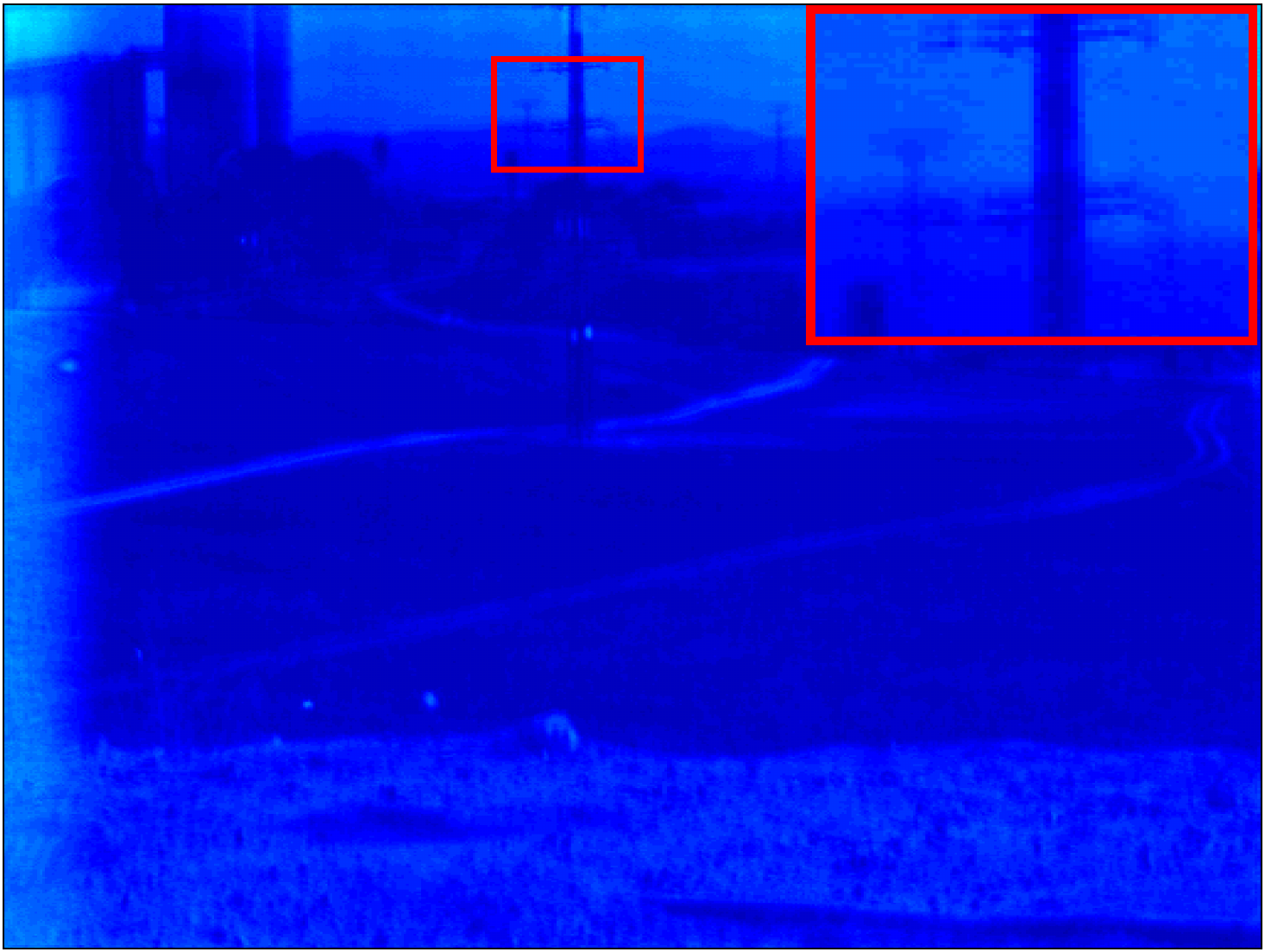}}
		\end{minipage}
		\hspace{1.4cm}
		\begin{minipage}{0.05\textwidth}
			\centerline{\includegraphics[width=0.94in,height=0.94in]{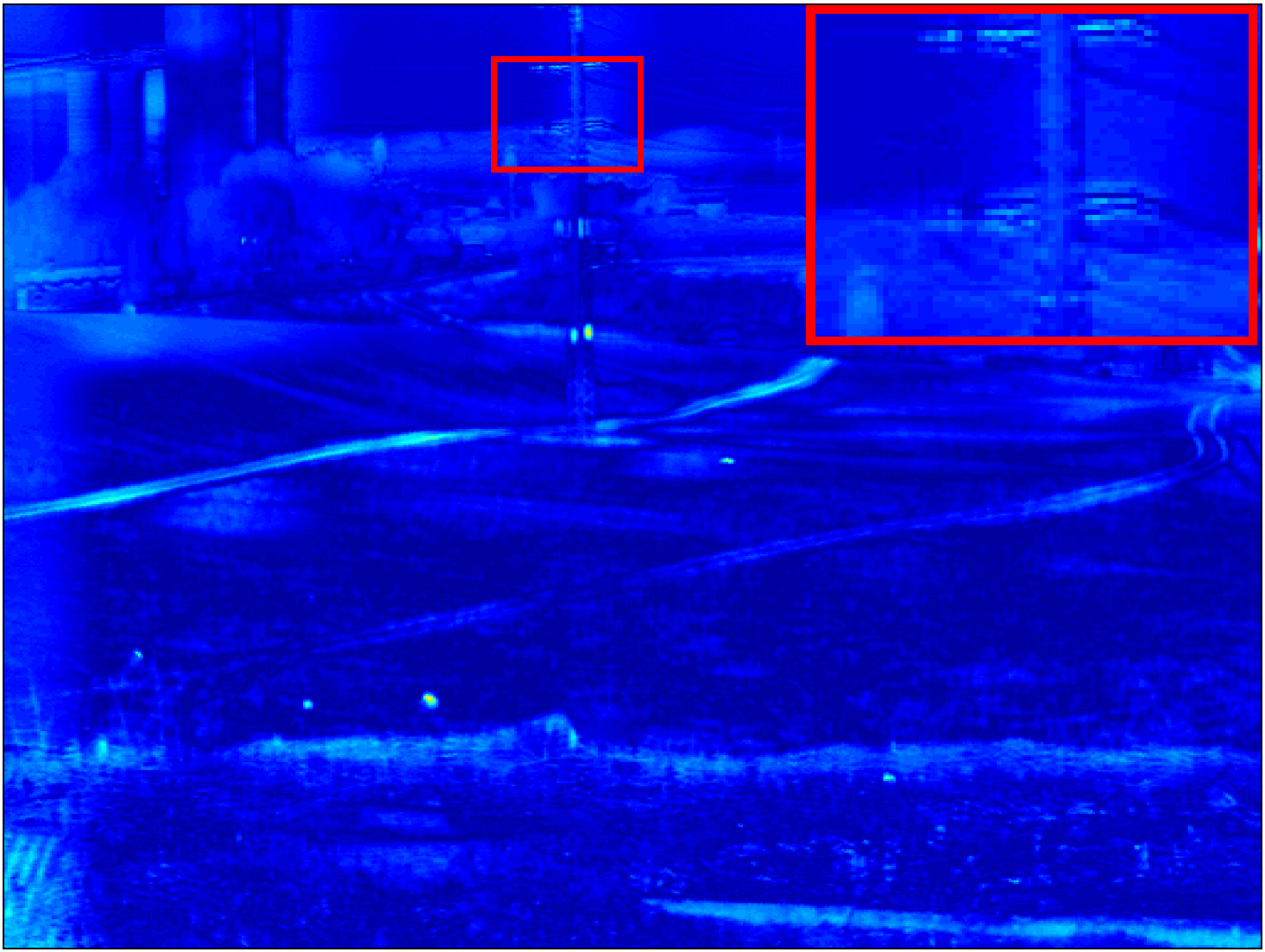}}
		\end{minipage}
		\hspace{1.4cm}
		\begin{minipage}{0.05\textwidth}
			\centerline{\includegraphics[width=0.94in,height=0.94in]{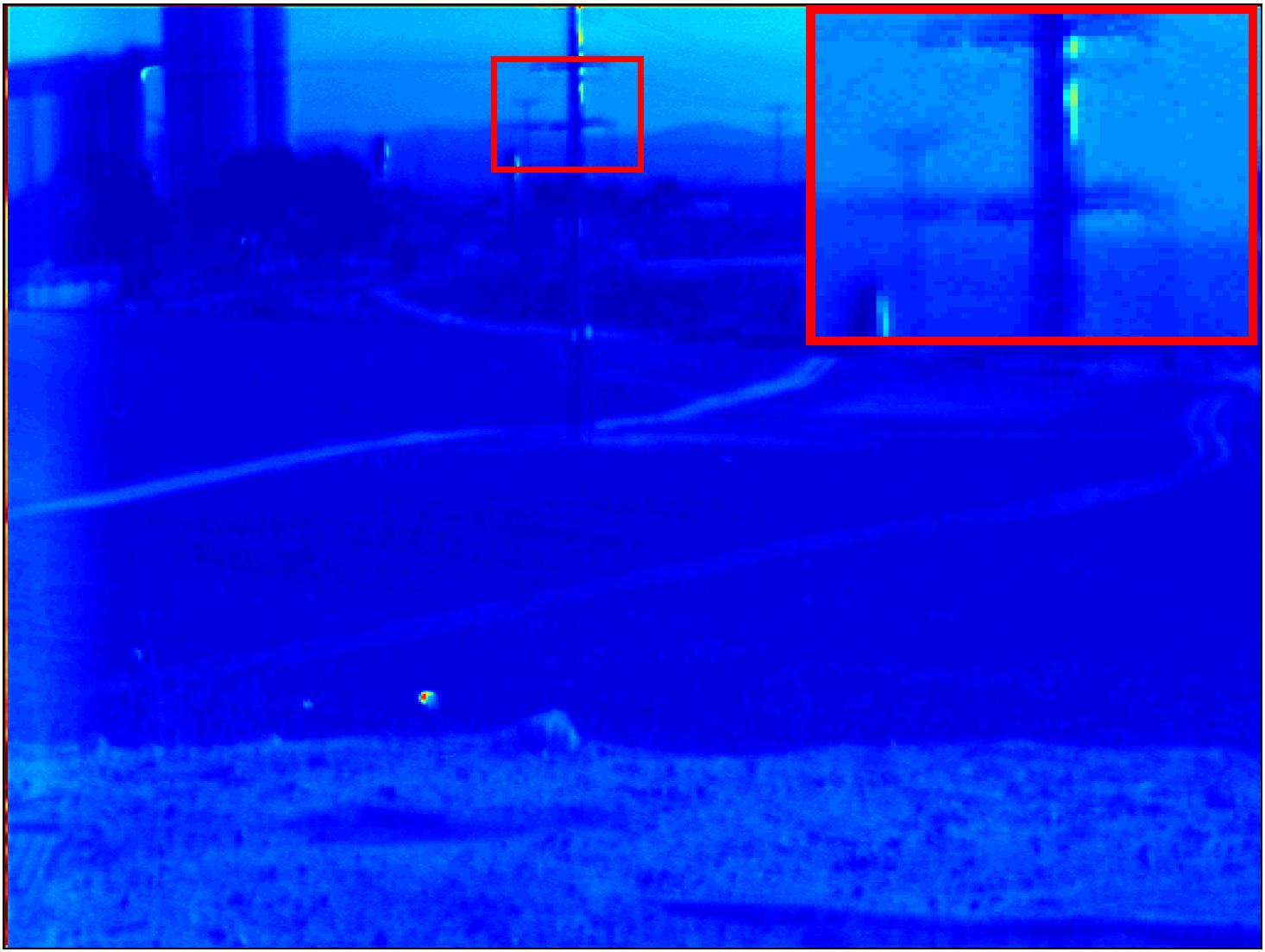}}
		\end{minipage}
		\hspace{1.4cm}
		\begin{minipage}{0.05\textwidth}
			\centerline{\includegraphics[width=0.94in,height=0.94in]{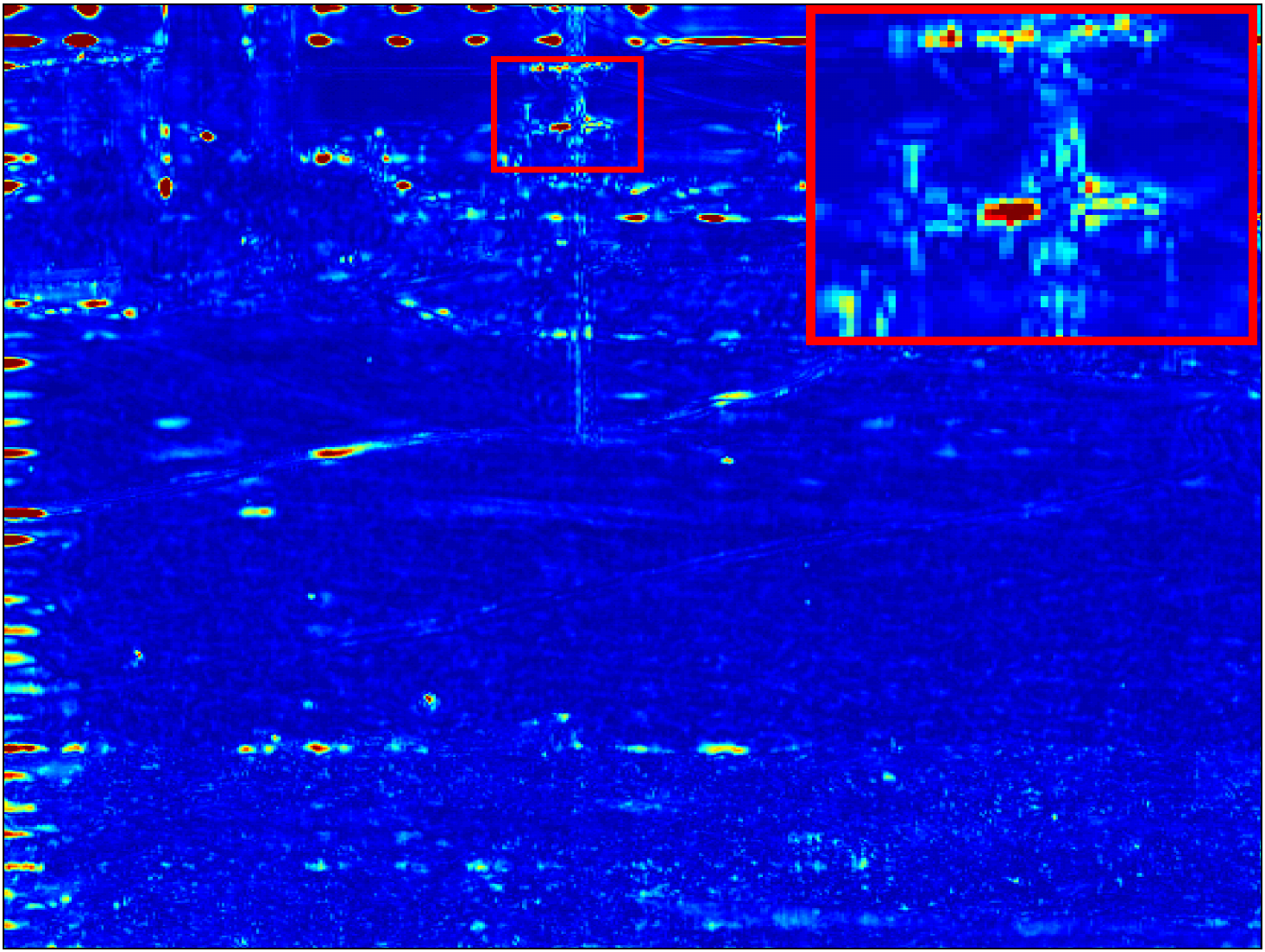}}
		\end{minipage}
		\hspace{1.4cm}
		\begin{minipage}{0.05\textwidth}
			\centerline{\includegraphics[width=0.94in,height=0.94in]{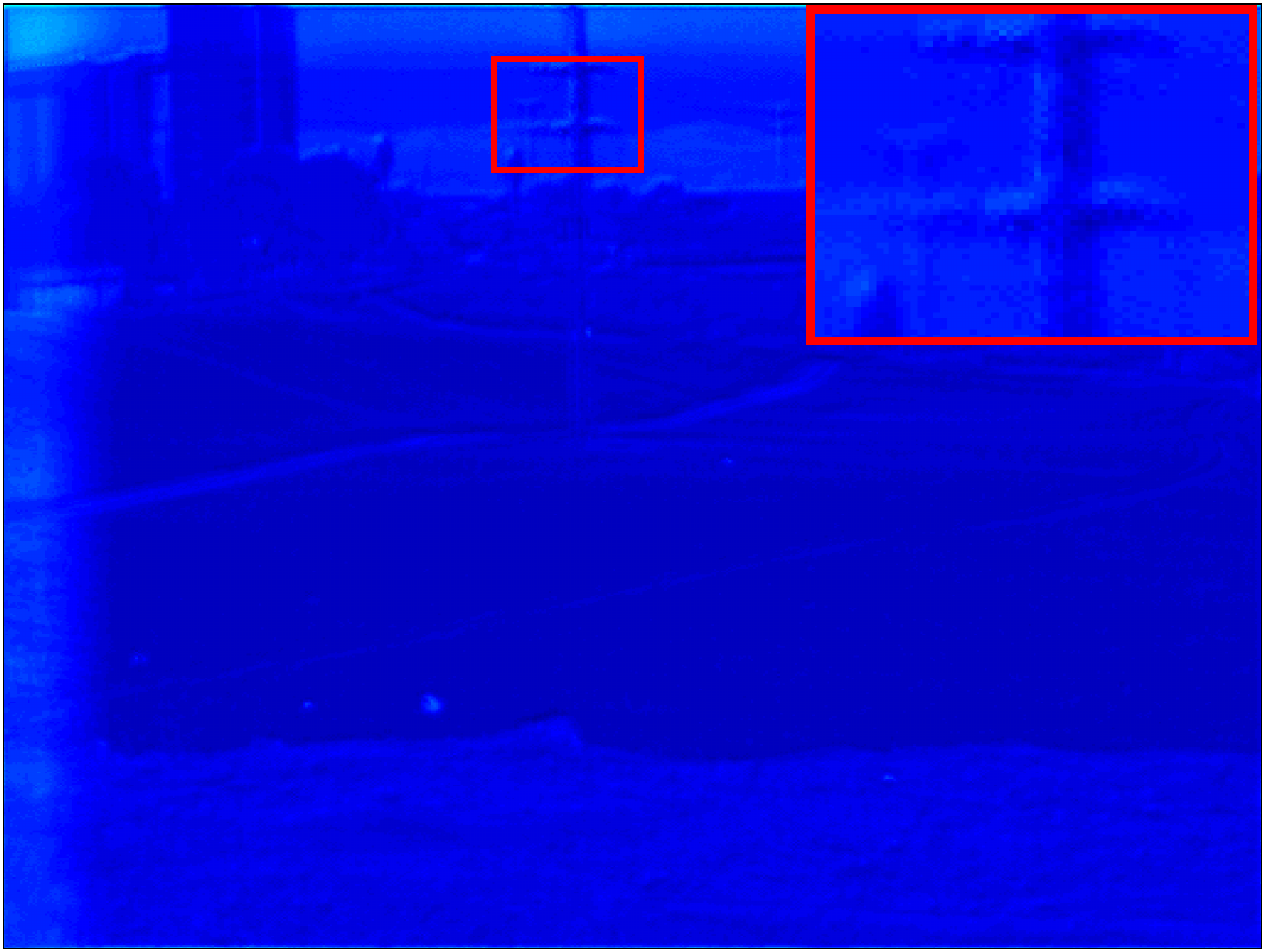}}
		\end{minipage}
		\hspace{0.8cm}
		\begin{minipage}{0.005\textwidth}
			\centerline{\includegraphics[width=0.15in,height=0.94in]{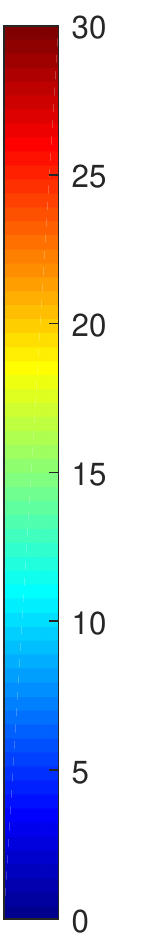}}
		\end{minipage}
		\vfill
		\hspace{0.55cm}
		\begin{minipage}{0.05\textwidth}
			\centerline{\includegraphics[width=0.94in,height=0.94in]{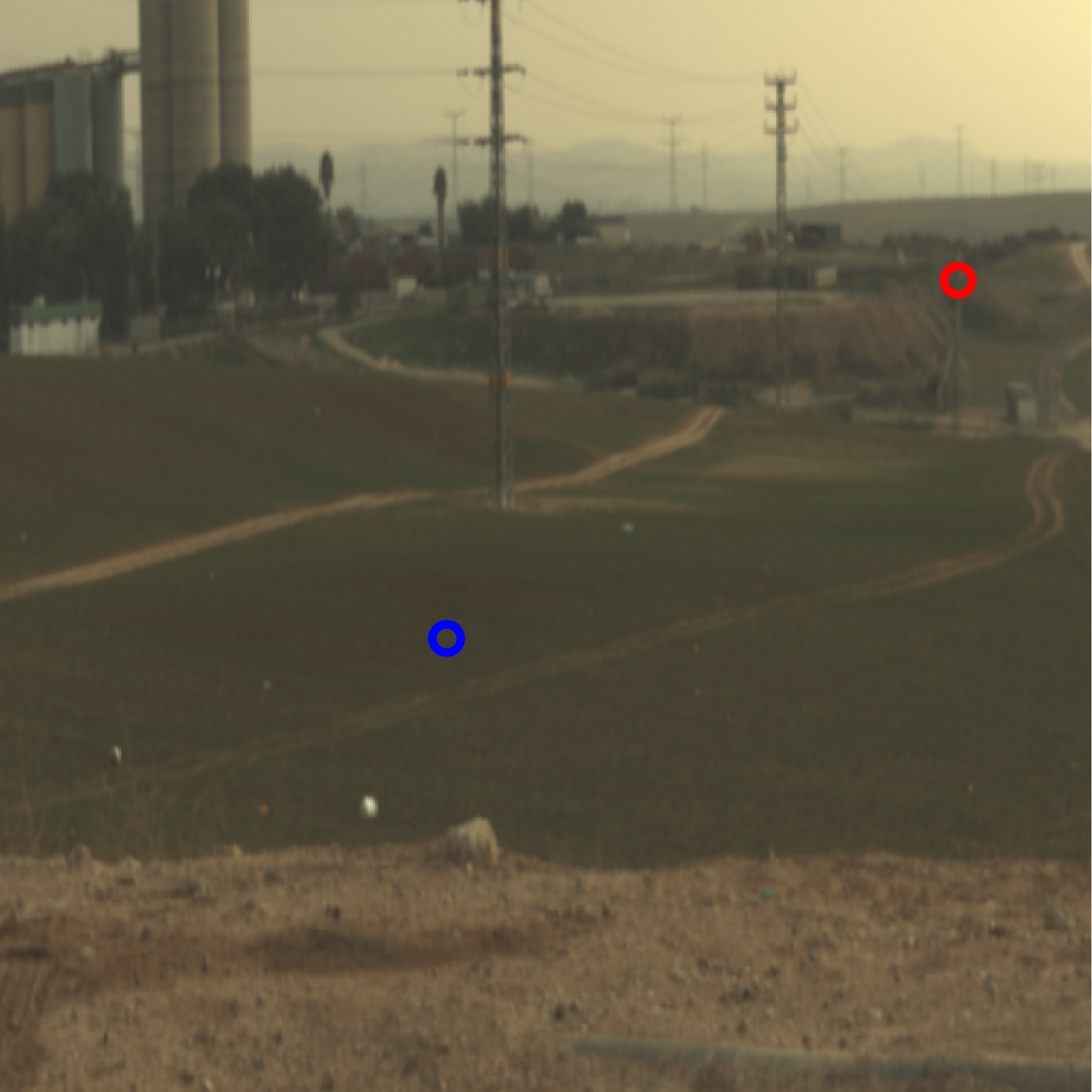}}
		\end{minipage}
		\hspace{1.4cm}
		\begin{minipage}{0.05\textwidth}
			\centerline{\includegraphics[width=0.94in,height=0.8in]{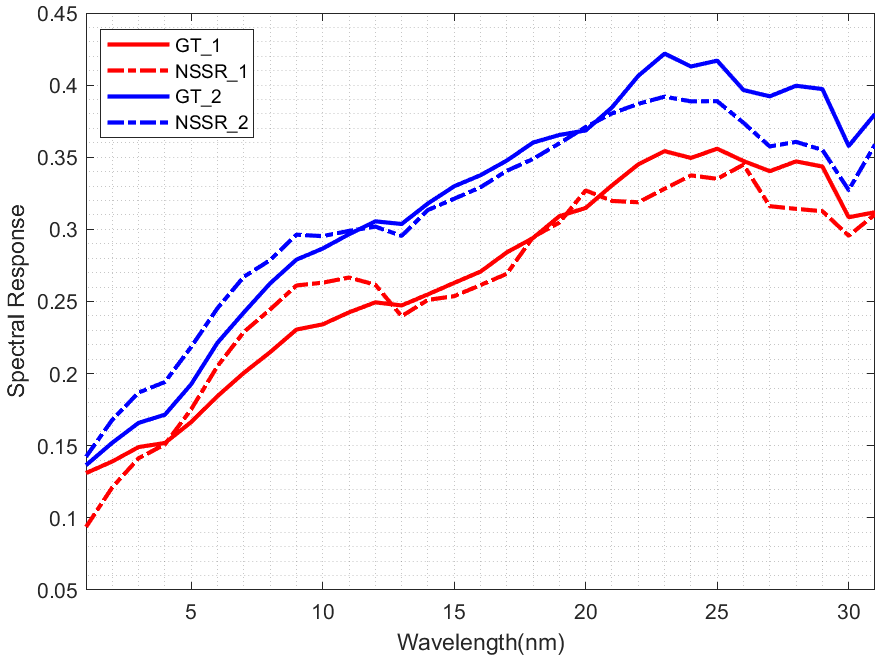}}
		\end{minipage}
		\hspace{1.4cm}
		\begin{minipage}{0.05\textwidth}
			\centerline{\includegraphics[width=0.94in,height=0.8in]{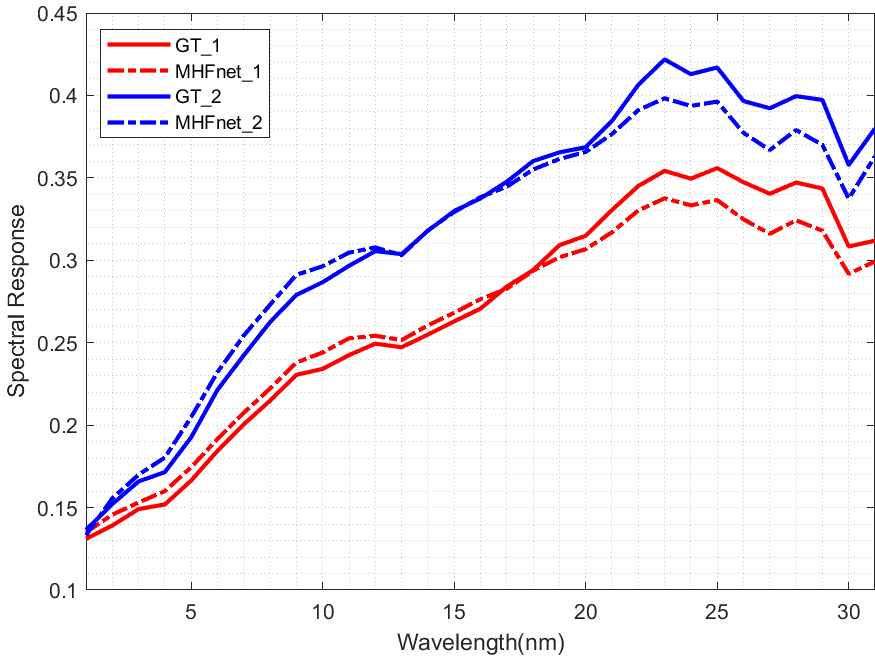}}
		\end{minipage}
		\hspace{1.4cm}
		\begin{minipage}{0.05\textwidth}
			\centerline{\includegraphics[width=0.94in,height=0.8in]{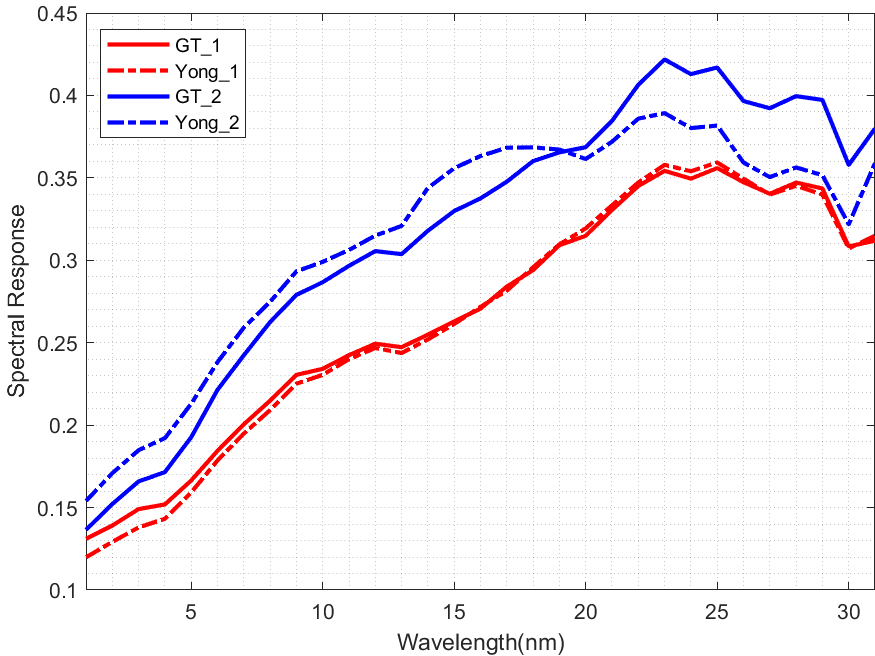}}
		\end{minipage}
		\hspace{1.4cm}
		\begin{minipage}{0.05\textwidth}
			\centerline{\includegraphics[width=0.94in,height=0.8in]{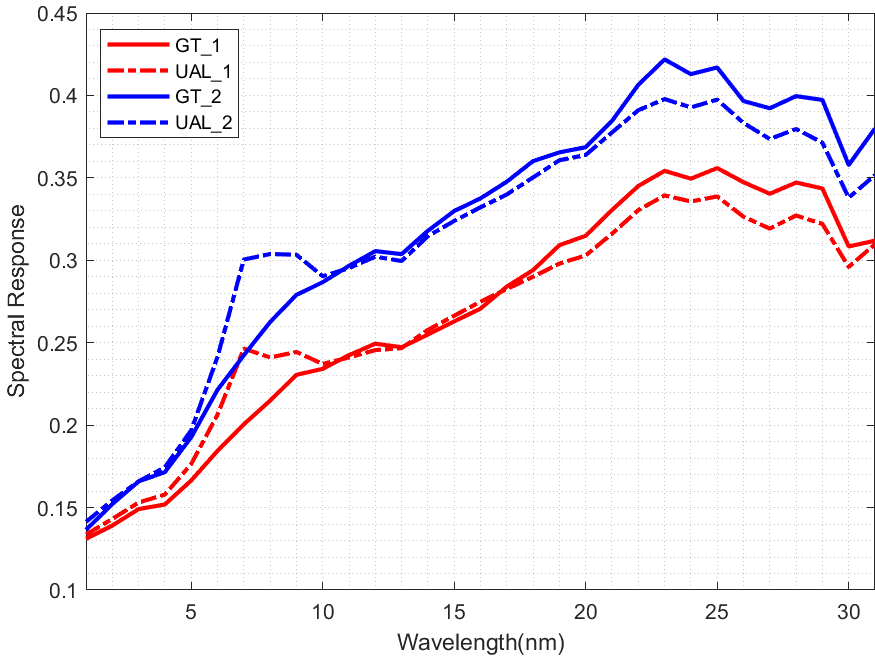}}
		\end{minipage}
		\hspace{1.4cm}
		\begin{minipage}{0.05\textwidth}
			\centerline{\includegraphics[width=0.94in,height=0.8in]{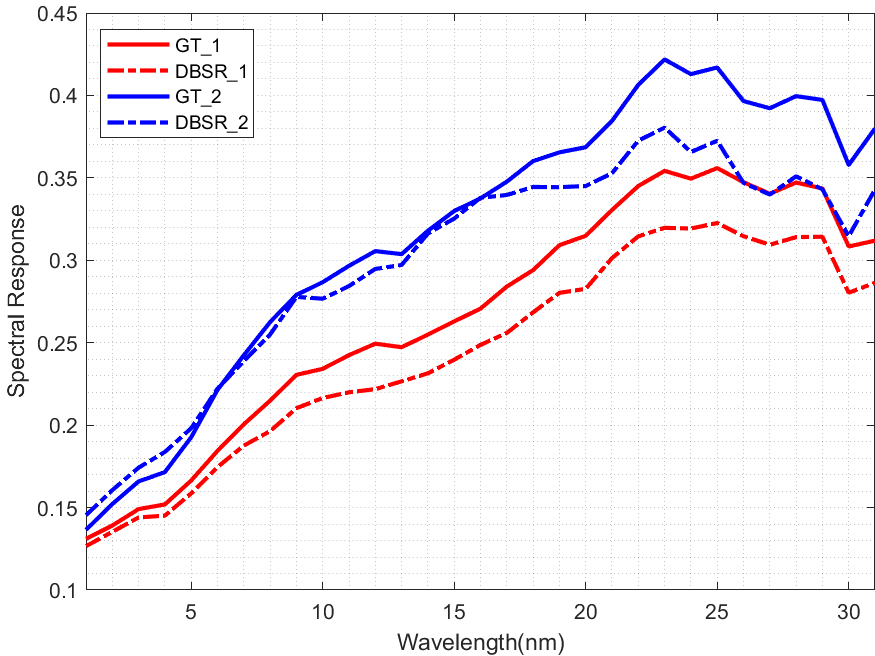}}
		\end{minipage}
		\hspace{1.4cm}
		\begin{minipage}{0.05\textwidth}
			\centerline{\includegraphics[width=0.94in,height=0.8in]{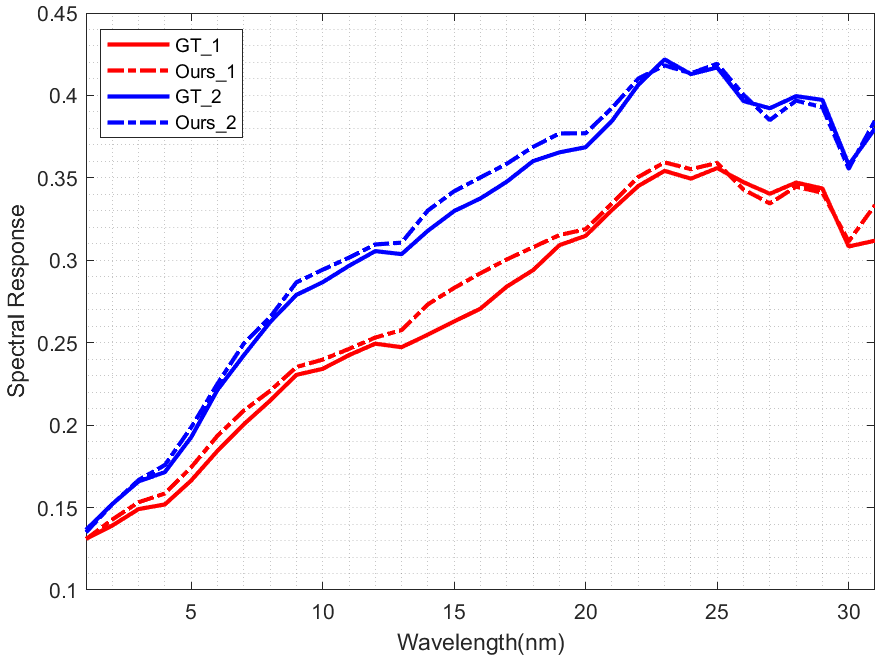}}
		\end{minipage}
		\hspace{0.8cm}
		\begin{minipage}{0.005\textwidth}
			\centerline{\includegraphics[width=0.15in,height=0.8in]{./Picture_Result/ICLV_K1_S32_4040-0.01/BGU_HS_00243_RGB/colorbar_placeholder.png}}
		\end{minipage}
		\vfill
		\hspace{0.55cm}
		\begin{minipage}{0.05\textwidth}
			\centerline{{\scriptsize (a) GT}}
		\end{minipage}
		\hspace{1.4cm}
		\begin{minipage}{0.05\textwidth}
			\centerline{{\scriptsize (b) NSSR~\cite{7438864}}}
		\end{minipage}
		\hspace{1.4cm}
		\begin{minipage}{0.05\textwidth}
			\centerline{{\scriptsize (c) MHFnet~\cite{8953470}}}
		\end{minipage}
		\hspace{1.4cm}
		\begin{minipage}{0.05\textwidth}
			\centerline{{\scriptsize (d) Yong~\cite{8019510}}}
		\end{minipage}
		\hspace{1.4cm}
		\begin{minipage}{0.05\textwidth}
			\centerline{{\scriptsize (e) UAL~\cite{Ours_CVPR2020}}}
		\end{minipage}
		\hspace{1.4cm}
		\begin{minipage}{0.05\textwidth}
			\centerline{{\scriptsize (f) DBSR~\cite{9136736}}}
		\end{minipage}
		\hspace{1.4cm}
		\begin{minipage}{0.05\textwidth}
			\centerline{{\scriptsize (g) Ours}}
		\end{minipage}
		\hspace{0.8cm}
		\begin{minipage}{0.005\textwidth}
			\centerline{}
		\end{minipage}
		\caption{The visual SR results of all methods on the ICLV dataset. The observed LR HSI and HR MSI generated by $\mathbf{k}_1$ and $\mathbf{P}_{0.01}$, respectively. SNRs of both two observed images are 40dB, and the SR scale is 32. }
		\label{Fig_ICLV_Visible}
	\end{figure*}
	
	\noindent \textbf{The experiment results on the NTIRE dataset}\quad
	In this experiment, we further compare the proposed method with the other competing methods on the NTIRE dataset with different SR scales. The observed LR HSI and HR MSI are generated by $\mathbf{k}_1$ and $\mathbf{P}_{0.01}$, respectively. SNRs of both these two observed images are 40dB. We summarize the detailed numerical results in the Table~\ref{Tab_NTIRE_Result_Diff_Scale}. Same as the results on the CAVE and Harvard datasets, the proposed method also has obvious advantage over the other comparison methods. In addition, we also provides the visual results of all methods for further comparison. In Figure~\ref{Fig_ICLV_Visible}, the visual results of the reconstructed HSI and the corresponding spectrum curve are provided. It can be seen that the proposed method obtains better performance than the other comparison methods. The proposed method can well reconstruct the latent HSI in both the spatial and the spectral domains.

	{\small
		\bibliographystyle{ieee_fullname}
		\bibliography{egbib}
	}
	
\end{document}